# Field Theory of Macroeconomics

Heribert Genreith[1], Germany
IFARA Institute[2], Switzerland

# Abstract

In this article we will show that the macro-economy and its growth can be modelled and explained exactly in principle by commonly known field theory from theoretical physics. We will show the main concepts and calculations needed and show that calculation and prediction of economic growth then gets indeed possible in correct "Dollars and Cents". As every field theory it is based on an equation of continuity, which in economic terms means the full balance of all sources and sinks of Capital (Assets) and real Goods (GDP) in the bulk. Uniqueness of field theory of macroeconomics then can be derived from adapting Noether's Theorems, which is based on the notion of invariants to derive unique field equations. We will show that the only assumption which is needed for a self-consistent non-linear macro-economic theory is that the well known Quantity Equation, used in corrected formulation, holds at least locally in time.

For the theoretical and practical calculations shown herein we will use preferably the available time lines of economic data from 1950 – 2010 for the FRG[3]. Germany is a very good usable sample, as it started after the world war in 1948 with introduction of the Deutsche Mark (DM) from a real point zero, both in GDP and Capital. Also until today there was no other war which could have disturbed the data very much, with the exception of the integration of the GDR (German Democratic Republic, DDR) after the end of the cold war in 1990. This event but will give us a very good document on the behavior of economies to external effects, here the effect of population growth.

Another well founded criteria for Germany as a sample economy is the very high quality and completeness of the data in international comparison, regarding GDP and, which is very much important, for the the bulk of any kind of Assets as well. They are provided by the German central bank, the *"Bundesbank"*, and the governmental statistical bureau *"Statistisches Bundesamt"*. Those data are available freely on the internet and will be used for most of the sample calculations needed to check and to control theory against statistical econometric measurements. This article is an advanced extraxt from formerly published articles and books [Genreith[4], 2011,2012; Peetz/Genreith, 2011a,2011b].

---

1      Email: heribert.genreith@t-online.de; homepage: http://genreith.de
2      Ifara.eu : The Institute For Applied Risk Management (IFARA) is a nonprofit, non-partisan, independently funded and interdisciplinary research organization devoted to serving the public. By means of a research-based approach IFARA generates viable, effective public policy responses to important economic problems and financial interdependencies that profoundly can affect our society.
3      FRG means Federal Republic of Germany, the German abbreviation is BRD Bundesrepublik Deutschland. GDR means German Democratic Republic. The German abbreviation is DDR Deutsche Demokratische Republik (1948- 1990).
4      This article is written by a theoretical physicist working on macro-economics since 2009. This must lead to some unavoidable frictions with economists, as field theory and theoretical analysis techniques are not commonly known in economics which is traditionally more leaded by econometrics and statistics. And visa versa a physicists will not be aware of every construction, definition and even wording done in theoretical macro-economy since today. This may sometimes lead to some misunderstandings. I tried to avoid this by "over-explaining" one or another thing mostly in footnotes. So either physicist or economist will find some explanation really dispensable but it is done due to the wide spread of different scientific background of possible readers. I apologize for this inconvenience. For any question or support please contact me.

# Contents

# 1   Definition of the Main Problem

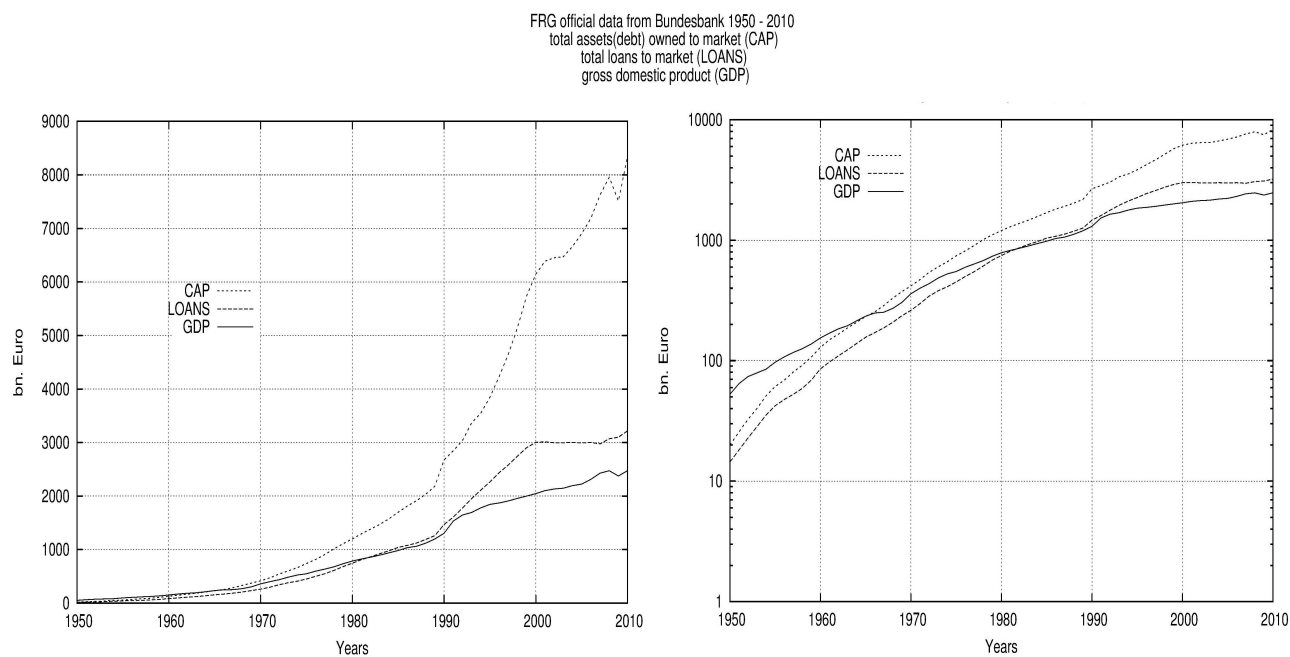

FRG official data from Bundesbank 1950 - 2010
total assets(debt) owned to market (CAP)
total loans to market (LOANS)
gross domestic product (GDP)

*Fig. 1, left side:* GDP and total of financial capital [CAP] stock in the FRG from statistical data of the Bundesbank and the Statistisches Bundesamtes (destatis.de). Data from years 1950 to 2010 in billions of €. Due to the fresh start in 1948 "from point zero", the German economy is a very clean and undisturbed sample. The jump in GDP around 1990 arises from the sudden inclusion of the people of the GDR which added significantly to the population of the FRG.

*Fig. 1, right side:* Same data in logarithmic plot in y-axis. In the logarithmic plot one can see more clearly the fact, that in 1967 the financial capital stock grows beyond 100% of GDP, and from 1982 even the loans to GDP begin to exceed 100% of GDP. While in the linear plot one could think of an exponential growth in CAP but a linear growth in GDP, the logarithmic plot but seemingly shows, that all the functions are exponentials but with in time decreasing slopes. The time-dependent slopes of Capital and Loans both are but unfortunately always steeper than the slope of the GDP.

Fig. 1 illustrates the official main figures of the macro-economy of the FRG from years 1950 to 2010, as there is the whole of all assets, the amount of loans to the real economy and the GDP itself. The evolution proceeds relatively steadily with the first strong step can be seen in 1990. It is the results from the integration of the GDR and its population (13 Mio.) into the FRG. The considerable increase in population produced the offset in financial capital stock and GDP at the same time. The next unusual step (here a drop in slope of capital) can be seen around the year 2000: The strong growth of capital from that time on could no longer be maintained. It was the era of so-called dotcom crash, when the speculation in the then still quite fresh Internet technologies proved to be exaggerated. The next break-in, the Lehman crisis is noticeable from 2008 both in a negative bend of the asset development and the development in the GDP as well.

What is immediately noticeable is the enormous increase in the spread between the development of the totality of all financial capital stocks[5] *K* and the gross domestic product *Y*. The ratio *K/Y* was in 1950 about 0.38, yet it rose to 3.4 in 2010 by a factor of nearly 10. For as we can see in Fig.1 of the real numbers, despite the enormous growth of capital since 1990, no appropriate economic recovery[6] has taken place. On the contrary, growth has actually slowed steadily, and got eventually even negative. Even the enormous financial support in wake of the Lehman crisis could change this situation not sustainable.

---

5     With "financial capital stock" is meant the whole of all assets in the financial industry of an economy. It is not equal to the usual notion on "capital stock" as the sum of fixed assets in real goods like machinery or buildings etc. Here it means only "paper" assets: Time series BBK01.OU0308: Total liabilities / All categories of banks; Assets and liabilities of banks (MFIs) in Germany/Assets/Balance sheet total; at : http://www.bundesbank.de/Navigation/EN/Statistics/Time_series_databases /Macro_economic_time_series/

6     The 1990s boom was an exceptional effect based solely on the sharp increase in the population by the GDR inclusion.

In Fig. 2 we have entered the same numbers in relation to the GDP (left, which means relative GDP=1) and also in relation to the whole of all assets (right, which means relative CAP=1). We can see that in 1967 the total capital stock already exceeded the entire[7] GDP, from the mid-1980s then even the direct use of financial capital stock by loans exceeded 100% of the GDP. In the DotCom hype the absorbed amount of loans but reached a maximum value of nearly 1.5-times the GDP, which is quite amazing. Since then, despite considerable expansion of money supply, it but decreases continuously again to about 125%.

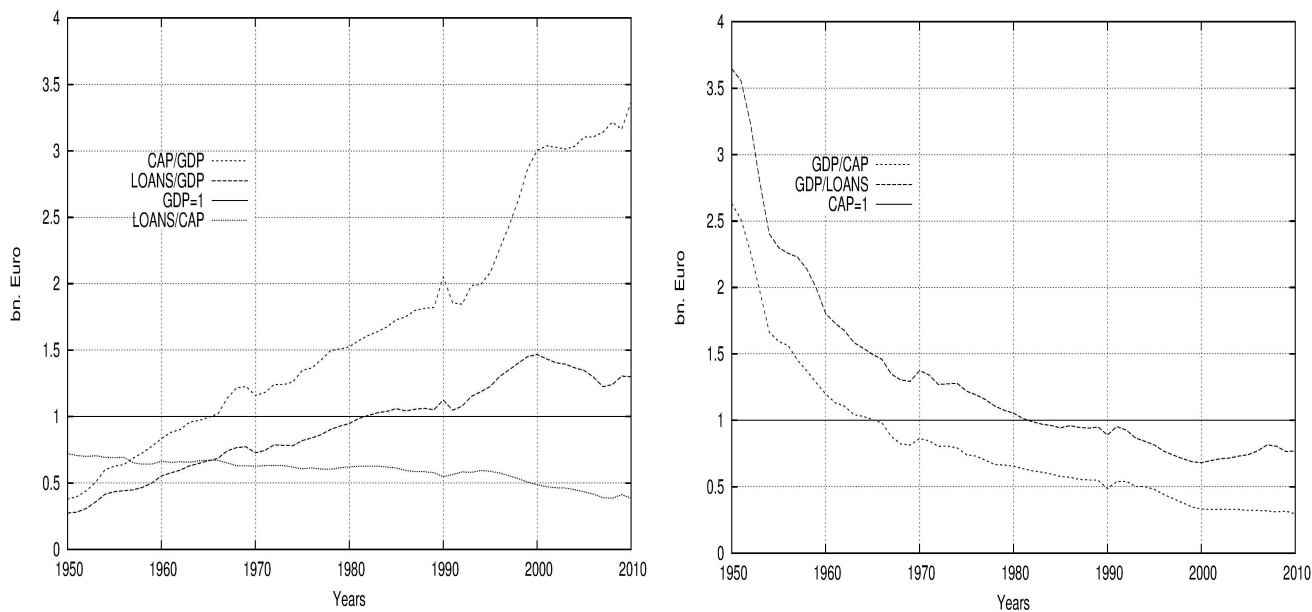

*Fig. 2, left:* Total financial capital ratio (top) and percentage of direct capital investment to GDP (center) in relation to GDP in the Federal Republic of Germany from years 1950 to 2010 according to figures from the Bundesbank. The very lower curve describes just the relationship between the topmost two: The relation of loans to GDP to the total business of the financial institutions. From about 2000, this ratio falls below 0.5 = 50%. For the official tables see OU0308 and OU0115 from Bundesbank (Appendix). *Fig 2, right:* GDP and LOANS now in relation to the total financial capital stock CAP. From the late 1960ies the GDP gets less than CAP, and from the mid 1980ies even less than the summed loans. From the mid 1990ies the GDP gets even lower than 50% of the financial capital stock. Thus the ability to absorb loans and assets decreases dramatically over time.

Remarkably is that classical growth models still predict here a further increase in growth by power of more capital, although this is not the case[8] apart from short-term effects. Important in this context is the concept of capital efficiency, also called capital productivity or *marginal productivity of debt*. This indicates how much new GDP is generated by each monetary unit of new capital. This is the coefficient of the marginal rates

$$k_i := \frac{dY/dt}{dK/dt} \tag{1.1}$$

So we have a look at the real numbers in Figure 3. As we see this coefficient decreases continuously in time. At the beginning of the FRG we have generated an average of more than one euro for every euro of fresh capital. But this capital productivity decreased in 2010 to virtually zero, and even falls further in the medium term. An effect that we can see clearly in the USA and other countries too. Fresh capital is thus de facto in the later time even counterproductive. How can that be? One can thus also study the inverse of the capital productivity, ergo the GDP with respect to productivity of capital:

$$y_i := \frac{dK/dt}{dY/dt} \tag{1.2}$$

This indicates how much additional capital per € GDP growth needs to occur. Because of $dY/dt \to 0$ this coefficient however gets possibly singular, and is therefore running against uncomfortably large numbers up to infinity For each additional € of GDP there has seemingly to be "produced" if not hundreds, but finally thousands or billions of euros of capital.

---

7     Marked 1 = 100 % in Figure 2.
8     Except for short-term cyclical fluctuations without medium-or long-term significance.

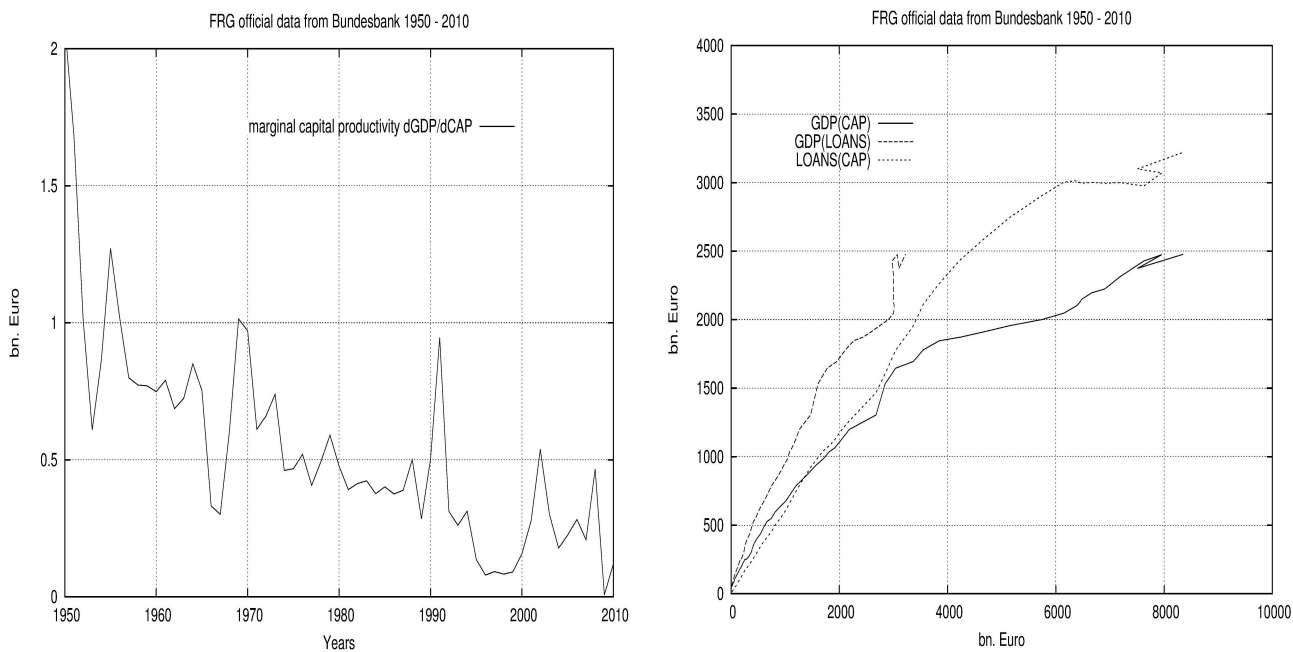

*Fig. 3, left:* The marginal capital productivity $k_i = \dfrac{dY/dt}{dK/dt} = \dfrac{\Delta GDP}{\Delta CAP}$ in the FRG from 1950 to 2010, according to official figures from Bundesbank. At the start of the economy every new Euro of total capital created about one or even more new Euro of GDP, later this margin is decreasing. In the end it comes to zero or eventually gets even less than zero. *Fig. 3, right:* The implicit functions Y(K), Y(L=LOANS), LOANS(K) form official data from Bundesbank. The effectiveness from capital input Y(K) decreases in slope with growing capital and in the end hits a seemingly hidden barrier: All those function get crippled by building up retrograde "crisis slopes" since the 2000ties. The phenomenon of such crisis slopes can be observed in several countries like the EU, USA and Japan too.

This strange-sounding effects of the variables *Y* and *K* are international progressed similarly in a phenomenological way, particularly in the already well-developed western democracies. Another part of the problem are the curves over the period of public debt and inflation. Even this, internationally empirically anywhere to observe phenomena, run off to the seemingly same-shaped verifiable laws. The often represented populist assumption that these hostages of community would be based on incompetent politicians is certainly used far too short, as these phenomena are consistent over all in the world and can be observed regardless of the political or economic orientations of the nations.

## 2  Introduction and Definitions

As a first step, we clarify some definitions of our most important terms used in here.

**Definition**: The *__total financial capital stock__* is the value of K from the U0308 series data of the Federal Bank of the FRG (*Bundesbank*). In this text the amount **K may** sometimes in short just be called *"capital stock"* which should not be confused with the notion of "fixed assets". The Bundesbank notion is *"Time series BBK01.OU0308: Total liabilities / All categories of banks"*.

**Definition**: The term Loans *__total financial capital stock__* is the total amount of money lending from banks to the real economy for any purpose. The Bundesbank notion is *"Time series BBK01.OU0115: Lending to domestic non-banks (non-MFIs) / Total / Including negotiable money market paper, securities, equalisation claims / All categories of banks"*.

**Definition**: The *total capital coefficient* is the ratio

$$k_t(t) := \frac{K(t)}{Y(t)} \quad (2.1)$$

of the sum of all national bank assets *K* to gross domestic product *Y*.

**Definition**: The *total GDP-coefficient* is the reverse ratio

$$y_t(t) := \frac{Y}{K} \quad (2.2)$$

of the gross domestic product *Y* to the sum of all national bank assets *K*.

**Definition**: The *classical capital coefficient* is the ratio of the immediate use of capital in the form of loans into the real economy. The sum of this money can be seen in detail at the Federal Bank data series U0115 and we refer to this aggregate as $M_K$. The term $M_K \equiv L$ is similar to the notion of loans which may sometimes be abreviated with the letter *L*. . So the classical capital coefficient is defined as:

$$k_c(t) := \frac{M_K}{Y} \quad (2.3).$$

**Definition**: We define *the immediate capital investment* into the real economy as the share of total capital stock, which is introduced in a direct manner to the GDP, and was defined above with the abbreviation $M_K$. We define this kind of banks business the *Commercial Banks Business* (CBB).

**Definition**: The *indirect use of capital* is thus the difference from the total sum of all bank assets *K* (OU0308 series) and (OU0115 series), so defined as

$$M_m(t) := K(t) - M_K(t) \quad (2.4)$$

Indirect capital investment means that it is used as an investment vehicle mainly in the interbank market. We define this kind of banks business associated with $M_m$ the *Banks own Business* (BoB) or also as *investment banking*.

**Definition**: The *indirect capital coefficient* is thus defined as

$$k_m(t) := \frac{K - M_K}{Y} \quad (2.5).$$

# 3  Derivation of Basic Equations

Even simple correlations of systems are often already non-linear in nature. This must result at last in a self-consistent model of <u>nonlinear</u> dependencies. The *special field theory of macroeconomics* is for easeness a linear theory in the first step, which means important micro-economic parameters, such as the interests of capital must be approximated from measured values. A self-consistent fully nonlinear model however will be able to explain even these microscopic parameters by itself from very few fundamentals. An introduction into the nonlinear *general field theory* will be given in the second part of this article.

In the special field theory, we but ask first the somehow simpler question:

***"What kind of development in time do the functions of total financial capital stock K and gross domestic product Y take under a given return on investment?"***.

A physical *field* may be in the general case any abstract arbitrary dimensional mathematical space of an observable. For any viable field theory *continuity equations* (balance-equations) are immanent. For this the overall balance[9] of sources and sinks naturally plays an important role. From economic reasoning it is clear, that on the one side that the production in GDP gives reasons to financial wins[10], and also on the other side that finance gives reason for GDP growth e.g. by credits for consumption and investments. Thus we are now looking[11] for two seemingly interdependent functions Y and K.

We will at first develop the main linear equations of motion in an intuitive way, by balancing them correctly, and thus bring them about. We come in, because the two functions must be sought in mutual dependency

$$\frac{dY}{dt} = F(t, Y(t), K(t), p_i(t), ...) \quad \text{and} \quad \frac{dK}{dt} = G(t, Y(t), K(t), q_i(t), ...) \quad .$$

The functions $F$ and $G$ are ad hoc however unknown. What is certain is that the main macroeconomic functions $Y(t)$ and $K(t)$ must be dependent not only on time $t$ but also on each other. In addition they will be dependent on some microscopic secondary parameters $p_i, q_i$ which can be possibly time-dependent functions like e.g. interest percentages and other parameters like population growth and so on. The two differential equations will have the usual basic structure

$$\frac{dY}{dt} = Sources_Y + Sinks_Y \quad \text{and} \quad \frac{dK}{dt} = Sources_K + Sinks_K \quad ,$$

and we may use the secondary micro-economic parameters $p_i, q_i$ for the macro-economic balance accounting. The GDP $Y$ has initially two main propellants: Once the population growth, because a significant cause of GDP growth by itself is out to increase the number of consumers and entrepreneurs $p_B = p_B(t)$, caused by the natural population growth and from migration as well, which possibly can also be negative. Secondly, there are loans for consumption and investment-credits to real economy, that will drive it. The parameter function $p_Y$ is therefore the rate of capital investment into the GDP, described by their share from the total financial capital stock. The principal equation for GDP $Y$ therefore can be found as

$$\frac{dY}{dt} = b_0 + p_B Y + p_Y K \qquad (3.1\ a).$$

Accordingly, the increase on the capital stock is mainly due to savings $p_S$ and on capital returns $p_K$:

$$\frac{dK}{dt} = a_0 + p_S Y + p_K K \qquad (3.1\ b).$$

We add here two free parameters $a_0$, $b_0$ which may arise in <u>open</u> economies through external[12] inputs.

---

9  The required main balance or continuity equation as we shall see, will be given by the long-known economic balance equation $MV = HP$ : the so-called quantity equation (or *Fisher*-equation) of macroeconomy. However, until today this equation in macro-economy is mostly used just as a time-snap rule-of-thumb, we but will have to define this equation finally much more precisely and in a general sence.
10  And sometimes of course gives also reasons for losses. But what counts is that it gives reasons for any changes *d/dt* which can be positive or negative.
11  From pure mathematical reasons it is also clear that to success we need as a minimum requirement *at least a system of two coupled linear independent differential equations* for such a determination.
12  For GDP this could be given e.g. through donations to third-world countries without any return in capital. The parameter $a_0$ are Capital in- and outflows to an economy which are not balanced from an inner trade of any product. Such an outflow or inflow may also produced by self buying states debt or by dpreciation of money by bankruptcy.

$p_S$ is the income that is saved from GDP thus increasing the total financial capital stock. Furthermore it grows with the average interest rate $p_K$. Just from balancing reasons it then must hold

$$p_K = -p_Y := p_n \qquad (3.2)$$

with our new definition of the *net business rate* $p_n$ as the effective excange rate between capital and real economy:

$$\frac{dY}{dt} = b_0(t) + p_B(t)Y - p_n(t)K \qquad (3.3\text{ a})$$

$$\frac{dK}{dt} = a_0(t) + p_S(t)Y + p_n(t)K \qquad (3.3\text{ b}).$$

**Definition:** We will call the function $-p_n$ the net investment ratio or $p_n$ the net (financial) business rate of the banking industry.

This special parameter[13] function must be determined in the next section in more detail. The above differential equation system is an initial[14] value problem.

## 4  Determination of the Net Business Rate

For the formulation of the net business rate $p_n(t)$ in detail, we need a statement about the ratio of investment into the real economy and to the transactions of the financial system to itself. So let us consider again the differential equation (3.3 b) for capital. The coefficient $p_K$ is the effective interest on the total capital *this* year. These interests stems from the two components of direct and indirect use of capital. So we can split it to

$$\frac{dK}{dt} = a_0 + p_S Y + p_{ve} K_E - p_{vr} K_R \quad \text{(with then} \Rightarrow \frac{dY}{dt} = b_0 + p_B Y + p_{vr} K_R - p_{ve} K_E \text{ )} \qquad (4.1)$$

because the share of capital stock which is sold within the financial industry immediately gains returns $p_{ve}$. The proportion which is given initially in the real economy reduces the availability of capital stock this year but receives its returns later, presumable[15] next year. The interest rates $p_{ve} \neq p_{vr}$ however can also be different. With the simple substitution $K_R = K - K_E$ we now get:

$$\frac{dK}{dt} = a_0 + p_S Y + (p_{ve} + p_{vr})K_E - p_{vr} K \qquad (4.2)$$

One may easily operate at least numerically further on with these equations.

But due to the complex analytical derivations needed further on, this equations may be simplified a little bit more, such that the main actors of the economy can be studied from analytical solutions more decise and convenient.

---

13  The choise of the plus and minus sign is mandatory but only a matter of taste. We chose the plus and minus sign here in the sense that a negative $p_n$ means that financial industry distributes "wealth" to GDP and a positive sign means the visa versa.

14  This means, that the values of the functions Y and K are to be determined from their initial values $Y(t_0) = Y_0$ and $K(t_0) = K_0$ and the values for later times $t > t_0$ are computed by adding the integrated change rates. This can happen in the ideal case analytically or if not possible by numerical methods. In numerical procedures but the dependencies of the solutions on the input parameters get lost and can be determined only approximately by laborious variations of the input parameters. Thus analytical procedures are always preferred, because in contrast to numerical methods mathematical analysis will take all the inner dependencies considered further on to the thus derived equations.

15  The small but important difference means that some functional dependencies **must be retarded in time**. It is the analogy to the fact that **fields have limited velocities and take their time to spread over space and react with the observables**.

For a further approximation we may assume that investments into the real economy and in Banks own business don't differ very much in average interest rates. So we can now summarize this flat in one rate by

$$p_K := p_v - p_r \tag{4.3}$$

defined as the difference between the results of expenditure and revenue, with parameter function $p_v$ then as the <u>average nominal interest rate on all types of assets</u>. The interest proportion which fails this year due to loans then shall be $p_r$. This approach will simplify[16] calculations thereafter significantly. For this reason we define the relative value $p_{rel}$ as the share which represents the direct use of capital to the total capital stock as

$$p_{rel} := \frac{M_K}{K} = k_c \tag{4.5}$$

which is identical to the classical capital ratio[17]. Since we have to specify this relative proportion not absolute but as a share of total interest income, we now get the net business ratio $p_n$ in terms of $p_{rel}$ as:

$$p_n := p_v(1 - p_{rel}) - p_v p_{rel} = p_v(1 - 2 p_{rel}) \tag{4.6}$$

If we look at the real data of $p_{rel}$ (Fig. 4), then we see its regular course stemming from the Bundesbank's data. We can start with a simple phenomenological[18] approximation, namely that the share of reinvestment decreases slowly according to a simple exponential form:

$$p_{rel} = \frac{p_{rel0}}{e} \exp\left(-\frac{t - T_h}{T_h}\right) \tag{4.7}$$

This function starts at $t=0$ with the value $p_{rel0}$ and after the exponential half-lifetime[19] $T_h$ it falls off to $1/e = 0.3679$ and then very smoothly goes towards zero. Our formula is therefore justified phenomenologically as

$$p_n = p_{v0}\left(1 - \frac{2 p_{rel0}}{e} \exp\left(\frac{T_h - t}{T_h}\right)\right) \tag{4.8}$$

with

$p_{v0}$ average nominal interest rate over all assets
$p_{rel0}$ Initial direct capital investment as a share of total capital stock
$T_h$ The period after the direct capital investment has dropped to about 1/e=36.7% (exponential half-lifetime).

In the interest of generality and purely <u>phenomenological</u> analytical consideration, it can be assumed further, that at the ultimate beginning of an economy almost 100% of the capital[20] is used for loans into the real economy (Fig. 4 upper fit curve with $p_{rel0} := 1$) giving

---

16      But of course may be calculated in full detail, if very different interest rates apply.
17      According to figures from the Bundesbank (Time series OU0115, loans to domestic non-banks) this value was in 1950, two years after the introduction of the German mark, at about 73%. But it decreased continuously to less than 40% in 2010.
18      "Phenomenological" means that we get the real course of a function due to obvious assumptions. Instead an exponential, you can of course take any other customized functions (eg polynomials) or even take the measurements themselves. For a customized analytic view but a phenomenological function is much more substantial, because the causal link will go along with analytical analysis then. For numerical considerations the kind of the adjustment function however is of lesser importance.
19      The Euler constant is $e \approx 2.7183$. After the Time $T_h$ the ratio drops to $1/e \approx 0.3679$ which is about 37 % percent. The exponential half-lifetime is always greater than the decimal half-lifetime which accounts for a higher level of $1/2 = 0.5 = 50\%$.
20      Thus we can get rid of the start constant $p_{rel0}$. Logically, as the basic problem is an initial value problem, we can lay back the virtual starting point to the time this constant equals unity. For special calculations of course it can be conducted with analytical if wanted.

$$p_n(t) = p_{v0}\left(1 - \frac{2}{e}\exp\left(\frac{T_h - t}{T_h}\right)\right) = p_{v0}\left(1 - 2\exp\left(\frac{T_0 - T}{T_h}\right)\right) \qquad (4.9)$$

with $T_0$ the Date af the start of integration and $T$ the running variable of the actual date of years.

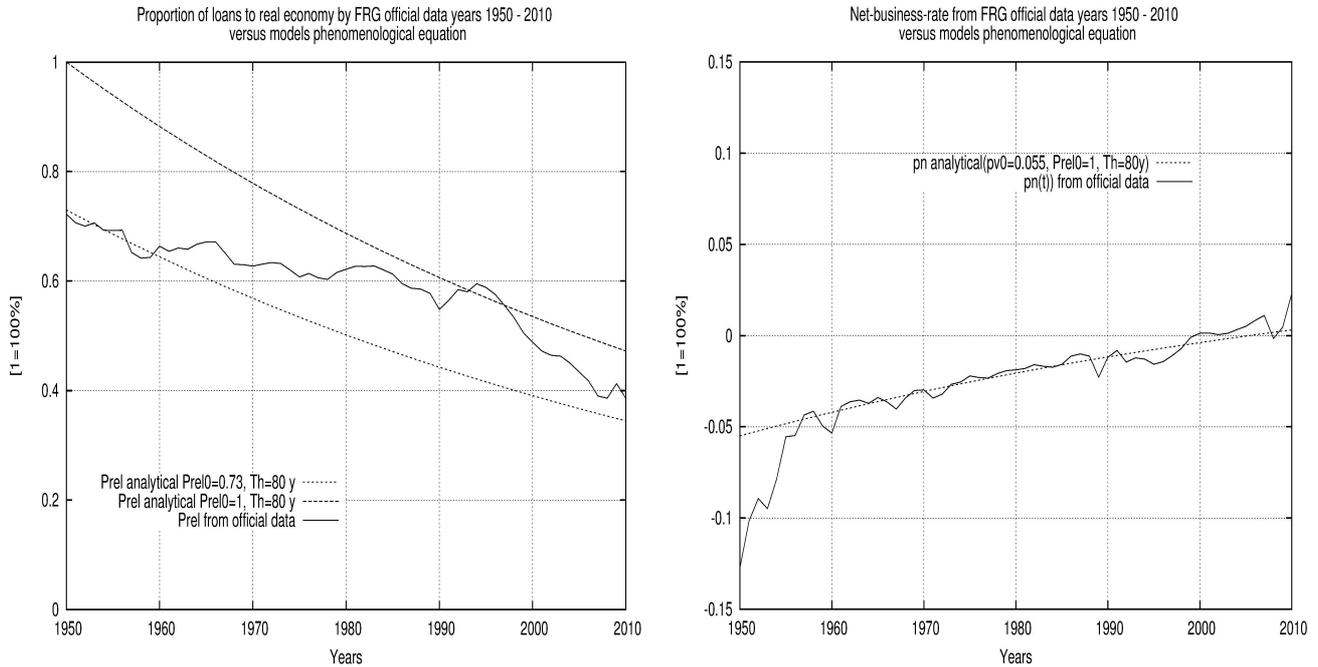

*Fig. 4, left:* The two phenomenological fits compared to the official data of the share of commercial banks business to total business (consumptive and investment-loans, the Bundesbank data series OU0115 relative to the sum over all assets OU0308). *Fig 4, right:* Net business rate official data versus phenomenological formula. The differences at the beginning results from a lot foreign money at the beginning of the FRG, e.g. from the Marshallplan in 1948 to 1953.

For a first analytical demonstration of the effectiveness of the so-accounted system, we make some further simplifications which are generally permitted.

First we assume that the foreign balance is null, and therefore set $a_0 = 0 = b_0$. Furthermore, the relatively low population growth in the FRG will be set to $p_B = 0$ zero. The savings rate in the BRD leveled most of the time around 10% annually, so we can start with a flat constant $p_S = 0.1$ rate. $T_h$ can be easily determined from the real data series and from the same source we can get $p_{v0}$ as the average long-term nominal interest rate over all kinds of assets. Thus we can write now

$$p_n(t) = p_{v0}\left(1 - \frac{2}{e}\exp\left(\frac{T_h - t}{T_h}\right)\right) \qquad (4.10)$$

and we get the <u>most simplified basic ODE system</u> as

$$\frac{dY(t)}{dt} = -p_{v0}\left(1 - \frac{2}{e}\exp\left(\frac{T_h - t}{T_h}\right)\right) K(t) \qquad (4.11\ a)$$

$$\frac{dK(t)}{dt} = p_S Y(t) + p_{v0}\left(1 - \frac{2}{e}\exp\left(\frac{T_h - t}{T_h}\right)\right) K(t) \qquad (4.11\ b)$$

which now only contains the core effects and is therefore suited best to a first analytical examination of the most important relationships.

For a first examination of the found system and in anticipation of some of the coming chapters, we will do now a just numerical integration of the equations.. As we can see, with choosing the constant values $p_S=0.1$, $p_{v0}=0.055$ and $T_h=80$ we get the already very good agreement between theory in its simplest form and reality (Fig. 5).

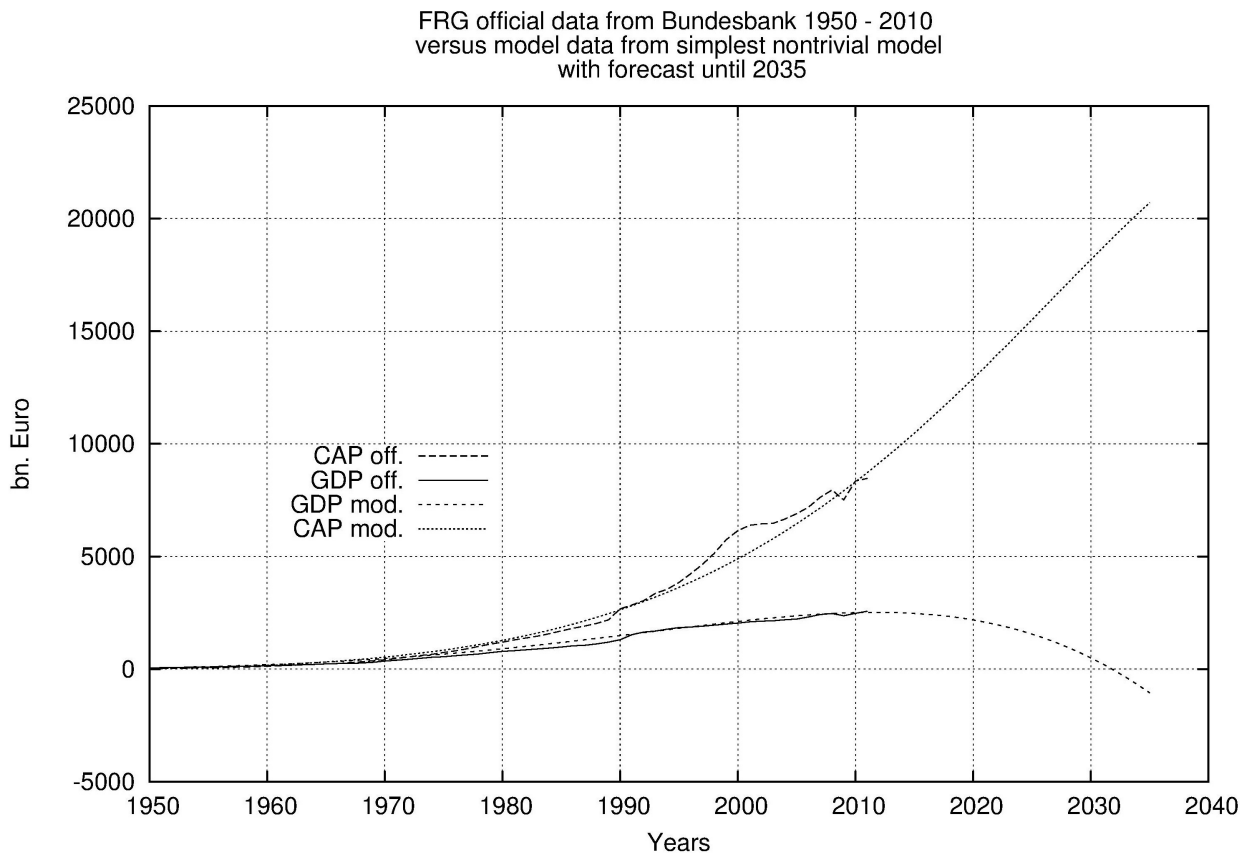

*Fig. 5:* Development of the total financial capital stock CAP and GDP from 1950 to 2010 in the FRG according to official figures. The "buckle" of (abroad) money from 1the 1990ies to 2000ties is due to the so called "DotCom-Bubble", which correspond also to a small underperformance of the real GDP relative to the modeled GDP in this years. The output of our phenomenological basis model until year 2035 is shown as well. Remark: The simplified model is an **initial value problem**, which means it is in full **predictive** given just the starting values Y(T=1950) and K(T=1950) in year 1950.

In contrast to classical growth modelling[21] the prediction[22] of this initial value problem is already very close to reality. The deviations still occuring are at the GDP around 1990, which was the inclusion of the GDR, and a large buckle in assets around 2000, which is due to the socalled DotCom-Bubble which was accompanied by large inflow of external money. The output of the phenomenological basis model (4.11) until year 2035 is shown as well. It predicts, that without affecting the economy by unusual monetary politics (like e.g. the socalled *"Quantitative Easing"* or other externals), the economy would steadily go down until completely vanishing in 2032.

From theory of a single closed nation it is not possible to derive external inputs which are dependent on seemingly[23] unpredictable external decisions. Also such data, especially regarding assets, are not really[24] adressed in today statistical data. In our numerical sample here but we can include at least the official data on population growth, which is an external to GDP itself. In Fig. 6. we then see the effect of the solution for *Y* with and without the official population growth taken into account.

---

[21] Which give for practice just some phenomenological assumptional rules with very limited benefit. Which also showed in recent financial crisis not any warning on the coming crisis condition $dY/dt \approx 0$.

[22] It must be emphasized here that the intergration <u>just needed the starting values</u> of *Y*(1950) and *K*(1950) in the years 1950 and applying then just, with the fixed average interest rate and fixed average savings rate, the phenomenological fitted share of CBB to BoB (smooth exponential).

[23] This issue we will adress later in this article. Externals, at least the most prominent ones, but could be calculated theoretically if one builds up a full system of ODE's for the whole world economy.

[24] As for the effect on the national economy the question on any asset is in principle *"who pays the interest rates?"*, which means are they paid from local market players or from foreigners?

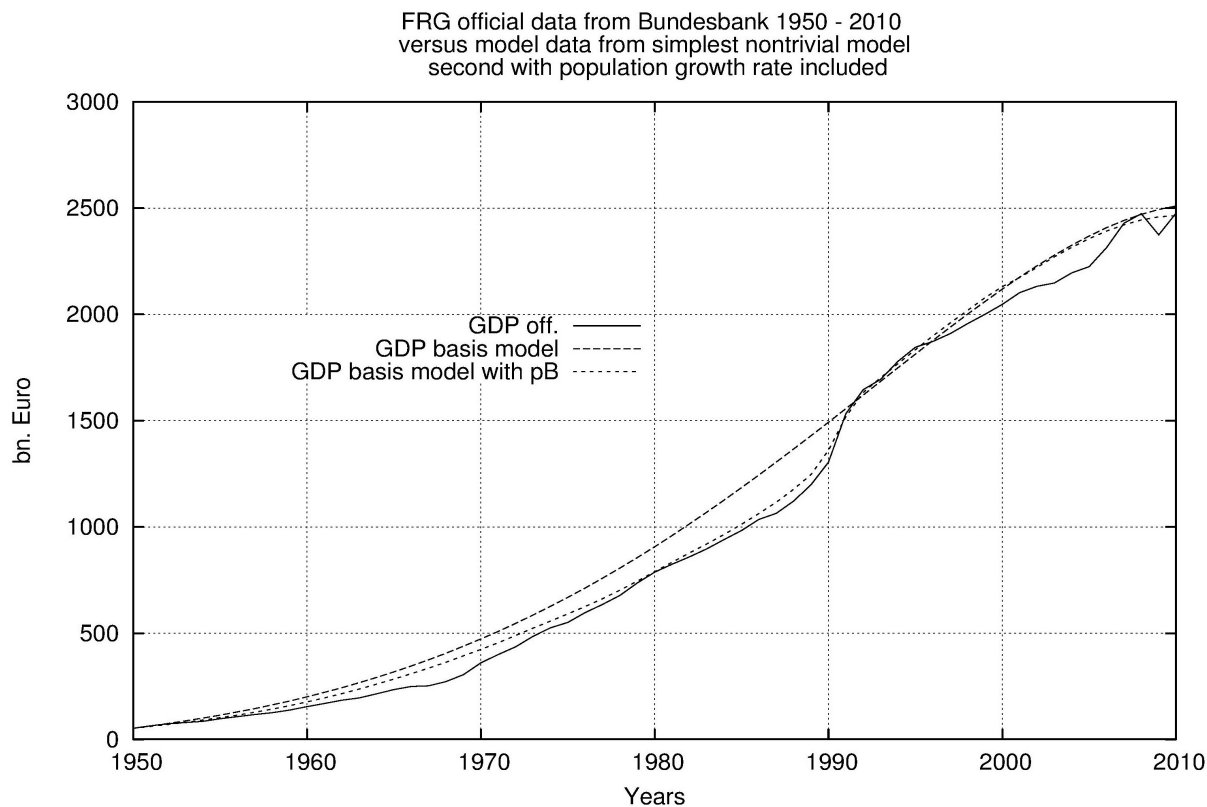

*Fig. 6:* GDP from Bundesbank official data (full line) and model calculations basis model (long dotted line) and the same model including official population growth in comparison (short dotted line). One can see that the sudden population growth due to the inclusion of the GDR to the FRG in 1990 is responsible for the jump in GDP at this time.

This gives already a pretty global well fit between the numerically integrated model GDP and real GDP measured by the Statistisches Bundesamt in Germany. It predicts obvious in Euro and Cents how the inclusion of additional 13 Mio. inhabitants gave reason for the rise in German GDP around 1990. How good the overall fit is, even in the most simplified form of the equations used for integration presented here, it must be considered that the officially measured values we compare with are already affected by a lot of statistical and systematic errors, which arise from different reasons[25]. The last deviations between modeled and measured data from the mid-90ies on are also induced by externals: so the dotCom-Bubble (approximatly 1995-2005) and finally the Lehman-bankruptcy in 2008 which both influenced the German GDP negatively.

## 5 Some Remarks about Units

In mathematical analysis units do not play a big role, but in technical and physical contexts, however, very well. In particular, equations can be checked to see if there apply the same units in a sum[26] and on the left and right hand side of an equation as well. If the balance of the units is not correct, then there is probably an inconsistency in the formula. So we have to control units at some important points. Here we look at the just derived basic equations

$$\frac{dY}{dt} = b_0 + p_B Y + p_Y K \quad \text{and} \quad \frac{dK}{dt} = a_0 + p_S Y + p_K K$$

by writing them down in their pure units, which seemingly do not fit:

$$\frac{Cur/y}{y} = b_0 + \frac{1}{y} \cdot \frac{Cur}{y} + \frac{1}{y} \cdot Cur \tag{5.1 a}$$

$$\frac{Cur}{y} = a_0 + \frac{1}{y} \cdot \frac{Cur}{y} + \frac{1}{y} \cdot Cur \tag{5.1 b}$$

---

25   So not only from unavoidable accounting flaws, needed assumptions on e.g. black markets etc., but also from changing the econometric rules of accounting from time to time since 1950. Witthin this circumstances and limitations, and considering the in this sample used model simplifications (which but are permitted from good reasons), the prediction of the initial(!) value problem is already astonishingly close.

26   In sums this fact is important, but never in products. One has but to imagine that functions such as the exponential functions may be represented as a polynomial sum, so holds $e^x = \sum \frac{x^n}{n!}$ . So the physical unit of the argument will always have to be dimensionless. For example with $x=t$ as time in seconds does not work, as it returns in units $sec + sec^2 + sec^3 + ...$ which can not been added up correctly. Correct therefore is $x = t/T_e$ , which means a dimensionless argument of time *t* in respect to some unit of time $T_e$ . Mostly $T_e$ is chosen as 1 year.

The still vague terms $a_0$ and $b_0$ we do not have to look at more closely, because if they are used they will be defined in the appropriate units. The apparent problem is, however, that GDP generally is in currency units per year *(Cur/y)*, and the capital stock is measured in the contrast as total assets = liabilities in units of currency *(Cur)*. For this we must remember that we always calculate here on an annual basis $\Delta T = 1$, ie years. Therefore, one can write in full:

$$\frac{d \Delta T \cdot Y}{dt} = b_0 + p_B \Delta T \cdot Y + p_Y K \tag{5.2 a}$$

$$\text{and} \quad \frac{dK}{dt} = a_0 + p_S \Delta T \cdot Y + p_K K \tag{5.2 b}$$

Now the units equations are correct in terms of units:

$$\frac{y \cdot Cur/y}{y} = [b_0] + \frac{1}{y} \cdot y \cdot \frac{Cur}{y} + \frac{1}{y} \cdot Cur = \frac{Cur}{y} \tag{5.3 a}$$

$$\text{and} \quad \frac{Cur}{y} = [a_0] + \frac{1}{y} \cdot y \cdot \frac{Cur}{y} + \frac{1}{y} \cdot Cur = \frac{Cur}{y} \tag{5.3 b}$$

Because of the unity $[dt]_J = [dT] = [\Delta T] = 1$ year, one can simply omit the factor on an annual[27] basis. Capital is a stock variable, measured in monetary units, such as € or $. The gross domestic product GDP, however, is meant as a current unit, measured in € or $ per year. In purely formal terms the capital coefficient *K/Y* then is not dimensionless, but has the unit year[28]. Here we need some thoughts regarding the appropriate unit of GDP. Ideally there should be the totality of tradable goods and services in the denominator to have a dimensionless capital coefficient.. That means not only the current GDP *Y* this year, but also the sum of the parts $a_i$ of the past few *i*-years GDP which are not fully depreciated and still trade-able :

$$\hat{Y} = [Y + \sum_i (a_i Y_i)] \Delta T \tag{5.4 ?}$$

But it is now that the calculated annual GDP

$$Y_{official} = \hat{Y} \tag{5.5}$$

is actually already this sum above. This is the fact, because the official actual GDP contains also all of GDP that has been generated by the trading of used items. So, trades about second-hand-automobiles or real estate or the flea market[29], etc. For this reason, the current GDP value is justified as well as a stock to use in units of euros or

---

27  If we, however, compute on a different time basis e.g. monthly, then we will have to introduce such a correction factor if necessary. It should be noted also that the unity of a percentage, eg the interest rate, is not truly percent, but percent per year. So mathematically $1/y$. In the lingo of the general, even among economists, the correct unit is often omitted, as of course is usually assumed an annual basis.
28  Which may be interpreted as the time virtually needed to compensate the whole debt to market by new GDP.
29  In addition, it also contains as an important side note, the proportion of GDP which is introduced by the banking industries itself. Another side note is: The black market is not or just approximately included but is often also a large non-vanishing part of the real economy with the same engagemant as legal actions of trade in macro-economics balance. So estimates of illegal drug sales see there between 5% and 15% of worlds GDP. A rough estimate of black-market activities may be made by the sum of circling pure cash $M_0$, which is in modern times of electronic payments largely used in (semi-)illegal trade and thus not always correctly accounted in the official estimates of national GDP.

## 6 Chain-Correction and Calibration

The integrated values of *Y* and *K*, as well as the real values can be summarized conveniently in the total capital coefficient *K/Y*. As therein inflation is shortened out economic effectiveness is always given by this coefficient[30]. Currency is purely nominal, and the effective value of currencies is not determined by the pure nominal but by what one could possibly buy[31] with it. What we have not considered in the linear[32] basis model in detail until now is the impact of inflation on this nominal monetary expression of *Y* and *K* in units of currency. In the linear model we may compute *Y* and *K* in units of points, as we start with dimensionless numbers

$$Y(t_0) = Y_0 := 1 \quad \text{and} \quad K(t_0) = K_0 := \frac{K_{t_0}}{Y_0} \tag{6.1},$$

which can be extracted from the real numbers, here e.g. for the FRG in 1950, which was indeed

$$K_0 = \frac{19.966\, bn.\, €}{52.582\, bn.\, €} = 0{,}38 \tag{6.2}.$$

Currency points and currency units differ only by a conversion factor. The inflation or chain-correction would be without inflation ( $p_i = 0$ ) a constant (*c=const.*) over time. But since we always have inflation[33] this conversion factor will be a variable in time. As inflation has exactly the same effect on capital and GDP it creates no relative[34] gain or loss in the coefficient $k_t = K/Y$ . Therefore, the currency values of the individual functions *Y* and *K* in the linear model need a chain-correction-function which includes the non-linear[35] effect of inflation[36].

If one wants to calculate such an inflation adjustment from the official empirical data one needs to bear in mind however, that these official values (CPI) are calculated by a variety of different specific group of buyers and their temporally variable commodity baskets, and thus are only approximately valid[37] for the economy over all. The time-varying baskets $P_{WK}$ can be described mathematically as

$$P_{WK_i}(t) = \sum_{ij} h_{ij}(t) a_{ij}(t) P_{ij}(t) \tag{6.3}.$$

This means that the price of the *i-th* commodity basket of shopping for the specific consumer group *i* (eg average income people, entrepreneurs, industries, etc.) is the weighted product price from the sum over all products *j*. Their proportions $a_{ij} \in [0,\ 1]$ however vary seriously according to specific consumer groups between 0 and 1. In addition there is an additional summation factor $h_{ij} \neq 1$ eventually different from unity by the use of hedonic[38] statistical methods.

---

30  Due to the fact that both GDP and financial capital stock CAP are measured in units of currency.
31  At least theoretically. In practice this may not be possible, due to a lack in enough goods to buy.
32  Inflation is a non-linear effect of the economy. We will discuss Inflation further on in this article in more detail.
33  As we will see later, this is an intrinsic symmetry principle and a predictable effect of the economy.
34  This but does the interest rate $p_v$ very effectively as it applies only to money. While the money (or assets or debt) amount steadily grows in time, GDP is effected vica versa with more or less strong and fast depreciation.
35  We will show the non-linearity of inflation later in this article.
36  In the usual official statistics it is done by chaining indices with inflation due to a special reference year which is often the year 2000. We will do monetary corrections starting at the beginning of an integration of our system of differential equations. How to make these handy conversion will be explained here.
37  Another rejection is the sometimes poor quality of official data, as from the 1980s, this statistical figure is also increasingly bent by hedonic methods in some western countries. Using hedonic methods in the national price statistics, the United States is playing a pioneering role. Since the mid-1980s, a hedonic price index e.g. for computers has been implemented and until today there is hedonic quality adjustment methods in use for many other products. E.g including since 1987 for rentals, since 1991 in apparel, since 1993 in multi-family homes, since 1997 in digital telephone systems, and television sets since 1999. To make things worse in addition to that, also the official growth rate of GDP today is also effected by hedonic methods in the US. (citation source: www.destatis.de)

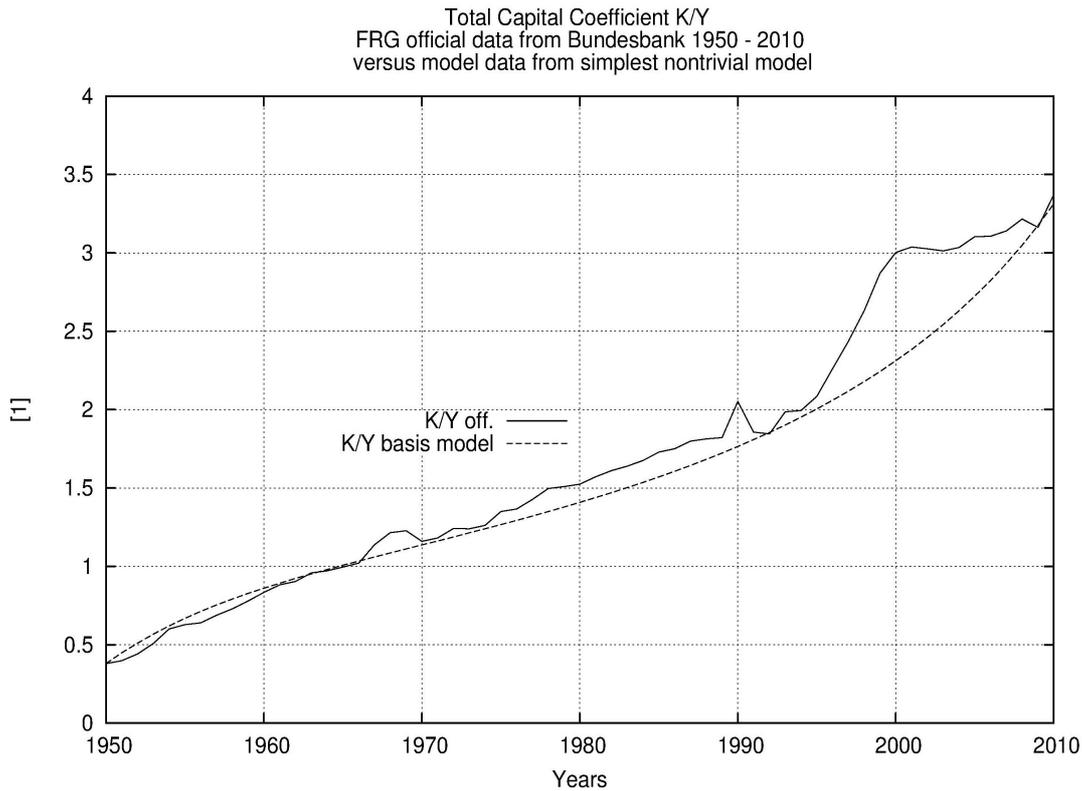

*Fig. 7: Total capital coefficient of the FRG according to official figures compared to the calculations by the simplest basis model. The differences result from the assumption of fixed values, not considering population growth and others and the lack of international balance sheets. Indeed the capital coefficient grows seemingly linearly for a very long time (around 1960-1990) but comes to exponential growth in the end. Overall it is fundamentaly a lying S-shaped curve.*

The relevant for the consumer group *i* calculated inflation rate is then the temporal change in the price of the *i-th* commodity basket, so:

$$I_i := \frac{\dot{P}_{WK_i}}{P_{WK_i}} \qquad (6.4).$$

The importance of the capital coefficient *K/Y* is that it establishes the proportion between the in principle value-free money to the valuable trade-able goods and services. It is also free from the effects of inflation explained, as both the capital and the GDP is calculated in units of currency and inflation cancels conveniently out in this coefficient.

The graphical representation of the single function $k_t(t) = \frac{K}{Y}$ *(2.1)* thus offers the easiest and fastest way to compare different economic data. This applies both for the comparisons with other nations as of course the comparison against the various model calculations.

---

38  This means that an assumed product quality improvements will be adopted with a "better" price resulting from $h_{ij} < 1$ which then in effect will reduce the nominal inflation. All those factors are variable in time and the choice is not given here by international clear rules. In addition in some countries like the USA or UK also the GDP is hedonised. Then accordingly for GDP a $h_{ij} > 1$ is used to include the fictitious quality improvements as an addition to the nominal GDP. Such non-technical but philosophical bendings of the real data may influence the usefulness for economic computing devastatingly.

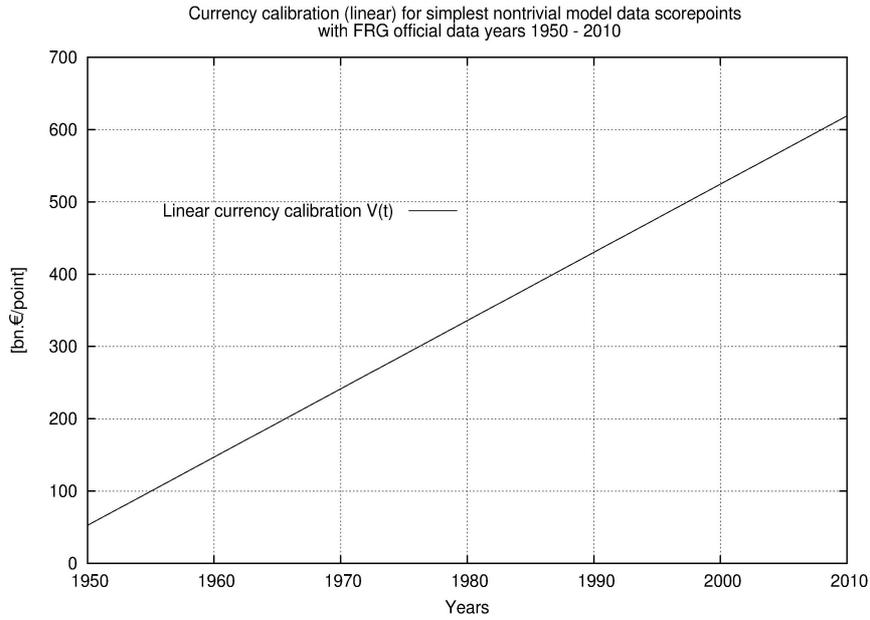

*Fig. 8:* *Linear Currency Inflation calibration from model points to real chained Euro for the FRG. Appropriate for modeling reasons is the linear calibration, since it can not enter any real values through the back door.*

To convert the values calculated in points into actual monetary[39] data, it is necessary to determine the conversion factor. The GDP *Y* is fundamentally affected by its growth $p_w = \dot{Y}$ plus the *nominal* effect of inflation $p_i$ too. The financial capital stock *K* but in addition is effected by the average interest rate $p_v$ too. This last effect[40] is of course fully contained in our basic computation.

Thus the point scores of *Y* and *K* have to be converted into real monetary units by including the additional effect of inflation. There are several possibilities to do this. The most elegant method would be now with the inflation calculated self-consistent analytically. This is indeed possible but from its non-inear nature it is far from trivial, so it will been explained later on at some chapters in this article. The second possibility on hand would be to use the statistical official rate of inflation (CPI), which is derived from the evaluation of a special shopping basket. However, among those already described and unfortunately very serious limitations, such an approach may never give accurate overall results.

Much better for the linear case and model-theoretically permitted is a just simple linear interpolation here. The chain-ratio $V^{ch}(t)$ of score points to nominal currency is determined at the beginning *i* and at the end time *e*:

$$V_i^{ch} := \frac{Y_{Whr}(t_i) + K_{Whr}(t_i)}{Y_P(t_i) + K_P(t_i)} \tag{6.5}$$

$$V_e^{ch} := \frac{Y_{Whr}(t_e) + K_{Whr}(t_e)}{Y_P(t_e) + K_P(t_e)} \tag{6.6}$$

---

39      Currency is purely nominal, whether Dollars, Rubles, Euros or Pounds. Or even points, it makes no difference for computing except the multiplication with a currency conversion function *c*.

40      In classical macroeconomics is, however, distinguish between nominal and real interest rates. The nominal interest rate is calculated correctly by the model. The so-called real interest rate, however, is what remains after deducting the inflation determined statistically from the shopping basket. Although the amount of the real interest rate is important for savers and investors, however for macroeconomics only the nominal rate does matter. The real interest rate is just a derived number from two separate calculable effects. The inflation rates calculated by the statistical institutes using shopping baskets data are already in barely comprehensible manner changed data. They are more or less good "house numbers", but largely useless for exact analytical computations. You may try to calculate in return the raw data from already statistically processed data. However, this is usually only approximately possible because the algorithms used by the statistical institutes usually are not bijective.

of the integration.
Now one can linearly interpolates the intermediate values with:

$$V^{ch}(t) = V_i^{ch} + \frac{V_e^{ch} - V_i^{ch}}{t_e - t_i} \cdot (t - t_i) \qquad (6.7)$$

The model values in currency units are obtained then as

$$Y_{Mod}(t) = Y_P(t) \cdot V^{ch}(t) \qquad (6.8\ a)$$
$$K_{Mod}(t) = K_P(t) \cdot V^{ch}(t) \qquad (6.8\ b)$$

which are now functions with the same linear conversion factor provided both to *Y* and *K*. This linear approximation will do very well for the demanded fit.

# 7 Analysis of the Basic Equations

Following our numerical investigation we will now do the epistemological much more demanding analytical one. The most fundamental basic system

$$\frac{dY(t)}{dt} = -p_v (1 - \frac{2}{e} \exp(\frac{T_h - t}{T_h})) K(t) \qquad (7.1\ a)$$

$$\frac{dK(t)}{dt} = p_S Y(t) + p_v (1 - \frac{2}{e} \exp(\frac{T_h - t}{T_h})) K(t) \qquad (7.1\ b)$$

allows at hand of its structure already some analytical statements. We consider in the following the idealized situation of a closed economy, i.e. the foreign terms $a_0 \approx 0 \approx b_0$ are assumed zero or close to zero. So obviously, GDP growth stops if $dY/dt = 0$ and capital growth at $dK/dt = 0$. Further on, GDP growth stops when at least one of the three right hand side factors are zero:

$$p_v = 0 \quad \text{or} \quad K = 0 \quad \text{or} \quad \frac{e}{2} = \exp(\frac{T_h - t}{T_h}) \qquad (7.2).$$

The case *K=0* is trivial, the case $p_v = 0$, however, is a little surprising at first glance. It means that the flow of credit dries up, if the return on investment is zero. It is of course a self-explaining fact from the view of an investor. From the last factor, we can now estimate the time $t_{max}$, when the GDP growth stops non trivially:

$$\ln \frac{e}{2} = \frac{T_h - t_{max}}{T_h} \Leftrightarrow t_{max} = T_h (1 - \ln \frac{e}{2}) = 0,693 \cdot T_h \qquad (7.3).$$

With the realistic choice $T_h = 80$ for the FRG this results in a critical evolutionary time of

$$t_{max} \simeq 55 \quad \text{years.} \qquad (7.4)$$

After this time GDP in the medium term falls ever[41] further. An effect which now shows[42] up in developed economies like the U.S. or the EU as well. It can be mitigated, however, due to the effect of population growth

---

[41] GDP then will fall ever on if the development is left on its own without unusual monetary politics.
[42] If we start from 1950 this means year 2005 for the FRG. This is indeed very close to the data in reality: The crisis began with the End of the DotCom-Hype in 2001 first not very much recognized. It grew up with the real estate bubble peaked mid 2006 (summarize maybe found at http://en.wikipedia.org/wiki/Subprime_mortgage_crisis ) in the USA and culminated into the Lehman bankruptcy in late 2008 followed by still unsolved financial crisis in Dollar and Euro.

since then the additive term can feed continuing growth:

$$\frac{dY(t)}{dt} = p_B Y(t) - p_v(1 - \frac{2}{e}\exp(\frac{T_h - t}{T_h})) K(t) \quad (7.5)$$

and from that rules:

$$p_B = -p_v(1 - \frac{2}{e}\exp(\frac{T_h - t}{T_h})) \frac{K(t)}{Y(t)} \quad (7.6).$$

Population growth is a phenomenon that but can be controlled[43] very badly through policy measures. On the capital side, it looks a little better. There does the savings rate obviously the same task as population growth in GDP:

$$p_S = -p_v(1 - \frac{2}{e}\exp(\frac{T_h - t}{T_h})) \frac{K(t)}{Y(t)} \quad (7.7)$$

Finally, we take a look at the complete system of equations:

$$\frac{dY(t)}{dt} = b_0 + p_B Y(t) - p_v(1 - \frac{2}{e}\exp(\frac{T_h - t}{T_h})) K(t) \quad (7.8a)$$

$$\frac{dK(t)}{dt} = a_0 + p_S Y(t) + p_v(1 - \frac{2}{e}\exp(\frac{T_h - t}{T_h})) K(t) \quad (7.8b)$$

The foreign contributions $a_0$ and $b_0$ can therefore be used at any time to compensate for deficits. Because when considering the case $p_B \approx 0 \approx p_S$, then $b_0$ takes over the role of $p_B$ and $a_0$ takes over the role of $p_S$. Due to the complex mutual dependencies of all parameters, the possibilities of correction in practice are unfortunately not so simple.

From the equations (8.6) it follows, that for the debt burden *dK/dt* it should hold:

$$p_S Y(t) = -p_v(1 - \frac{2}{e}\exp(\frac{T_h - t}{T_h})) K(t) = p_v(t) K(t) = \frac{dK(t)}{dt} \quad \Rightarrow \quad \dot{K}! \approx p_S Y \quad (7.9)$$

and by using the official data for the FRG we may see, that this indeed is the case in all the "good" years, until financialisation of the 1990ies starts to undermine this balanced situation. Until the mid-90ies the debt burden was always balanced by national savings $\dot{K} \approx p_S Y$.

But then after debt burden got up to 2.5 times the savings which resulted in crisis. The buckle around the 90ies is pointed out, as it is just a statistical effect of the GDR inclusion. Only depreciating and expelling capital abroad saved German GDP in the past Lehman years. Balance but is not given anymore after the mid-90ies and will presumably also never come back without unusual monetary politics.

From (8.5) but it also follows that $\dot{K} \approx p_B Y$ also a strong population growth can compensate for the growing debt burden. Thus in times of crisis a migration can balance the economic situation

$$\dot{K}(t)! \approx [p_S(t) + p_B(t)] \cdot Y(t) \quad (7.10)$$

at least for some additional time. In words we can say that the debt burden overall should be always balanced by the combined effect of savings plus migration/population growth to come not into crisis. The small effective population growth in Germany but does obviously not give rise to such a compensation, while in other countries, e.g. the US, this compensation effect may be much stronger.

---

[43] As one can see a stable population was important in the beginning, and the actually slightly declining population of the German could never hurt much GDP growth. After the onset of the crisis years, however, only a significant population growth could avoid decline in the GDP in the medium-term. Especially for the USA the migration of people to the states is an important part of the US-growth. In the last decade the US got about 2 Million new inhabitants per year just out of migrations.

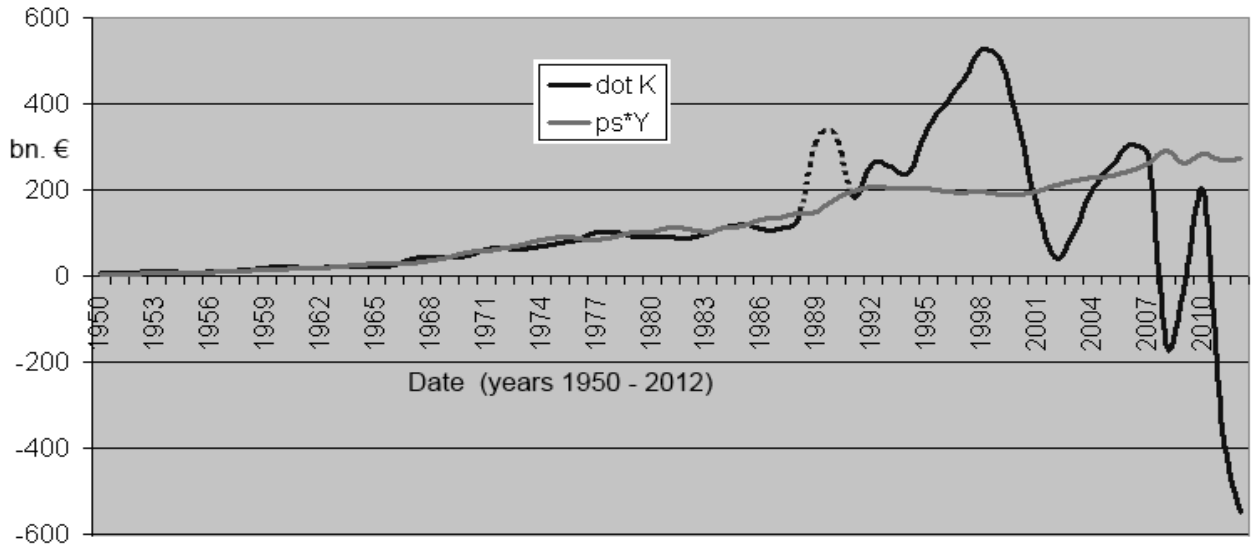

*Fig.9: The debt burden dK/dt of the FRG in relation to its savings $p_S Y$ from year 1950 to 2012*

Including foreign contribution (open system) it follows from (8.7) that for open system balance holds

$$\dot{K} ! \approx (p_S + p_B) \cdot Y + b_0 - a_0 \qquad (7.11)$$

where all functions and parameters are functions of time in principle. It says, that a GDP donation strengthens the compensation power, but foreign money coming in means that it weakens the compensation power of the peoples savings. This but means that in any financial crisis an actually unbalanced debt burden can eventually be fixed by foreign (e.g. food-)donations $b_0$ inflow or foreign assets $a_0$ outflow.

Assuming for simplicity the usual $b_0 = 0$ and $p_B \approx 0$ (which means $\dot{K} + a_0 \approx p_S \cdot Y$) we come for the amount of needed outflow/inflow

$$a_0^{Inflow} \approx p_S \cdot Y - \dot{K} \ \wedge \ \dot{K} \leq p_S Y \quad \Leftrightarrow \quad a_0^{Outflow} \approx p_S \cdot Y - \dot{K} \ \wedge \ \dot{K} > p_S Y$$

(7.12)

of assets-burden. So we will need foreign assets to come into a country if the savings quota is to large and an outflow (e.g. selling[44] to abroad) of assets if savings are to small for the overall balance.

After this introductory treatment of the structure of our system of differential equations, we now come to the necessary integration[45]. The analytical integration even easier systems of differential equations are often not trivial. So already our simplified model can not been integrated in closed form.

In the alternative but it can be integrated analytical at least piecewise to gain the information of the analytical behavior of the solution functions. Thus we may regard $p_n$ as piecewise constant, where the integration runs only over a limited number of years. It shows out that, depending on whether the frequency expression generated

$$\Phi := p_n(4p_s - p_n) \tag{7.13}$$

is positive, negative or zero, different solutions must be discussed within their respective local scope. The following **Basis-Solution in uninflated Currency-terms** can be found with the help of an algebraic analytical solving software:

$$Y(t) = \begin{cases} [Y_0 \cosh(\frac{\sqrt{-\Phi}}{2}t) - p_n \frac{(Y_0 + 2K_0)}{\sqrt{-\Phi}} \sinh(\frac{\sqrt{-\Phi}}{2}t)] \exp(\frac{p_n}{2}t) & \wedge \ \Phi < 0 \\ [Y_0 - p_n(\frac{Y_0}{2} + K_0)t] \exp(\frac{p_n}{2}t) & \wedge \ \Phi = 0 \\ [Y_0 \cos(\frac{\sqrt{\Phi}}{2}t) - p_n \frac{(Y_0 + 2K_0)}{\sqrt{\Phi}} \sin(\frac{\sqrt{\Phi}}{2}t)] \exp(\frac{p_n}{2}t) & \wedge \ \Phi > 0 \end{cases} \tag{7.14a}$$

$$K(t) = \begin{cases} [K_0 \cosh(\frac{\sqrt{-\Phi}}{2}t) + \frac{2p_s Y_0 + p_n K_0}{\sqrt{-\Phi}} \sinh(\frac{\sqrt{-\Phi}}{2}t)] \exp(\frac{p_n}{2}t) & \wedge \ \Phi < 0 \\ [K_0 + (p_s Y_0 + \frac{p_n}{2} K_0)t] \exp(\frac{p_n}{2}t) & \wedge \ \Phi = 0 \\ [K_0 \cos(\frac{\sqrt{\Phi}}{2}t) + \frac{2p_s Y_0 + p_n K_0}{\sqrt{\Phi}} \sin(\frac{\sqrt{\Phi}}{2}t)] \exp(\frac{p_n}{2}t) & \wedge \ \Phi > 0 \end{cases} \tag{7.14b}$$

---

44        Indeed for the US population growth and selling debt give reason to the today still existent (small) growth.
45        With numerical methods this is not a significant problem, however, there will be lost inevitably important analytical relationships. This can only be avoided by the analytic integration. Analytical solutions are so important because they only allow universally valid quantitative results. Analytical programs can save days on computing to try the "crack" of the integrals. As a tool for integration the computer algebra system Maxima version 5.22.1 (or http:wxmaxima.sf.net http:maxima.sf.net) was used, which is freely available on the Internet under the GNU GPL license. As a fee-based program Maple 9 was used to.

The $\Phi$ function has the dimension of a frequency and is crucial for the kind of growth. The growth rate $\Phi := p_n(4p_s - p_n)$ is zero for $p_n = 0$ or $p_n = 4p_s$. Therefore, we need to make at this point a further case distinction:

$$Y(t) = \begin{cases} Y_0 \exp(\frac{p_n}{2}t) & \wedge \ \Phi = 0 \wedge p_n = 0 \\ [Y_0 - (2p_s Y_0 + 4p_s K_0)t] \exp(\frac{p_n}{2}t) & \wedge \ \Phi = 0 \wedge p_n = 4p_s \end{cases} \tag{7.15a}$$

$$K(t) = \begin{cases} [K_0 + p_s Y_0 t] \exp(\frac{p_n}{2}t) & \wedge \ \Phi = 0 \wedge p_n = 0 \\ [K_0 + (p_s Y_0 + 2p_s K_0)t] \exp(\frac{p_n}{2}t) & \wedge \ \Phi = 0 \wedge p_n = 4p_s \end{cases} \tag{7.15b}$$

As any currency is just an arbitrary scaling factor, the formulas can be written with the <u>normalized specific growth factors</u>[46]

$$\delta Y := Y_0 \exp(\frac{p_n}{2}t) \quad \text{and} \quad \delta K := K_0 \exp(\frac{p_n}{2}t) \tag{7.16}$$

from which now are following the specific forms

---

46  Which are in principle nothing else than the solutions from the oversimplified IWF 2005 growth model $dY/dt = gY$ and $dK/dt = gK$ with $g := p_n/2$. The IWF 2005 thus in principle is close to the quasi-balanced case with $\Phi = 0$ (no crisis any time) and $p_n = 0$ (perfect balance between both banking models CBB and BoB).

$$\frac{Y(t)}{\delta Y}=1 \quad \text{and} \quad \frac{K(t)}{\delta K}=1+p_s\frac{Y_0}{K_0}t \quad \text{for} \quad p_n=0 \wedge \Phi=0 \tag{7.17}$$

or

$$\frac{Y(t)}{\delta Y}=1-2p_s(1+\frac{2K_0}{Y_0})t \quad \text{and}$$

$$\frac{K(t)}{\delta K}=1+2p_s(1+\frac{Y_0}{2K_0})t \quad \text{for} \quad p_n=4p_s \wedge \Phi=0 \tag{7.18}.$$

When $p_n=0$ gets zero the upper turning point of the GDP development is reached. At beginning of the national economy, the "good times", $p_n$ is negative, and after the reversal point it runs into the positive range ("crisis times"). The results in specific terms are:

$$\frac{Y(t)}{\delta Y}=\begin{cases} \cosh(\frac{\sqrt{-\Phi}}{2}t)-p_n\frac{1+2\frac{K_0}{Y_0}}{\sqrt{-\Phi}}\sinh(\frac{\sqrt{-\Phi}}{2}t) & \wedge \; \Phi>0 \\ \cos(\frac{\sqrt{\Phi}}{2}t)-p_n\frac{1+2\frac{K_0}{Y_0}}{\sqrt{\Phi}}\sin(\frac{\sqrt{\Phi}}{2}t) & \wedge \; \Phi<0 \end{cases} \tag{7.19a}$$

$$\frac{K(t)}{\delta K}=\begin{cases} \cosh(\frac{\sqrt{-\Phi}}{2}t)+p_n\frac{1+2\frac{p_s}{p_n}\frac{Y_0}{K_0}}{\sqrt{-\Phi}}\sinh(\frac{\sqrt{-\Phi}}{2}t) & \wedge \; \Phi<0 \\ \cos(\frac{\sqrt{\Phi}}{2}t)+p_n\frac{1+2\frac{p_s}{p_n}\frac{Y_0}{K_0}}{\sqrt{\Phi}}\sin(\frac{\sqrt{\Phi}}{2}t) & \wedge \; \Phi>0 \end{cases} \tag{7.19b}$$

To clarify the principle structure of the derived formula we may use the abreviations

$$\alpha_0:=\frac{Y_0}{K_0}\equiv y_t(T_0) \;;\; \beta_S:=\frac{p_S}{p_n} \quad \Rightarrow \quad \alpha_0^{-1}=\frac{K_0}{Y_0}\equiv k_t(T_0):=A_0 \;;\; \Phi\equiv p_n^2(4\beta_S-1)$$

$$\alpha:=1+\frac{2}{\Phi}A_0 \;;\; \beta:=1+\frac{2}{\Phi}\beta_S\alpha_0 \;;\; \gamma:=\frac{1}{2}+A_0 \;;\; \eta:=\frac{1}{2}+\frac{1}{4}\alpha_0 \tag{7.20}$$

and thus **finally we can write using the specific growth forms** $\Delta_S(f):=f(t)/\delta f$ :

$$\Delta_S Y(t)=\begin{cases} \cosh(\frac{\sqrt{-\Phi}}{2}t)-\alpha\, p_n\sinh(\frac{\sqrt{-\Phi}}{2}t) & \wedge \; \Phi<0 \\ \cos(\frac{\sqrt{\Phi}}{2}t)-\alpha\, p_n\sin(\frac{\sqrt{\Phi}}{2}t) & \wedge \; \Phi>0 \\ [(1 \;\wedge\; p_n=0)\; \vee\; (1-\gamma p_s t \;\wedge\; \beta_S=\frac{1}{4})] & \wedge\,\Phi=0 \end{cases} \tag{7.21a}$$

$$\Delta_S K(t) = \begin{cases} \cosh\left(\frac{\sqrt{-\Phi}}{2}t\right) + \beta\, p_n \sinh\left(\frac{\sqrt{-\Phi}}{2}t\right) & \wedge\ \Phi < 0 \\ \cos\left(\frac{\sqrt{\Phi}}{2}t\right) + \beta\, p_n \sin\left(\frac{\sqrt{\Phi}}{2}t\right) & \wedge\ \Phi > 0 \\ [(1+\alpha_0 p_s t\ \wedge\ p_n = 0)\ \vee\ (1+\eta p_s t\ \wedge\ \beta_S = \frac{1}{4})] & \wedge\ \Phi = 0 \end{cases} \qquad (7.21b)$$

or in just other terms[47]:

$$\Delta_S Y(t) = \begin{cases} \cosh\left(|p_n|\frac{\sqrt{1-4\beta_S}}{2}t\right) - \alpha\, p_n \sinh\left(|p_n|\frac{\sqrt{1-4\beta_S}}{2}t\right) & \wedge\ \beta_S < \frac{1}{4} \\ \cos\left(|p_n|\frac{\sqrt{4\beta_S-1}}{2}t\right) - \alpha\, p_n \sin\left(|p_n|\frac{\sqrt{4\beta_S-1}}{2}t\right) & \wedge\ \beta_S > \frac{1}{4} \\ 1-\gamma p_s t & \wedge\ (p_n \neq 0 \wedge \beta_S = \frac{1}{4}) \\ 1 & \wedge\ (p_n = 0 \wedge \beta_S = \frac{1}{4}) \end{cases} \qquad (7.22a)$$

$$\Delta_S K(t) = \begin{cases} \cosh\left(|p_n|\frac{\sqrt{1-4\beta_S}}{2}t\right) + \beta\, p_n \sinh\left(|p_n|\frac{\sqrt{1-4\beta_S}}{2}t\right) & \wedge\ \beta_S < \frac{1}{4} \\ \cos\left(|p_n|\frac{\sqrt{4\beta_S-1}}{2}t\right) + \beta\, p_n \sin\left(|p_n|\frac{\sqrt{4\beta_S-1}}{2}t\right) & \wedge\ \beta_S > \frac{1}{4} \\ 1+\eta p_s t & \wedge\ (p_n \neq 0 \wedge \beta_S = \frac{1}{4}) \\ 1+\alpha_0 p_s t & \wedge\ (p_n = 0 \wedge \beta_S = \frac{1}{4}) \end{cases} \qquad (7.22b)$$

With the abbreviation $\omega_\Delta := |p_n|\frac{\sqrt{1-4\beta_S}}{2}$ we can now write for the total capital coefficient

$$k_t = \frac{\Delta_S K(t)}{\Delta_S Y(t)} = \begin{cases} \dfrac{\cosh(\omega_\Delta t) - \alpha\, p_n \sinh(\omega_\Delta t)}{\cosh(\omega_\Delta t) + \beta\, p_n \sinh(\omega_\Delta t)} & \wedge\ \beta_S < \frac{1}{4} \\ \dfrac{\cos(-\omega_\Delta t) - \alpha\, p_n \sin(-\omega_\Delta t)}{\cos(-\omega_\Delta t) + \beta\, p_n \sin(-\omega_\Delta t)} & \wedge\ \beta_S > \frac{1}{4} \\ \dfrac{1-\gamma p_s t}{1+\eta p_s t} & \wedge\ (p_n \neq 0 \wedge \beta_S = \frac{1}{4}) \\ \dfrac{1}{1+\alpha_0 p_S t} & \wedge\ (p_n = 0 \wedge \beta_S = \frac{1}{4}) \end{cases} \qquad (7.23)$$

usable for further analysis.

---

47  Which can be seen as the correction factors to the solutions of the oversimplified IWF 2005 growth model.

# 8 Discussion of the Analytical Solution

The piece-wise solutions of the simplest model equations give already a lot of approaches for a comprehensive discussion of the principle behavior of a credit-driven economy. First, we compare the properly integrated piecewise solution to the IMF 2005 formula[48]. Exemplary of the IWF-calculated GDP $Y_{IWF}$ at the beginning of the development ( $\Phi<0$ ) is:

$$Y_{IWF} = Y_0 \exp(g\,t) \tag{8.1}$$

We see that the analytically singular IMF 2005 model promises a constant growth rate *g*. The IMF model usually sets for *g* the general growth of the capital stock and thus provides a general exponential growth for Capital and GDP at the same growth rate. In fact but from (8.9a) it is a combination of exponential and harmonic growth

$$Y(t) = [Y_0 \cosh(\frac{\sqrt{-\Phi}}{2}t) - p_n \frac{(Y_0 + 2K_0)}{\sqrt{-\Phi}} \sinh(\frac{\sqrt{-\Phi}}{2}t)] \exp(\frac{p_n}{2}t) \qquad (\rightarrow 7.14a).$$

The hyperbolic functions now can be written as exponential function, because it holds in general

$$\sinh(x) = \frac{1}{2}(\exp(x) - \exp(-x)) \;\wedge\; \cosh(x) = \frac{1}{2}(\exp(x) + \exp(-x))\;.$$

For small *x* but also holds $\sinh(x) \approx 0$ and also $\cosh(x) \approx \exp(x)$. We can therefore write as an approximation for small arguments $x = \frac{\sqrt{-\Phi}}{2} t \ll \pi$ :

$$Y(t) \approx Y_0 \cosh(\frac{\sqrt{-\Phi}}{2}t) \exp(\frac{p_n}{2}t) \approx \exp((\frac{\sqrt{-\Phi}}{2} + \frac{p_n}{2})t) \tag{8.2}$$

The argument is

$$\sqrt{-\Phi}/2 = \sqrt{p_n(4p_s - p_n)}/2 = \sqrt{p_n^2 - 4p_n p_s}/2 = p_n/2 \cdot \sqrt{1 - 4p_s/p_n} \tag{8.3}$$

and thus applies to the beginning of the development we get

$$\sqrt{-\Phi}/2 \approx p_n/2 \tag{8.4}.$$

Hence we actually start early in the economy with approximately

$$Y(t_{start}) \approx Y_0 \exp(|p_n|t) = Y_{IWF}(t) \tag{8.5}$$

Analogously, this applies also for $K$, as is easily verified. At the beginning there is an exponential slope not only of the capital stock, but also of the GDP. However, not globally but only locally in time, and also are there not the same values $g = p_v$ for the slope rates, but $|p_n|$ for the GDP and $p_v$ for the capital stock. Since in the beginning almost all capital growth gains from direct investment in GDP, also these values are initially very similar. No wonder then that singular interpolating models may initially give the impression of a global scope. This fact explains easily why simple models based on extrapolating statistics in "good times" may lead to a seemingly correct "model" although it will be far away from reality at much later times.

Over time however the $\cosh$ and $\sinh$ terms develop a very different behavior as they change smoothly to cos and sin terms. Since they all are harmonic functions their argument $\frac{\sqrt{-\Phi}}{2}t$ is of particular importance.

---

[48] The International Monetary Fund standard model IWF 2005 states the solutions $Y = Y_0 \exp(g\,t)$ ; $I = I_0 \exp(h\,t)$ ; $K = K_0 \exp(j\,t)$ with usually chosen $g = h = j$ to be equal for GDP *Y*, Investment *I* and Capital *K*. Either way it states a singular uncoupled System of ODE's for *K,I,Y* which integrates for every function separately. For Capital *K* but is not taken the total financial, capital stock but merely the part of Commercial Banks Business which are Loans to the economy. Remark: Neither with $g = h = j$ nor $g \neq h \neq j$ the above ODE's are coupled. Regardless of this choice they always integrate separately and fully uncoupled.

The argument of a time-dependent harmonic function has the structure $\omega t$, with $\omega$ the so-called angular frequency with the relationships $\omega = 2\pi\nu$. Here $\nu$ is the frequency in Hertz. The Wavelength $\lambda$ and frequency are linked by the wave velocity $c = \lambda\nu$. The characteristic time $T_c$ of course, is the reciprocal of the frequency $T_c = 1/\nu$. So now we can determine these values closer.

The **characteristic frequency** is given by

$$\nu_c = \frac{\sqrt{-\Phi}}{4\pi} \tag{8.6}$$

and further, the **characteristic time** to

$$T_c = \frac{4\pi}{\sqrt{-\Phi}} = \frac{4\pi}{\sqrt{p_n^2 - 4 p_n p_s}} \tag{8.7}.$$

The time $T_c$ can be calculated if one uses the official figures of the statistics of the FRG. At the beginning of development $p_n$ was approximately -17.9% and $p_s$ about 4.2%. So it follows immediately the *rough* approximation

$$T_c \approx \frac{4\pi}{\sqrt{0.179^2 + 4 \cdot 0.042 \cdot 0.179}} \approx 50 \quad \text{years} \tag{8.8}.$$

as the characteristic time. This characteristic time is again of the same magnitude as found earlier (7.4) as $t_{max} \approx t_c$ in this text. However, it has a slightly different meaning.

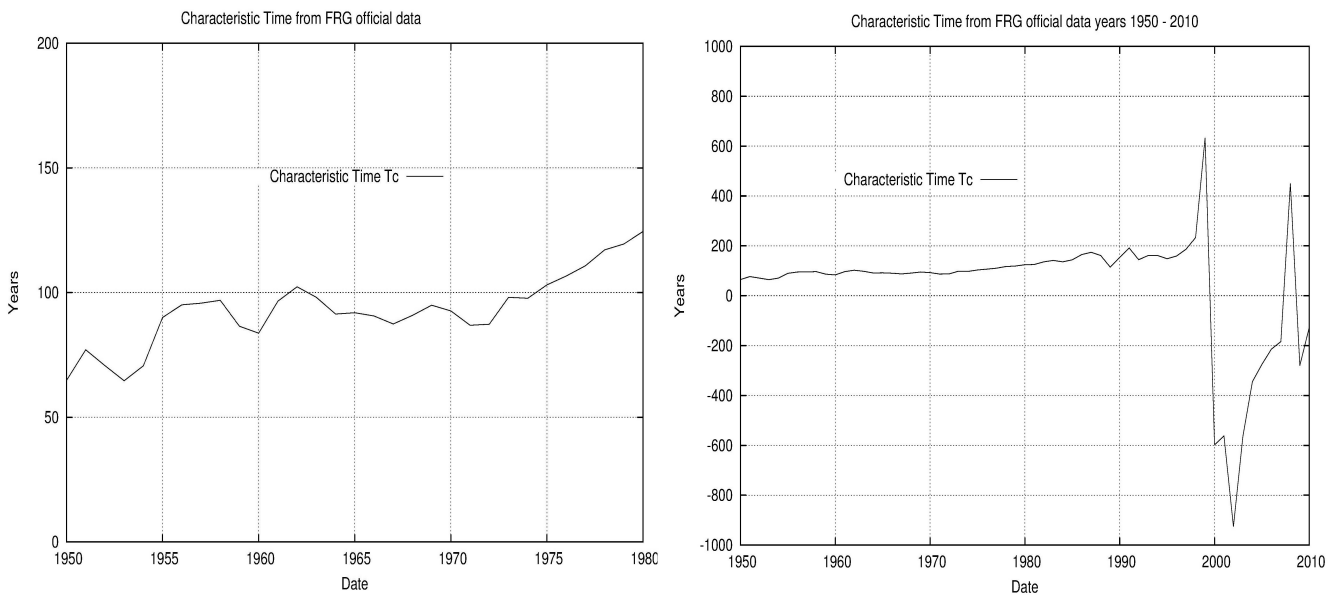

*Fig. 10, left:* Characteristic time Tc calculated according to the real data from the Bundesbank from 1950 to 1980. At the start in 1950 the characteristic time Tc equals 64.8 years, which points to the dates around 2014/2015. *Fig. 9, right:* Characteristic time Tc calculated according to data from the Bundesbank from 1950 to 2010. From 1999 to 2000 it changes sign and gets imaginary and chaotic. This was the date when the DotCom-Bubble began to burst.

The meaning of $T_c$ is basically the answer to the question: *"In which time it would be possible with the given initial conditions, to increase GDP by a factor of $e = 2.71...$ ?"*. With the time-dependent increasing of the capital coefficient but this gets always more difficult, and the characteristic time therefore gets larger. We can see these characteristic values in the fig. 10, calculated according to the official figures of the FRG.

For the range of initial conditions $(Y, K)=(Y_0, K_0)$ we get about $T_c \approx 50$ years. But after that $T_c$ grows strongly and ultimately turns sharply to negative values. This inflection point is given by the year 2000. For the growth path of the economy thus the parameter $p_n(4p_s - p_n)$ is crucial. It must be kept negative to stay in the growth-regime of the $\cosh$ and thus does not enter into the realm of the down-sloping $\cos$. Since 2001, but this value is positive in the FRG. The share of domestic lending business has declined since 1950 by almost 73% to less than 40% in 2010. Looking at the official data (fig. 5) the bend in the exponential growth of capital can clearly be seen beyond the year 2000, after then the former exponential growth could not been sustained any more.

Next we should shed some light on the question how one could eventually escape the turning point back into the regime of positive growth. To address the growth parameters we reflect again on the importance of individual variables.

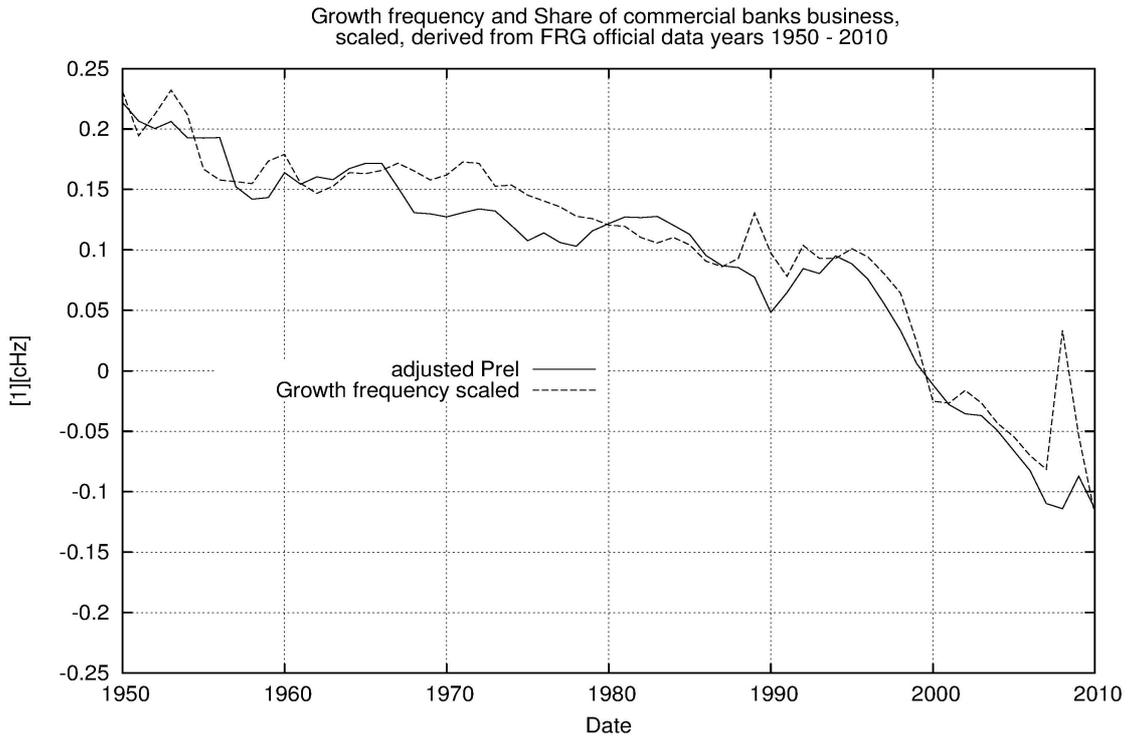

*Fig. 11: Growth frequency rate and share of commercial banks calculated (scaled for visibility) in accordance with the federal bank data. The data were normalized to illustrate the significant correlation between the two records. In the beginning of the 2000ties both functions change their sign, which marks entering crisis times.*

It is the condition

$$p_n(4p_s - p_n) = p_v(1 - 2p_{rel})(4p_s - p_v(1 - 2p_{rel})) < 0 \qquad (8.9)$$

to comply, which in turn gives the conditions for $p_v$ greater or less zero:

$$p_v > 0 :$$
$$[p_{rel} < \frac{1}{2} \wedge p_v(1 - 2p_{rel}) > 4p_s] \vee [p_{rel} > \frac{1}{2} \wedge 4p_s > p_v(1 - 2p_{rel})] \qquad (8.10)$$

$$p_v < 0 :$$
$$[p_{rel} < \frac{1}{2} \wedge 4p_s > p_v(1 - 2p_{rel})] \vee [p_{rel} > \frac{1}{2} \wedge p_v(1 - 2p_{rel}) > 4p_s] \qquad (8.11)$$

As we see it is in each of the four cases in turn critical if the relative share of investment in the real economy

$p_{rel}$ is larger or less than 50%, which is the same as adjusted growth rate $p_{arel} := p_{rel} - \frac{1}{2}$ is greater than or less than zero.

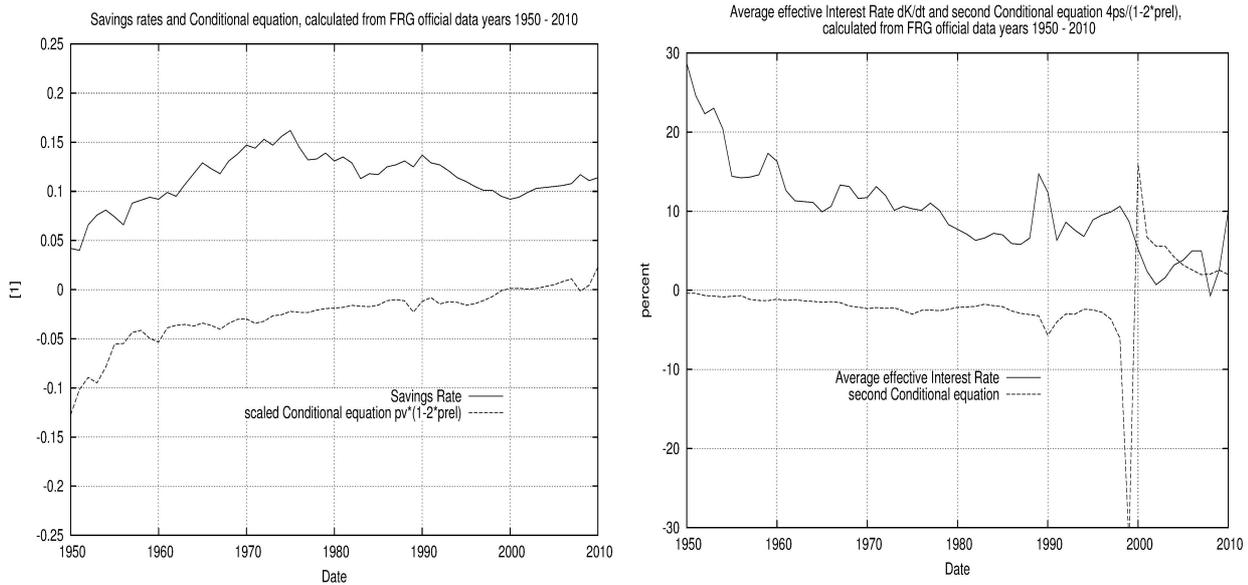

*Fig. 12, left: Savings rate and the scaled commercial business rate from Bundesbank data 1950-2010. Fig. 11, right: Average nominal interest rates of assets (dK/dt) and the second savings rate relative condition calculated from data provided by the Bundesbank.*

In the graph above we can see, calculated according to official figures from the Bundesbank, the growth frequency $\nu_c = \frac{\sqrt{-\Phi}}{4\pi}$ (scaled[49] by the factor 50 for visibility) and the share of real investment $p_{arel} = p_{rel} - 1/2$ (scaled by factor 4 for visibility).

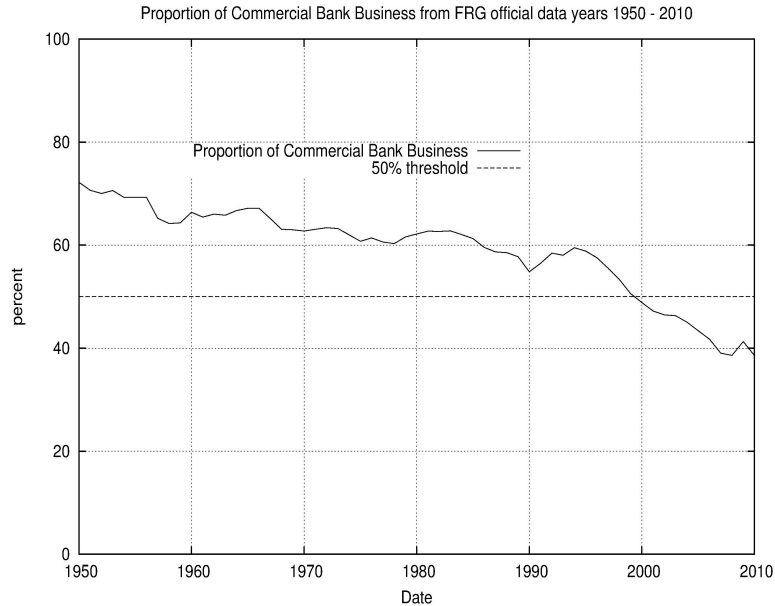

*Fig. 13: Proportion of the business of commercial banks to the total sum of bank assets business [%] according to the Bundesbank official data from years 1950 to 2011. In the year 2000 the 50% threshold falls.*

The latter value does however with the fall of $p_{rel}$ below the 50% level, and thus the preponderance of bank equity business, a sign change. So it will be unattainable with the FRG typical high savings rate in above condition. To return to the growth path with usual monetary politics thus the possibilities are very limited.

---

49  The linear scaling is only due to the better comparison of the functions in the graphic, it does nothing to the basic mathematical profile of the functions.

# A short so far model application sample for the FRG:

We may use the functions (4.1) for some detailed analysis. In the next graph we have taken the available data from Bundesbank and Statistisches Bundesamt into account as much as they are available. The main problem on statistical data is usually that the whole of all assets are not fully accounted for. In the FRG this problem does not occur.

But although all assets are accounted it is never accounted in today statistics the main question: *"Who pays the bill?"*. Which means the question, in which economy the interests are paid, so from market participants in the country itself or from somebody abroad? We could take here only the Target-2 credits which are assets for which the (mostly) southern European EURO-Countries pay interests to Germany. But this is just one of the not exact known amounts adding to the external capital $a_0$ .

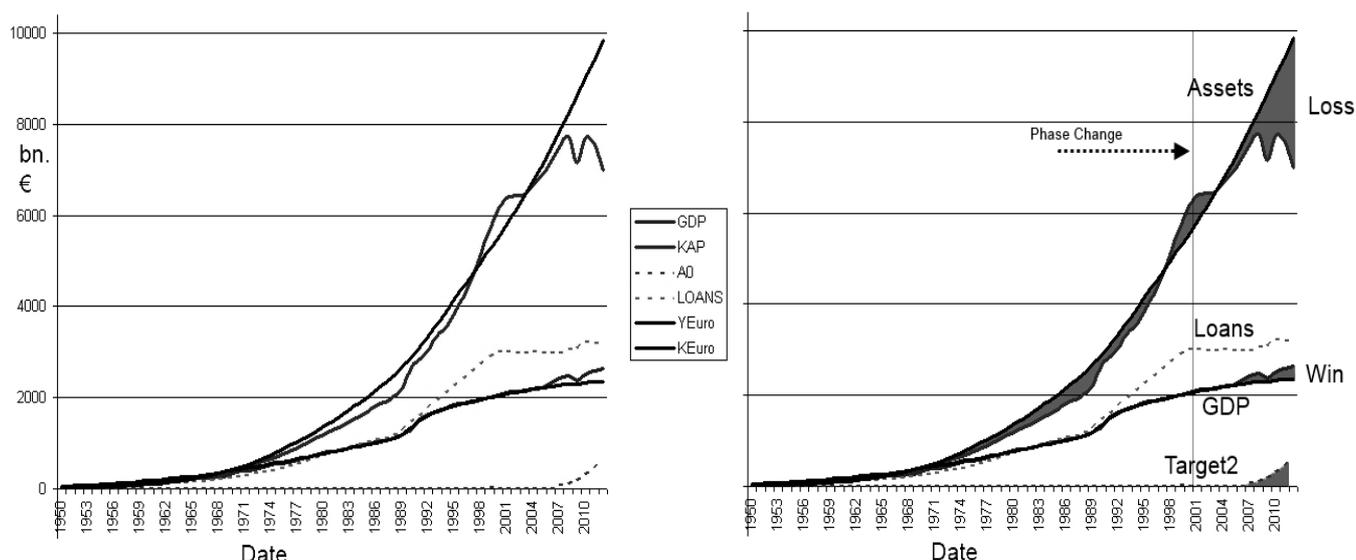

*Fig. 14: Official Data and Model Data from (4.1) in comparison. The amount of external money (a0) can only be estimated, as it is not accounted for in the official statistics. In this graph we used just the so called Target2-credits as a partial of the externals. This is because for this amounts for the direct debt of mostly southern European countries to the Bundesbank via the EZB, and thus the debt burden is paid by other countries.*

What we can see in the official data is that until around 2000 the overall capital gain was very exponential. In 2001 this trend broke down (DotCom-Crisis) and the same happened for loans to the economy. It was the time the total capital coefficient reached 3=300%. From this time on capital was "underperforming". With the then ongoing financial crisis, which reached another highlight with the Lehman-Brothers bankruptcy, finally the financial industry in the FRG lost a lot of assets.

As the most stable and well performing country in the EU Germany had to pay only very small interest rates to his debts. So not only by depreciation but especially by out-migrating of capital to the less stable, but more interests paying southern European countries (or even emerging market countries or the USA), the German financial part of the economy lost very much capital. But after phase change this was a large benefit (see 8.9 to 8.11), as it surpressed much the interest-burden of the real economy. Thus the underperformance of the official assets data to the model data relates ultimately to an overperformance of the GDP, as our model so far does not deal with migration of money but assumes a quasi-closed economy.

If we now have a look at the actual data of the FRG, we can see why in contrast to other European Contries Germany performed that relatively well and we may see also the four phases of the evolution of the German economy from 1950 to 2012:

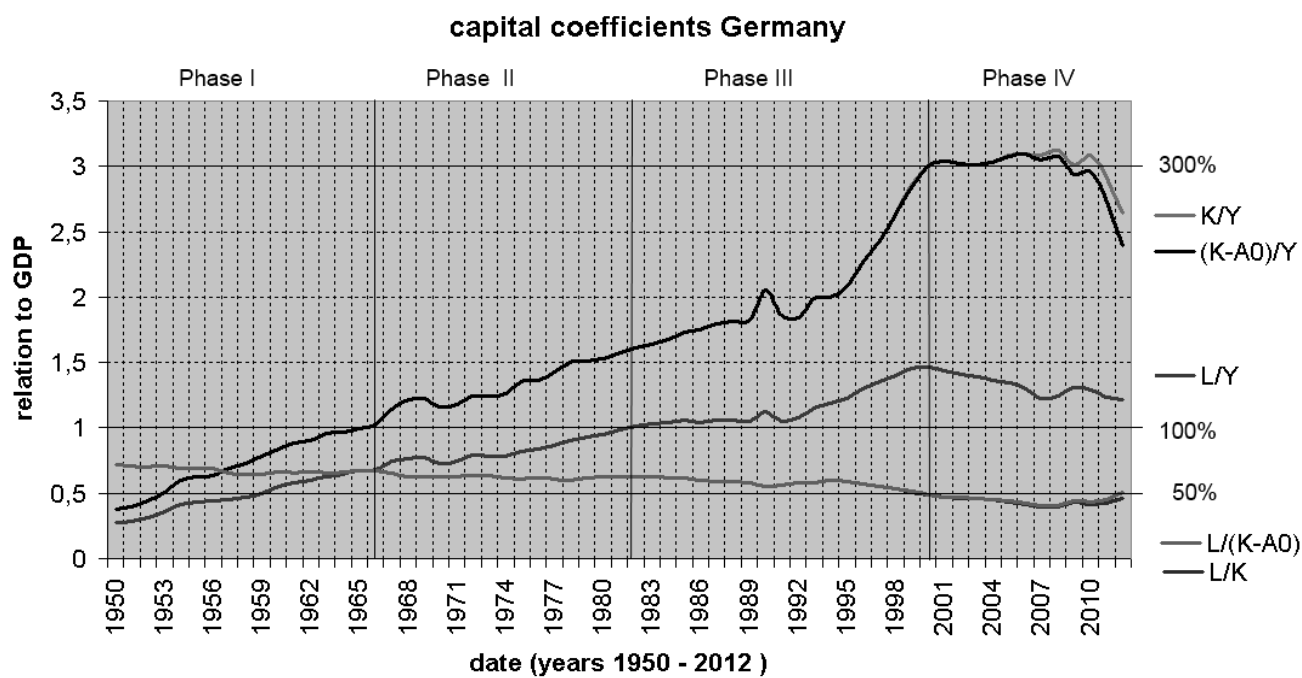

*Fig. 15: Capital Coefficients for Germany derived from the official data of Bundesbank for the years 1950 to 2012. K means total of Capital(Assets), L means Loans, Y means GDP, a0 assets debt payed abroad.*

The first phase of GDP-driven economy ended in 1966/1967, when the ratio of assets to the GDP climb up to *K/Y*=1=100% (und thus also *L/K=L/Y*). After that the economy but got capital driven. The next phase was reached in 1982/1983, as then the loans reached the limit of *L/Y*=1=100% also. In 1990 the GRD population was included formerly just "over night" to the FRG, which gives reason to the "buckle" around this date.

Following in the second half of the 90ies the high time of investment banking started all around in the globalised world, thus accelerating the climb in *K/Y* much more. Around the year 2000 finally the amount of *K/Y*=300% was reached while at the same time the amount of loans to GDP reached its all time high of nearly 150% of GDP. But at the same time the commercial banks business dropped below the crucial 50% line, followed by a large decline in the relation of Loans to GDP down to about 125%.

But in contrast to the mostly southern European states Germany again got into the region of at least moderate growth at about 2010. The reason is obviously the loss of to much financial assets to other countries. So the crucial 300% line for *K/Y* was left behind to much better values about 250% and at the same time the Loans to GDP ratio at least didn't drop further more. Since recent times indeed also the ratio of commercial banks business to investment business climbed again back to the crucial 50% line. The question if this positive evolution will be sustainable but is not easy to answer: It depends on the question of the direction of flow of the externals: If the ousted money would flow back then the positive trend will possibly brake again.

# 9   The Classical Quantity Equation

Already the French political theorist Bodin (1576) recognized basic ideas of the later quantity theory. The first workable formulation of the quantity theory comes from the English philosopher Locke (1689), where he introduced the term of monetary velocity. One of the most famous representative of the quantity theory of money was the U.S. economist Milton Friedman (1912-2006). In its essential expression the quantity theory asserts the relationship

$$M \cdot V = P \cdot H \tag{9.1}.$$

The significance is that the product of the circulating money supply $M$ times the velocity of circulation $V$ equals the product of the price level $P$ times the trading volume $H$, which is the frequency of transactions of real goods. The trading volume times the price level of the gross domestic product equals the GDP.

$$MV = PH = Y \tag{9.2}.$$

A fundamental problem is, however, that no doubt all four function are depends on time

$$M(t) \cdot V(t) = P(t) \cdot H(t) \tag{9.3}.$$

However this temporal relationship remains unknown, because of a conditional equation alone, the four solution functions are impossible to be clearly determined. Therefore, this well known connection is usually used only as a rule of thumb with just local validity in time. After all, holding two of the variables constant, one can imagine the influence of a change in the size of the third to the fourth variable. For local situations, which means if the formula is used for only a short time line, and so the variation of other sizes may be realistically considered negligible, then the quantity equation in this form can be already very helpful. Now this equation is also a typical balance equation, because it is merely a mathematical formulation of the principle of trade *"money for goods"*. In a higher valid theory, the quantity equation must be included and in that theory the time dependencies of the individual functions in principle should also be determinable analytically. Since it is usually treated only as a rule of thumb, it goes into the classical literature with a generous[50] treatment of the units. So first we need to ensure a workable definition of the four factors.

While the definition problems are relatively clear for *H, P* and *Y*, the problem is more difficult for what aggregate of money *M* represents, and how *V* should be defined[51] correctly. Units should be in fact both sides identical and due to *Y=HP* have the unit *currency/year*:

$$[M][V]=[H][P] \Leftrightarrow Cur \cdot \frac{1}{y} = \frac{Buy}{y} \cdot \frac{Cur}{Buy} \Leftrightarrow \frac{Cur}{y} = \frac{Cur}{y} \tag{9.4}.$$

Since money is measured in *Cur* the unit equation above is necessarily for a solution. Here the unit *Buy* is for a purchase or a number of purchases. Now the units on both sides agree. *H* is the number of purchases per year and the price level is the average price per purchase. So is the right side

$$[H] \cdot [P] = Buy/y \cdot Cur/Buy = Cur/y = [M_i] \tag{9.5}.$$

---

50  Sometimes one equates *Y* with *Y*, which is not true, and *P* is often identified with inflation, which also is not right. In addition, *M* is identified usually with one of the three monetary aggregates $M1, M2, M3 \subset K$, and finally the velocity *V* is set to any size, e.g. between 0 and 1, for non-existent or inaccurate definition of its unit.

51  *V* is the speed for revolving the liquid money in the economy, and we would suggest a similar motion of velocity like *m/sec* in mechanics, so the unit is actually *currency/year*. The *M* funds should be, at least it is often thought so in literature, just be that liquid money supply aggregate. But then one gets $[M] \cdot [V] = Cur \cdot Cur/y = Cur^2/y$ as a badly defined unit for the left hand side of the quantity equation. On the right hand side one often takes the price level in *currency* and the volume of trade as an unspecified number of the unit *1*. This results on the right side at the unit $[H] \cdot [P] = 1 \cdot Cur = Cur$, what is not properly absorbed. One can twist and turn it in different ways, but in each case either the units on both sides do not match, or the unit of a single size does not make sense.

The left side is not as fast to make clear. For this it needs some considerations. The product *(MV)* should of course be the annual money supply in GDP. Classically one assumes as an approximation one of the money aggregates $M_1$ to $M_3$ which have in fact a different degree of determination (cash and short-term investment pays off as distinct from long-term frozen money or assets). But these are actually only approximations, since these money aggregates are present in the balance certainly, but whether they actually run around and be used for purchases is by no means certain. But since we need a complete balance equation for the overall economy, we must proceed differently. For not the aggregates $M1, M2, M3 \subset K$ are to be set for *M*, but the total financial capital stock *K* as a whole. Multiplied by the money velocity *V*, we now get the actual amount of money in circulation $M^v$ that is

$$K \cdot V =: M^v \tag{9.6}$$

which has now the correct unit

$$[K] \cdot [V] = Cur \cdot 1/y = Cur/y = [M^v] \tag{9.7}$$

The now well defined continuity equation is written in expanded notation:

$$M^v = KV = HP = Y \tag{9.8}$$

The balance is alright because the circulating money (wages and revenues, but also duties and taxes, or withdrawals or loans from the Bank for investment and consumption) is of course the GDP *Y* as a whole. The measured monetary aggregates $M_1, M_2, M_3 \subset K$ are only a subset of the financial capital stock and just provide an approximation of $M_i$ and the associated money current amounts $M_i^v := K_i V_i$.

As an example we can now compute the balance equation for the the year 2008 in the FRG: the total capital stock *K* was then about 8000 billion Euros [€]. *V* was the proportion of this as available money, here about *0.3/year = 30%* [1/y]. *H* is the volume of trade, i.e. the number of purchases in 2008, which can by now only be roughly estimated: With the number of inhabitants in the FRG of about 83 millions and an average of one purchase a day, we assume about 83 million times 365 = 30 billion purchases per year [Buy/y]. The price level is then the GDP divided by the number of purchases, so *P* can then be calculated with a GDP at that time of about 2400 billion € to 80 € per purchase [€/Buy]. As is easily seen, then the equation is balanced:

$$M^v = K(8000\,billion\,\text{€}) \cdot V(\frac{0,3}{y}) = 2400\frac{billion\,\text{€}}{y}$$
$$Y = H(\frac{30\,billion\,Buy}{y}) \cdot P(80\frac{\text{€}}{Buy}) = 2400\frac{billion\,\text{€}}{y} \tag{9.9}$$

The monetary velocity estimated here is also obtained from the real numbers of the Bundesbank, so by the monetary aggregate $M_3$ or the total credit amount into the GDP. These statistical numbers are only an approximation, but a pretty good one. Therefore we compute for the past year 2007 now specifically out of official data: In December 2007, the GDP € was 2975.7 billion and the total capital stock was € 7625.7 billion. The monetary aggregate $M_3$ was communicated to 2187.8 billion € and the amount of outstanding loans to domestic non-banks in the FRG amounted to 2975.7 billion €.

Thus we get from $M_3$ a velocity of $V_{M3} = 0,287/y$ and by the loans to domestic non-banks we get $V_{OU0115} = 0,39/y$, which are both approximatly by the nature of the underlying data. The population in 2007 was 73.941 million inhabitants. Multiplied by 365 days gives 26.988 billion purchases that year. So is calculated the price level in this specific example to about

$$P[\text{€}/Buy] = \frac{7625,7[billion\,\text{€}] \cdot 0,3385[1/y]}{26,988[billion\,Buy/y]} = 95,65\frac{\text{€}}{Buy} \tag{9.10}$$

in the national economic average. The problem of the quantity equation, however, remains that there is still an under-determined equation system. Indeed, the calculation of the exact numbers are lacking further defining equations.

# 10 Conformity to Classical Quantity Equation

For this purpose we first consider the numerical integration of the simplified model. So we will first check whether our model actually withstands the requirements of the quantity theory. Then follows a theoretical foundation of the relationship. We consider again the basic model and we calculate the money supply. This consists on the one hand from the GDP of money in circulation $(1-p_s)Y$, i.e. the amount of money not saved. In addition there are contributions that are deducted from interest rates on savings, if one assumes that most market participants are interested[52] to preserve their nominal assets. So we get

$$K \cdot V = (1-p_s)Y + (1+p_s)\frac{dK}{dt} \tag{10.1}.$$

The product $P \cdot H$ is of course the GDP $Y$ in our model. Due to the quantity theory $KV/PH=1$ must now obtain trivially to our basic model. Thus in our numerically integrated model should rule

$$\frac{(1-p_s)Y(t)+(1+p_s)\frac{dK(t)}{dt}}{Y(t)} =! 1 \tag{10.2}.$$

We can now easily calculate these value over time. In fact the result shows, that the quantity equation is indeed respected. Over the entire non-critical area of development over time, it remains closely at the value 1, except for small deviations from numerical integration. Alike can also be calculated the velocity of money, which is so important. This is $V=\frac{PH}{M}$, with our total assets $K$ constitute a superset of what the money supply $M=cK$ is only one component. Thus, the monetary velocity is computed as

$$V(t) = \frac{PH}{M} = \frac{cY}{K} = c \cdot 1 \cdot \frac{Y}{K} = c \cdot ((1-p_s)Y(t)+(1+p_s)\frac{dK(t)}{dt})\frac{Y(t)}{K(t)} \tag{10.3}.$$

In a first approximation $c \leq 1$ can be assumed to be constant and set here to c=1. In the next graph, the corresponding values from the numerically integrated *K(t)* and *Y(t)* were obtained. The same formula was applied to the official time series of Bundesbank (for $p_S$, $Y$ and $K$) and re-chained to model points by our currency correction function.

In fig. 16 we see that from about 2007, the velocity begins to leave the liquidity area. Nonlinearities are beginning to be felt after 2010. At the latest by 2020, a simple linear approximation is no longer justified. One can see clearly that only after the crisis years in the vicinity of the collapse, the quantity equation in our simplified model loses its generality. Especially since external sources will play in crisis years an increasing role. About the occurring nonlinearities, we will also need to make more considerations later in this book. Significant is also the course of the resulting velocity. They tend to behave exactly as it is known empirically from official data. The ratio *KV/HP=1* is obviously an invariant of the economy. A deeper foundation receives this very old empirical fact of economics by comparison with physical problems. Systems with conserved quantities always satisfy a corresponding equation of continuity. These are basically on the structure

$$\frac{\partial \rho}{\partial t} + \text{div}\,\vec{j} = 0 \tag{10.4}$$

which is a usual continuity equation as we know from physics. Here it means that the charge $\rho$ is a conserved value and it is the source of a current $\vec{j}$. One can write such an equation for short as a four-vector: $\%DIV(\vec{j},\rho)=0$. The four-divergence is then defined as

$$\mathrm{DIV}(\vec{j},\rho):=(\frac{\partial}{\partial x},\frac{\partial}{\partial y},\frac{\partial}{\partial z},\frac{\partial}{\partial t})\cdot(j_x,j_y,j_z,\rho)=(\vec{v_e}\cdot\vec{j}+\dot{\rho})=0 \qquad (10.5)$$

where $v_e$ is the unit-speed.

---

52  In fact, the state plays a major role. For he draws from $p_S$ effective. See in later chapters on government debt.

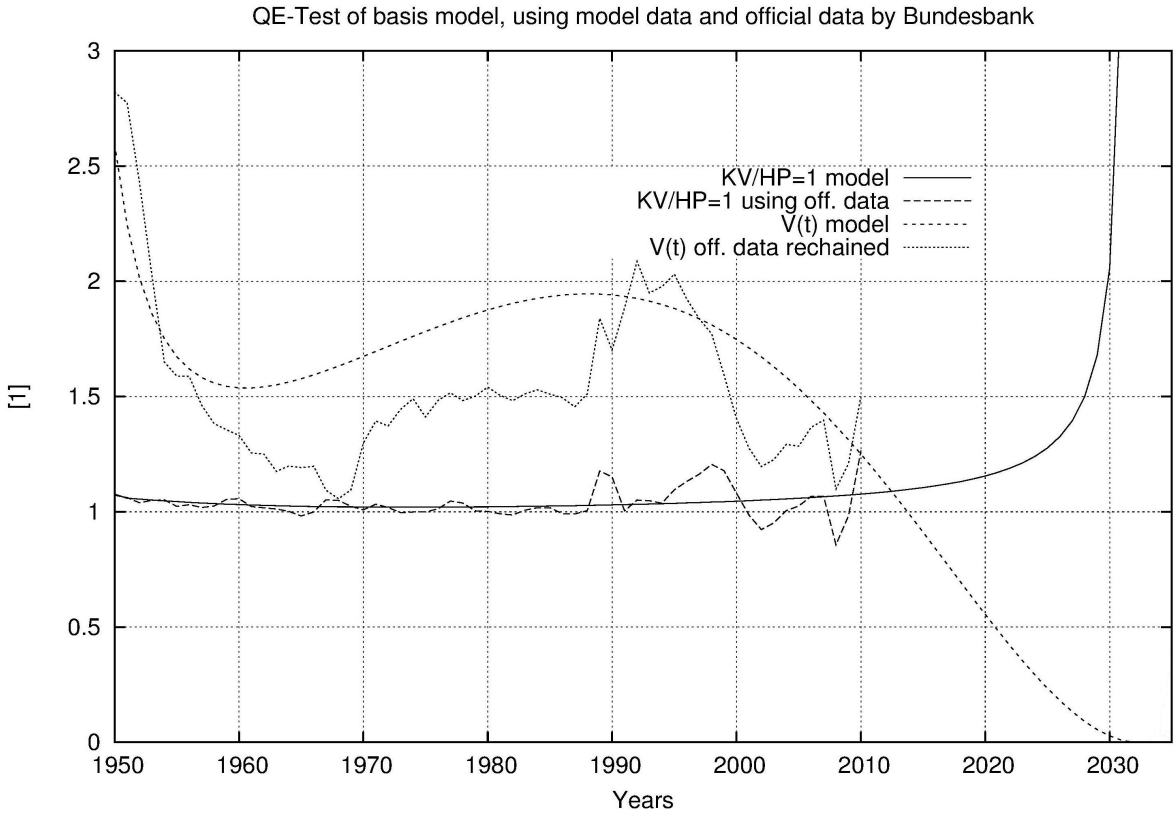

*Fig. 16: Quantity equation and monetary velocity in dimensionless units from 1950 to 2010 according to the basic model and to the official data applying the same approximate formula. From the 2000ies the monetary velocity is leaving the liquidity area around 1.5. Nonlinearities as a departure of the QE from 1 make itself strongly felt in the 2020ies at the latest.*

In economics, there is now trivially

$$K\cdot V = P\cdot H \Leftrightarrow K\cdot V - P\cdot H = 0 \qquad (10.6)$$

Using this simple description one can now see at the heart of the classical quantity theory. Because $V$ is a speed and $H$ the frequency of transactions, which we can regard as a time derivative. We can therefore write in one-dimensional analogy:

$$\mathrm{DIV}(K,-P)=0 \qquad (10.7)$$

The quantity theory is obviously just the necessary continuity equation, as with any well defined theory must exist such conservation values. It says:

*The prices or price levels are the sources of funds currents.*

If there are changes $\partial/\partial t$ to a source this results always automatically in resulting currents $\partial/\partial x,..$ . And vice versa rules:

*No currents can exist without an associated change in the sources.*

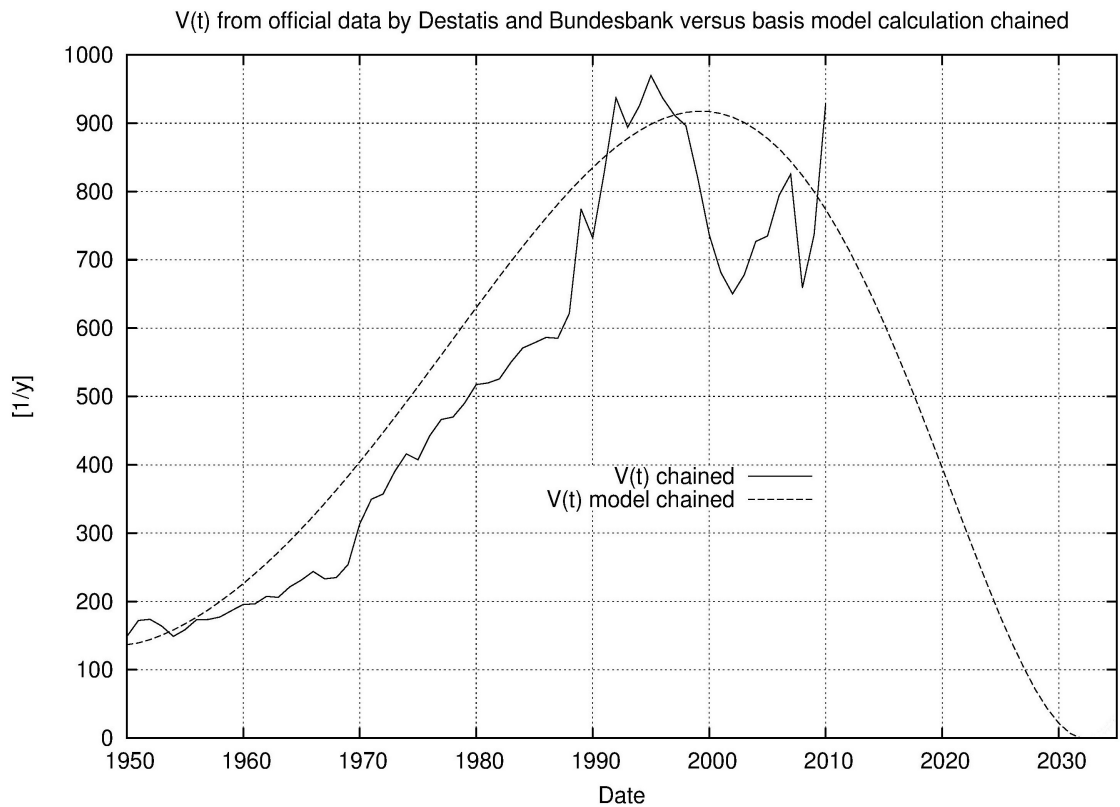

*Fig 17: Model data chained in comparison with official data*

Now we have to formulate this relationship, however, something more universal. The continuity equation can be studied analytically, as we have shown that the quantity equation of the economies

$$M^v = KV = HP = Y \qquad (10.8)$$

satisfies the usual conservation laws in field theory $DIV(K,-P)=0$. However, the quantity equation in this simple form is of little use. Because the velocity of circulation would be directly $Y/K=V$, so only the inverse of the capital coefficient. This would mean that the orbital velocity in about a sloping hyperbola, as would correspond to about *1/t*, and would constantly go to zero. This is true but, at best, very roughly, since statistics always also shows this typical U-shaped course.

Correctly we must look at various monetary velocities and price levels of different trading volumes and the associated amounts of money, because in its simplest form the quantity equation shows up only rough economic means. And just the different effect of various currents produce the U-shaped profile of the mean value of *V*. We have achieved this in the previous formulas by having used the basic equations of special field theory to derive the formula *V* in which the effects of two trade flows, namely the commercial banking business and the banks own business, are contained already implicit:

$$V = ((1-p_s)Y + (1+p_s)\dot{K})\frac{Y}{K} \quad \text{in contrast to} \quad V_{naiv} = \frac{HP}{M} = \frac{Y}{K}$$

Mean values are marked in the broad with a hat. The values without a hat are the ones, which in principle can be determined as pure data, even if the practical work for it can be very high. Thus, the amended quantity equation is:

$$K\hat{V} = \sum_i M_i V_i = \sum_j H_j P_j = \hat{H}\hat{P} = Y \qquad (10.9)$$

in which must be *i* and *j* not equal[53] in general. Strictly speaking, one would thus have to compute over the entire economic range[54], considering all specific price levels and the associated cash flows without any exception[55]. The continuity equation then could be written complete as $\text{div}_{v_i} M_i V_i = \partial_j H_j P_j$. For an analytical study, we have to take the sources and sinks that are suitable to the most important groups. The special problem lies in the importance of investment banking, whose share of the banking business in time increases sharply and eventually predominates all other business. This fact has to be mentioned at least in a moderately advanced form of balance equation:

$$K \hat{V} = M_R \hat{V}_R + M_I \hat{V}_I = \hat{H}_R \hat{P}_R + \hat{H}_I \hat{P}_I = Y \tag{10.10}$$

$$\text{with} \quad K = M_R + M_I \quad \text{and} \quad H = H_R + H_I \tag{10.11}$$

i.e., we split both the capital stock and the trading volume into their specific parts of its real economy and its investment shares.

## 11 Supply and Demand - Substitution

The total quantity equation is tantamount to the *economic supply and demand relation*. For on the one side $M_i V_i$ are the available cash flows and on the other side the respective trade volumes $H_j$ with their resulting prices $P_j$:

$$\text{econ. Demand} = \sum_i M_i V_i = \sum_j H_j P_j = \text{econ. Supply} \tag{11.1}.$$

For a single product, lets say for example PCs, this relationship is from a special business perspective: $M_{PC} V_{PC} = H_{PC} P_{PC}$. If the PC-company wants to increase its sales by another $\Delta H_{PC}$, he must either change the demand $M_{PC} V_{PC}$ that is, increase the flow of money, which can be achieved for example by better advertising. Or it has to change its price $P_{PC}$ and, therefore favorable has to reduce the price to get the additional sales. So holds:

$$(M_{PC} + \Delta M_{PC}) V'_{PC} = (H_{PC} + \Delta H_{PC}) P'_{PC} \tag{11.2}.$$

Economically, this is but in the overall balance to involve:

$$0 = \sum_i M_i V_i - \sum_j H_j P_j = \ldots$$

$$\ldots = \sum_i [M_{i-1} V_{i-1} + (M_{PC} + \Delta M_{PC}) V'_{PC}] - \sum_j [H_{j-1} P_{j-1} + (H_{PC} + \Delta H_{PC}) P'_{PC}]$$

(11.3)

Now the company has two options, the first of which is however extremely unattractive:

$$(M_{PC} + \Delta M_{PC}) V'_{PC} - (H_{PC} + \Delta H_{PC}) P'_{PC} = 0$$

or

$$(M_{PC} + \Delta M_{PC}) V'_{PC} - (H_{PC} + \Delta H_{PC}) P'_{PC} \neq 0 \tag{11.4}$$

in which

$$(H_{PC} + \Delta H_{PC}) P'_{PC} < (M_{PC} + \Delta M_{PC}) V'_{PC} \tag{11.5}$$

will be attempted. Because the goal is of course an increase in turnover and thus the operating profit. Because then the operating cost $(H_{PC} + \Delta H_{PC}) P'_{PC}$ is less than the cash flows derived, differentially expressed

---

53 If one breaks the range down to the smallest levels, then $i = j$ holds. Otherwise, this must not necessarily be the case, but it can be assumed in general.
54 Which of course includes all financial products and foreign transfers and also money creation through equity purchases of government bonds and other financial constructs.
55 Such a thing is at best an approximate number by detailed statistical works in practice.

$$\text{as} \quad \frac{d}{dt}(HP)_{PC} < \frac{d}{dt}(MV)_{PC} \quad \text{if successful} \tag{11.6}$$

This can only succeed economically if it clears the increased cash flows for PCs $(M_{PC}+\Delta M_{PC})V'_{PC}$ elsewhere in the GDP. In particular, this means increased revenue and an operating result in decreased revenue for one or more others must be recruited because the money flows the other one is no longer available.

In particular, this means increased revenue for one company results from reduced income from one or more other companies, because the funds raised money flows for the other companies is no longer[56] available. This leads to

$$0 = \sum_i M_i V_i - \sum_j H_j P_j = ...$$
$$\sum_i [M'_{i-1} V'_{i-1} + (M_{PC}+\Delta M_{PC})V'_{PC}] - \sum_j [H'_{j-1} P'_{j-1} + (H_{PC}+\Delta H_{PC})P'_{PC}] = 0$$

$$\textbf{if} \quad (M_{PC}+\Delta M_{PC})V'_{PC} - (H_{PC}+\Delta H_{PC})P'_{PC} \neq 0 \quad \textbf{holds}. \tag{11.7}$$

# 12 Deriving Marginal Utility and Gossen's Law

The marginal utility theory claims the empirically proven relation that

$$\lim_{t \to \infty} \frac{\partial U(t)}{\partial C_i} = 0 \tag{12.1}$$

the marginal utility of a utility function $U(t)$ of the goods $C_i$ goes to zero with time. This means that one privately used car produces high value, but the second already relatively less. The third and fourth car can slide into negative value. because of its high costs and only few additional benefits. However, we consider the case of very wealthy consumers well, that there will be purchased six or seven vehicles for private use, such as a number of different sports cars, even though their marginal benefit is actually negative. Empirically, this is explained with the fun factor as one intangible[57] benefit.

But the cause in the average of the national economy is that the average consumer is not able to buy an additional mobile to create fun, unless he saves it in a different position, so eventually to scrimp and save it. The diminishing marginal utility is explained in the following analysis of this competitive situation of cash flows. This also applies to the example of the affluent consumer. The affluent consumer will not scrimp and save it, but he will substitute it from his investments. As in place of the third sports car he could buy even more from the financial institutions promoted financial products.

Due to the validity of the continuity equation, we are dealing with a total economy that is a *substitution economy*. From this elementary law can be derived analytically the empirically known laws of marginal utility theory and e.g. also the second law of Gossen and much more. The second law of Gossen states that the amounts of consumer goods of the *n* goods available to an individual $x_1, ..., x_n$, with the differentiable utility function $u = u(x_1, ..., x_n)$ and the prices of goods $p_1, ..., p_n$, the second law of Gossen holds:

$$\frac{\partial u(x_1, ..., x_n)/\partial x_1}{p_1} = ... = \frac{\partial u(x_1, ..., x_n)/\partial x_n}{p_n} \tag{12.2}$$

For the analytical derivation, we differentiate now the quantity equation for the *j*-th consumer choice $H_j$:

---

56 This applies generally also in times with a positive economic growth, not always in absolute numbers but in relative terms. And the latter is here essentially.
57 In marginal utility theory the so-called "indirect utility function."

$$0 = \frac{\partial}{\partial H_j} \sum_i (M_i V_i - H_i P_i) = \frac{\partial}{\partial H_j}(M_j V_j) - P_j + \sum_{i-1}(M_i V_i - H_i P_i) \quad (12.3)$$

But now because of the enormous variety and thus numbers of products $i \rightarrow \infty$ in the last term, this is also given for the term

$$\sum_{i-1}(M_i V_i - H_i P_i) \simeq 0 \quad (12.4),$$

because we only have eliminated one but from millions of products, which changes the balance not significantly[58]. Then also for all j holds:

$$\forall_j : \frac{\partial}{\partial H_j}(M_j V_j) - P_j = 0 \Leftrightarrow \frac{\frac{\partial}{\partial H_j}(M_j V_j)}{P_j} = 1 = const. \quad (12.5)$$

and thus according[59]

$$\frac{\frac{\partial}{\partial H_j}(\sum_i^n M_i V_i)}{P_1} = \ldots = \frac{\frac{\partial}{\partial H_n}(\sum_i^n M_i V_i)}{P_n} \quad (12.6)$$

with the obvious conversion

$$\partial u(x_1, \ldots, x_n)/\partial x_n := \frac{\partial}{\partial H_n}(\sum_i^n M_i V_i)$$
$$\text{and} \quad p_n := P_n \wedge x_n := H_n \quad (12.7).$$

So the second law of Gossen results directly from the economic equation of continuity. The diminishing marginal utility now is derived from the economic situation of the *n*-competing products:

$$\frac{\partial}{\partial H_n}(\sum_i^n M_i V_i) = \sum_i^n (V_i \frac{\partial}{\partial H_n} M_i + M_i \frac{\partial}{\partial H_n} V_i) \quad (12.8).$$

The thus found Gossen utility function says, that the change of use of the *n*-th product $H_n$ only then does function, if either the quantity of the *i*-th product (represented by the amount of money $M_i$ available for it) with the amendment made by $H_n$ decreases, or the money has to circulate faster $V_i$. The latter, however, means that the consumer has to have more money per unit of time, ie a wage[60], at his disposal. Without a real general wage increase is therefore:

$$\frac{\partial}{\partial H_n}(\sum_i^n M_i V_i) = \sum_i^n V_i \frac{\partial}{\partial H_n} M_i \quad (12.9)$$

what can be change to:

$$P_n = \frac{\partial}{\partial H_n}(M_n V_n) - \sum_i^n V_i \frac{\partial}{\partial H_n} M_i \quad (12.10)$$

Has the purchase decision of the *n*-th product not come due to an improvement in income,

$$\text{then} \quad P_n = V_n \frac{\partial M_n}{\partial H_n} - \sum_i^n V_i \frac{\partial M_i}{\partial H_n} \quad \text{holds.} \quad (12.11)$$

---

58  As we will see later, with the exception of "system-relevant" products.

59  As the operator $\partial/\partial H_j$ picks out of the sum $\sum_i^n (M_i V_i)$ always the *j*-th element.

60  Even with wage increases, the approximation is usually justified as a general increase in the money supply via the levy on the goods is entitled inflated accordingly. The net effect will be regularly close to zero.

The *n*-th price so decides after the competitive substitution[61] of the *n*-th product to the *i* competing products. We can now look a little closer to this equation, which should indeed result in a diminishing marginal utility. Interesting there is of course the price, especially the price change. First we assume a lack of improvement in income and with

$$\frac{d}{dt}V_i = 0 \quad \text{results} \quad \frac{d}{dt}P_n = V_n \frac{\partial \dot{M}_n}{\partial H_n} - \sum_i^n V_i \frac{\partial}{\partial H_n}\dot{M}_i \tag{12.12}$$

Suppose even that the $V_i \approx V_j$ are all approximately equal[62], then we can write for simplicity:

$$\frac{d}{dt}P_n = V\left(\frac{\partial}{\partial H_n}\dot{M}_n - \sum_i^n \frac{\partial}{\partial H_n}\dot{M}_i\right) \tag{12.13}$$

The marginal benefit $\dot{P}=0$ is achieved then when

$$V=0 \quad \vee \quad \frac{\partial}{\partial H_n}\dot{M}_n = \sum_i^n \frac{\partial}{\partial H_n}\dot{M}_i \tag{12.14}$$

holds. This means in particular that the price change comes to a standstill as soon as the sum of the changed trading volumes (represented by their monetary equivalent $M_i$) of the *i* competing products cancels the change of trade of the *n*-th product.

## 13 Substitution Laws

We can now calculate such a utility function. Therefore we can invest the following considerations: When a new product is introduced or an existing product is promoted with the aim of increasing trade, we can estimate the resulting effects, provided by the trading volume with a prefactor:

$$H_i(t) = s_i(t)H_{0i} = \left(1 - (1 - h_{mini})\left(1 - \exp\left(-\frac{t - t_{0i}}{t_{shi}}\right)\right)\right) \cdot H_{0i}$$
$$\wedge \quad t \geq t_{0i}, \text{ otherwise } H_i = H_{0i} \tag{13.1}$$

The importance of this function is as follows: Each product has a particular significance $H_0$ to the purchaser. However, this product can be degraded up to a certain degree,

$$H_{min} = h_{min} \cdot H_0 \quad \wedge \quad h_{min} \in [0,1] \quad \wedge \quad H_0, H_{min} = const. \tag{13.2}$$

so that the money saved can be used for another product. This happens from the time $t_0$ on. As a rule, the substitution of a product is not abrupt, but spread over time, so we assume a half-life $t_{sh}$ of the substitution of the product. Such a half-life is particularly justified when we talk about major product groups and/or a large number of buyers. We can use our supply and demand-function that is now written as follows:

$$\sum_i M_i V_i = \sum_i s_i H_i P_i \tag{13.3}$$

where, particularly in commercial and private environment, through profit or loss or income change also an additional source or sink of $Q = \Delta(MV)$ may be present.

$$\Delta(MV) + \sum_i M_i V_i = \sum_i s_i H_i P_i \tag{13.4}$$

---

61 Remember, it is not only the competition of similar products such as cars from Ford or GM, but also the question of car or bread and butter?
62 With a sufficiently large number of products we can do this without great error.

Such a source $Q>0$ would allow the purchase of new products without compromising the residual volume, $Q<0$ in contrast, must automatically trigger a corrective response. Let us first examine the situation with $Q=0$ as it usually[63] exists in an economy. We consider the situation when two products *x* and *y* are in competition:

$$M_x V_x + M_y V_y + \sum_{i-2} M_i V_i = s_x H_x P_x + s_y H_y P_y + \sum_{i-2} s_i H_i P_i \tag{13.5}.$$

It is the product *y* to be advertised and the product or product group *x* should be taken ago for the necessary cash flows[64] first The remaining groups are again equal and can be subtracted. It is therefore
$M_x V_x + M_y V_y = s_x H_x P_x + s_y H_y P_y$ and with the function $s_i$, holds:

$$const. = M_x V_x + M_y V_y = H_x P_x + H_y P_y = ...$$
$$... = (1-(1-h_{minx})(1-\exp(-\frac{t-t_{0x}}{t_{shx}}))) \cdot H_{0x} P_x + ...$$
$$... + [(1-h_{minx})(1-\exp(-\frac{t-t_{0x}}{t_{shx}})) \cdot H_{0x} P_x + H_{0y} P_y] \tag{13.6}$$

This applies $H_{0i} := H_i(t_0) = const.$ and we may write with $P_{0i} := P_i(t_0) = const.$ the following equation:

$$(HP)_x(t) = (1-(1-h_{minx})(1-\exp(-\frac{t-t_{0x}}{t_{shx}}))) \cdot H_{0x} P_{0x}$$
$$(HP)_y(t) = [(1-h_{minx})(1-\exp(-\frac{t-t_{0x}}{t_{shx}})) \cdot H_{0x} P_{0x} + H_{0y} P_{0y}] \tag{13.7}$$

The product *y* thus takes the trade products shares $(HP)_x$ of *x*. So applies to the temporal changes

$$(\dot{HP})_x(t) = -\frac{(1-h_{minx})}{t_{shx}} \exp(-\frac{t-t_{0x}}{t_{shx}}) \cdot H_{0x} P_{0x}$$
$$(\dot{HP})_y(t) = \frac{(1-h_{minx})}{t_{shx}} \exp(-\frac{t-t_{0x}}{t_{shx}}) \cdot H_{0x} P_{0x} \tag{13.8}$$

and because of $(\dot{HP}) = \dot{H} P + H \dot{P}$ these changes are either be taken from $\dot{H}$ or $\dot{P}$ or both to increase or to reduce. So the question still remains, which can take place in the event of sources or sinks Q, if from $t_0$ holds:

$$Q + M_x V_x + M_y V_y = s_x' H_x' P_x' + s_y' H_y' P_y' \tag{13.9}$$

Now, Q may be invested in the *x* or *y* or both:

$$Q + M_x V_x + M_y V_y = (s_x H_x P_x + Q_x) + (s_y H_y P_y + Q_y) \wedge Q = Q_x + Q_y \tag{13.10}.$$

If the source Q is a one-time thing $Q_i \approx const.$, then this does not alter the rates of change, because constants $d/dt\, Q_i = 0$ vanish in the differentiation. If it is a permanent process, however, like the interest on savings, then this share does not disappear. These sources or sinks must therefore always be incorporated economically in $\dot{H}$ or $\dot{P}$ or both[65].

*This means that interest and compound interest is to generate only when trading volumes and prices rise accordingly. The same applies to income improvement or even decreases, which act according to their sign on trading volumes and prices as well.*

---

63 As long as the net foreign contributions are approximately zero.
64 The course can be taken also from financial products. Financial products are basically nothing more than another competitor. But special are the interests, which may represent an additional source Q, which will be examined in more detail later.
65 In fact, it is incorporated economically in both. This is due to the spiral symmetry of classical economics, which we treat in the general theory section. It is the symmetry under rotational-expansion.

# 14 Macro- and Microeconomic Substitution Rules

The general substitution results from the principle relation

$$\text{macroecon. demand} = \sum_i M_i V_i = \sum_j H_j P_j = \text{macroecon. supply} \quad .$$

Where we have seen in connection with the derivation of the law of Gossen after the *k-th* consumer choice $H_k$, we can now formulate the most general context of substitution:

$$\frac{\partial}{\partial H_k} \sum_{i,j} (M_i V_i - H_j P_j) = \frac{\partial}{\partial H_k} (M_k V_k) - P_k + \sum_{(i,j)-1} (M_i V_i - H_j P_j) \tag{14.1}$$

Because of the enormous diversity of economic products[66], ie $i, j \to \infty$, the last term provides the case of a particulate product decision for only one product

$$R_{VWL} := \sum_{(i,j)-1} (M_i V_i - H_j P_j) \simeq 0 \tag{14.2}$$

from which we get a zero contribution, since the removal of the *k-th* product makes no appreciable change in the total. The relation (15.2) is exactly the deciding factor as to whether the treatment of a system as macroeconmical (open system) or a microeconomical problem (closed system) is necessary or possible. The economic balance $R_{VWL} \simeq 0$ should disappear for microeconomical problems. So "systemically important" treatment is necessary, however, if this residue is significantly different from zero. Of "systemic importance" we can talk about when the changes of the *k*-products of a company leaving a residue $R_{VWL} \neq 0$, which approximates the rate of change of GDP's:

$$R_{VWL}^k := \sum_{(i,j)-k} (M_i V_i - H_j P_j) \geq \frac{dY}{dt} \tag{14.3}$$

Conversely, according to (15.1), we can write also:

$$R_{VWL}^k = \frac{\partial}{\partial H_k} \sum_{i,j} (M_i V_i - H_j P_j) - \frac{\partial}{\partial H_k} (M_k V_k) + P_k \geq \frac{dY}{dt} \tag{14.4}$$

and so after changing the elements

$$\frac{\partial}{\partial H_k} \sum_{i,j} (M_i V_i - H_j P_j) - \frac{dY}{dt} \geq \frac{\partial}{\partial H_k} (M_k V_k) - P_k \tag{14.5}$$

Because then always a relevance to the overall economy is given. One can thus express the last equations in words:

*Systemic importance is given if the total economic loss, given by the change of the trade of product k minus the change in GDP growth, is greater-equal to the damage caused by the change in trade of the product k for the producers themselve.*

The amount of *dY/dt has* relatively high levels by 10% at the start of a national economy and then will decreases continuously down to 0 or even will become negative. We can therefore say, that over time more and more producers get relevant to the system, because of the harm reduction effect -*dY/dt* in (15.5) will decrease. In this context, the effect of financial products is particularly interesting. Derivative financial products and the so-called bank's own business are characterized by the fact that money is traded essentially in exchange for money

---

66 $i=j$ is not mandatory, but can mostly be accepted.

or other monetary products (derivatives).

For a closer look, we also split up the capital side to the products of commercial banks (loans to domestic non-banks) $l$ and the products of the investment banks (banking own transactions) $m$ :

$$\sum_i M_i V_i = \sum_l M_l V_l + \sum_m M_m V_m = \sum_j H_j P_j$$

$$\text{with} \quad i = l + m \quad . \tag{14.6}$$

These $m$ financial products are themselves merchandise and you can write it thus:

$$\sum_l M_l V_l =: \sum_j H_j P_j - \sum_m V_m M_m \tag{14.7}$$

or

$$\sum_{a=1}^{i-m} M_a V_a =: \sum_{b=1}^{j+m} H_b P_b$$

$$\text{with} \quad H_m := V_m \quad \text{and} \quad P_m := M_m \tag{14.8}.$$

We now see the fundamental problem of the banks own business: Since this increases with time also $m$ increases. As a result, more and more financial products that come in substitution-competition with products of the real economy. In addition, the profits of these products have to come from the remaining capital stock of the real economy, which gets relatively smaller and smaller over time as with $a = i - m$ . This can be illustrated by again considering the quantity equation. For this purpose we differentiate now in respect to the $m_k - th$ trade (ie the *k-th* financial product of the number of $m$ ) of such a financial product:

$$\partial_{H_{mk}} \sum_{a=1}^{i-m} M_a V_a = \partial_{H_{mk}} \sum_{b=1}^{j+m} H_b P_b = P_{mk} + \partial_{H_{mk}} \sum_{b=1}^{j+m-(mk)} H_b P_b$$

and thus applies:

$$\partial_{H_{mk}} \sum_{a=1}^{i-m} M_a V_a - P_{mk} = \partial_{H_{mk}} \sum_{b=1}^{j+m-(mk)} H_b P_b$$

which reintegrates to:

$$\sum_{a=1}^{i-m} M_a V_a - \int P_{mk} \partial_{H_{mk}} = \sum_{b=1}^{j+m-(mk)} H_b P_b$$

$$\text{with} \quad \int P_{mk} \partial_{H_{mk}} \equiv \int M_{mk} \partial_{V_{mk}} \quad \text{per definition.} \tag{14.9}.$$

The net interest income of the *k-th* financial product is

$$Z_{mk} = \int P_{mk} \partial_{H_{mk}} \equiv \int M_{mk} \partial_{V_{mk}} \tag{14.10}$$

which will be withdrawn from the real economy available capital stock. All financial derivatives results in the overall balance of

$$\sum_{a=1}^{i-m} M_a V_a - \sum_{k=1}^{m} \left( \int P_{mk} \partial_{H_{mk}} \right) = \sum_{b=1}^{j} H_b P_b \tag{14.11}.$$

This means expressed in words:

*Capital investment of commercial banks minus net interest income in investment banking*
*= Capital available for real GDP goods trade* .

# 15 Comparison with Classical Economics

Macroeconomic Savings are of fundamental importance for the economic balance equation:

$$\frac{\partial \dot{M}_n}{\partial H_n} = \sum_i^n \frac{\partial \dot{M}_i}{\partial H_n} = \frac{\partial}{\partial H_n} \sum_i^n \dot{M}_i = \frac{\partial \dot{K}}{\partial H_n} =: \frac{\partial S}{\partial H_n} \tag{15.1}$$

which can be integrated easily in a natural way to $\dot{M}_n = \dot{K} = S$. We have used the fact that the sum of all the capital stock $M_i$ is represented by $K$ and $\dot{K}$ represents the annual increase in capital stock. In economic terms $\dot{M}_n = I$ is now investment, because we can regard $M_n$ as the sum of new products:

$$\dot{M}_n = \frac{d}{dt} \sum_k M_{nk} \tag{15.2}$$

So we obtain the well known relationship[67] in classical macroeconomics between investment $I$ and savings $S$

$$I = \frac{d}{dt} \sum_k M_{nk} = \dot{K} = S \tag{15.3}$$

and so finally

$$I = S \tag{15.4}.$$

<u>*Nota bene:*</u> **Savings $S$ are always including interest rates ( $\dot{K}$ ) on the whole capital stock $K$**, and Investments $I$ are into real economy and into financial products too. This is very different to common classical economy, where is stated $I=S$ to, but is meant as savings without interest rates and investment only to the real economy. This classical misconception leads to a lot of practical problems with predicting growth.

In classical macroeconomics applies to the economic savings of S the relation

$$S = I_{br} - D - F_{SA} \tag{15.5}.$$

This means that economic savings are equal to the gross investment $I_{br}$ less depreciation and any external borrowing $F_{SA}$. In the case of a balanced foreign trade balance is $F_{SA}=0$ and accordingly

$$I_n = I_{br} - D = S \quad \text{and} \quad I_{br} = I_n + D \tag{15.6}$$

for the net investment rate. $I_n$. It equals the gross investment less depreciation of $D$. In a closed economy savings should be fully absorbed by the net investment. For the economic savings but are usually set the amount of annual money put aside. So in our notation $S = p_s Y$ thus net saved money. It is, however as we have already verified, given the relationship

$$S = p_s Y = \frac{dK}{dt} - p_n K \tag{15.7}$$

(because $dK/dt = p_S Y + p_n K$ ). That means now that the economic savings at the beginning of development, where $dK/dt \approx 0$ is still relatively low, is actually given by

$$p_s Y = S = -p_n K = \frac{dY}{dt} = I_n \tag{15.8},$$

---

67 In a closed economy, there is a direct link between national saving and investment. This follows from the definition of the total demand function:Y=C+I+G. After conversion of the demand function it is obtained: I =Y-C-G. Substituting these with the national savings, then it follows: I = Y-C-G = (Y-C-T) + (T-G) = Sp + Sg = S and so I = S . (see e.g. http://de.wikipedia.org/wiki/Sparen )

So the classical context[68] is given.. At the end of the development but does the term $dK/dt$ take over leadership, since the capital coefficient, for example in the FRG is then increased by about 1/3 in the beginning to then even more than 3. It should therefore apply

$$p_s Y = S = \frac{dK}{dt} = p_s Y + p_n K = I_n \quad ??  \tag{15.9}$$

The contradiction is resolved when one recognizes that part of the economic savings are the interest $p_n K$ earned by savers of course also:

$$S = p_s Y + p_n K = I_n \tag{15.10}$$

This value gets very strong.with time, but no later than the crossing of the 50% mark at the bank corner office by the investment banking business. Thus the capital-output ratio in 1950 was just $K/Y \approx 0,3$ in th FRG. This means that in a typical savings rate of $p_S \approx 10$ % and a yield of about 5% brings the second term is just proportional $0,3 \cdot 0,05 = 0,015$, ie 1.5% of the 10% money put aside added. That may be neglected with a clear conscience. In 2008, however, was the capital coefficient $K/Y = 3,25$. In 2008, the second term that is brought about $3,25 \cdot 0,05 = 0,1625$, which is more than 16% added to the 10% and thus prevails the typical amount of money put aside clearly. But then the interest rates are no longer macroeconomically negligible.

*Important here is the realization:*

*In addition to the money put aside* $p_S Y$ *are to be counted the investment income from interest to the total economic savings. This difference in common opinion is of only marginal significance at the beginning of an economy. In the much later time of crisis but it is of great influence.*

## 16  Some Remarks on Technology, Productivity and Education

New products have to fight first for their part of GDP. This is for example well-known in the advertising industry, as a result of substitution economy. This fight is not abrupt, as with new and better products the therefore necessary additionally income is not generated automatically and at once. The classical growth theory expects, especially the already treated Cobb-Douglas production function, a noticeable jump[69] in the GDP by innovations. But one unfortunately cannot see this to be evident in the real numbers. Let us consider the above figure, where are placed in the Federal Statistical Office calculated GDP figures over a number of important events.

It is noticeable that GDP has, despite many technical and administrative innovation over time, rarely real jumps, but in general very continuously growing. So there was, above all, no jump after the introduction of the BAföG[70], which the first time allowed the university education of broad social strata in the FRG (keyword: human capital). Even the introduction of the small computers, the PC[71], which resulted in a rampant development in the computerization of offices and industry, but did not produced any discernible jump in GDP.

But the reunion with the German Democratic Republic GDR, which began in the run since 1989 and was largely completed by 1993, produced a very strong jump in the growth of the GDP. This jump resulted from the integration of the East German population in the FRG, which grew the population by about 25%. And thus induced an increase in the GDP's by 1200 to well over Euro 1500 billion.

---

68  Remember that $p_n$ is negative in our notation at the beginning and in the end gets positive.
69  A "jump" is characterized by an increase in slope of the function (which means in *df/dt*) and not just in the function *f* itself.
70  In 1971 was introduced BaFöG as student loans not paid back to full funding under Chancellor Willy Brandt for needy students. It guaranteed a wider circle of those entitled to receive a guaranteed and enforceable education funding. Under Chancellor Kohl, it was taken back to 1998 successively. Since 2001 it lives on in a reduced form.
71  First PC's were manufactured by Apple for the mass market. But only providing a freely available and not protected by patents cumbersome hardware platform from IBM, and a garage company called Microsoft for the operating system, made possible the triumph of the PC's from 1981 to today.

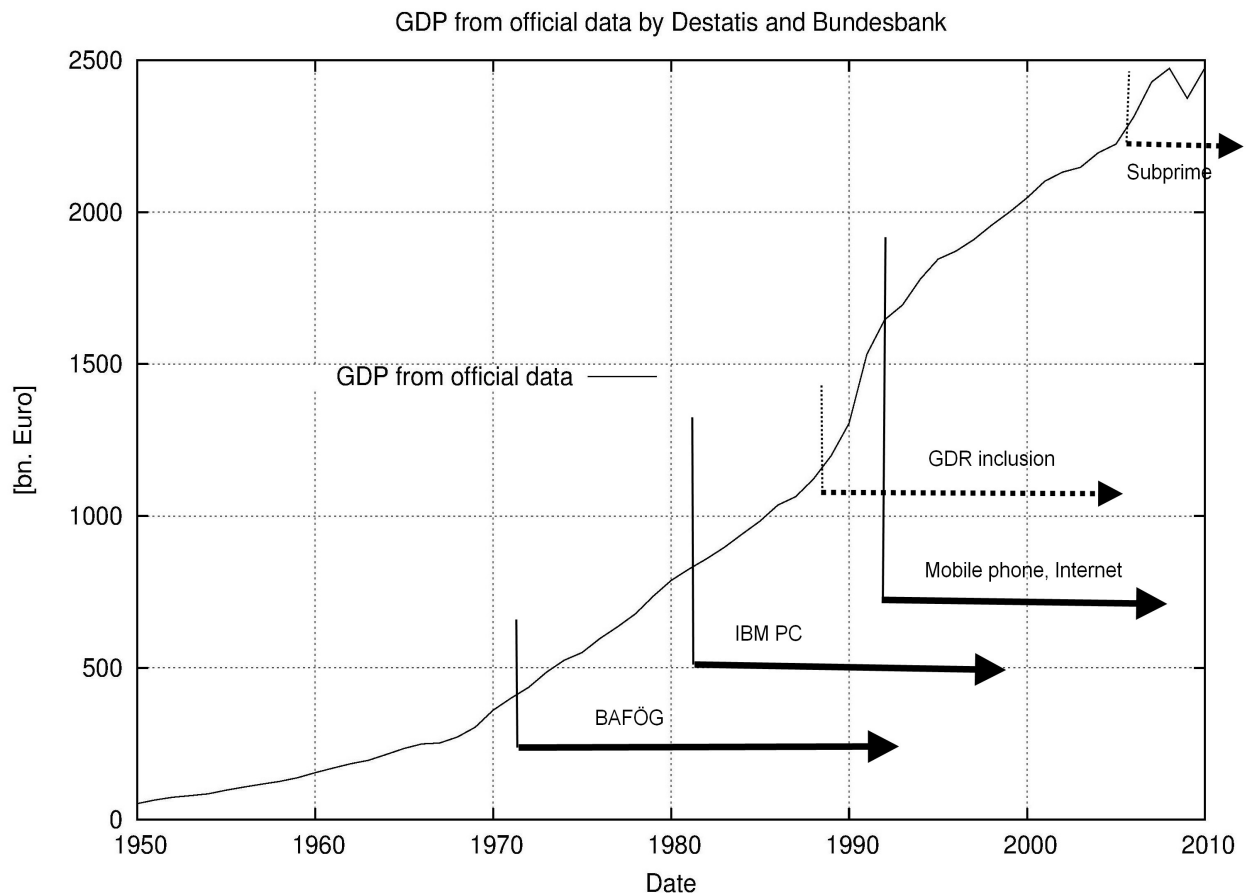

*Fig. 18: Evolution of GDP from 1950 to 2010 in the FRG. In addition to the official figure, the following markings are applied: Introducing first large program of grants for higher mass education [BAFöG], first IBM'PC with a Microsoft operating system, mobile phone boom (D-Net), Internet's first Web browser Mosaic. The largest externals to German GDP are marked with dotted lines: Integration of the GDR [DDR], and the subprime crisis which lead to the Lehman bankruptcy 2008.*

By contrast to common opinion, even the most important technical innovations did not induce any such jump. Nor did the PC, nor did it the mobile phone and not the internet too. On the contrary, they even flattened the slope of the GDP rather off. First in 2005, despite the lack of technology or large-scale educational innovations, we see a first real step. The reason might be sought in the tax reform begun in 2000 under Chancellor Schroeder which released additional funds for consumption. Of utmost importance is the Investment Modernization Act in force since 2004, but which allowed for the first time in the FRG so-called hedge funds.

With that an increasing job cut at the companies in the FRG took place in connection with the accompanying increase in the purchase and sale of all businesses and companies. This led to a significantly improved short-term revenue side of the business, but on the other side to a worsening decline of the demand to real and financial products. Especially since the large base component of consumption, the good earning middle class, has been disproportionately affected.

A little later the Lehman bankruptcy in 2008 led to a worldwide slump in the financial and economic markets too. Jumps in the GDP will only be possible if an increase in the money supply, in particular the wages of consumers, is associated. This is automatically and quickly rarest the case, but a medium-term effect of supply and demand for the product Labor, which of course is also a part of the full quantity equation. Since modern technology, so most of the PC and the Internet, however, made an increase in productivity without a boost in demand for the product Labor, this fizzled out obviously the effect of this so very sustainable technology innovations in the overall balance of the GDP's.

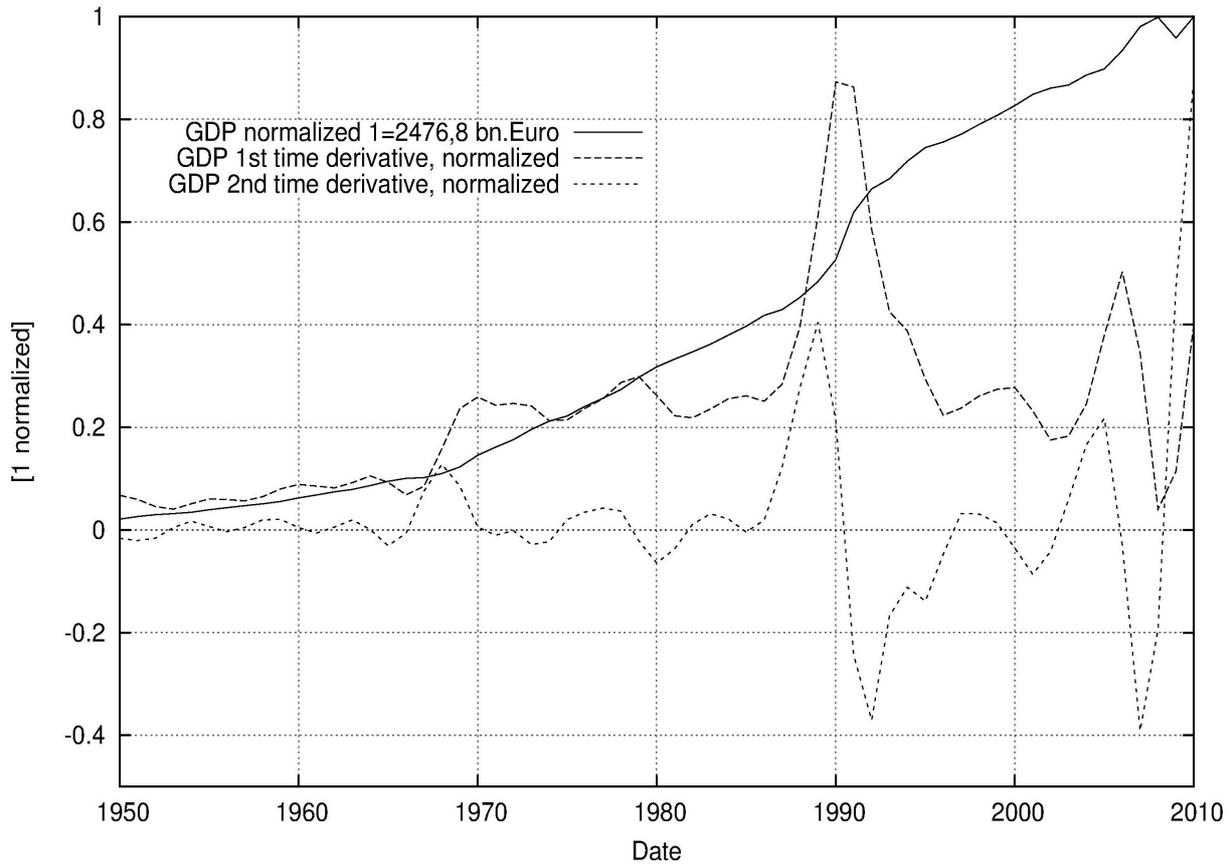

*Fig. 19:* GDP and its 1st (growth) and 2nd derivative (accelaration of growth) in respect to time. Normalized to max=1 for better comparison. Remark: The large Jump around 1990 is solely due to the effect of GDR inclusion. The next jump around 2008 is due to the Lehman-Brothers bankruptcy and the following depreciations of capital.

Due to the general substitution effect innovations are indeed no guarantee of an increase in the GDP's. But they are, just like the improvement of "human capital" through education, of course an indispensable condition for the competition between the economies to remain "tuned". Whether an economy grows more by agriculture or is able to grow by technology products, that is less obvious from the product, as from the availability of capital on the one hand and enough consumers of products on the other hand.

Productivity can be increased e.g. by advances in technology and/or in the education of the working people. To introduce the role of productivity into the equations, one can do this with an approximate or an exact ansatz. The approximate ansatz was first introduced by Bürkler and Besci [Besci, 2012] and does the following:

$$\frac{dY}{dt} = (p_B(t) + p_P(t))Y - p_n(t)K \qquad (16.1)$$

where $p_P(t) := ...$ is the function of *effective* productivity increase (decrease) which has to be modeled by e.g. using statistical data. This phenomenological ansatz will work, as the measured effective productivity increase/decrease is incorporated. But it will give no distinct answers to the question, *how* productivity works economical, e.g. in which way investments into productivity increase are effective for the supply of assets, labor and wages and at last to the GDP. To address such questions, one has to consider the retroactivity between investments into technology or education on the one side, and the substitutional effects with other products on the other side. This can only be done by constituting more dependent differential equations into the system. This will be addressed in the second part of the article (general theory). Introducing further parts, e.g. like labor *L* and wages, investments to real economy I, investments into technology T, investments into education E and so on will give further interdependent differential equations, e.g. the very complex system of six coupled differential equations:

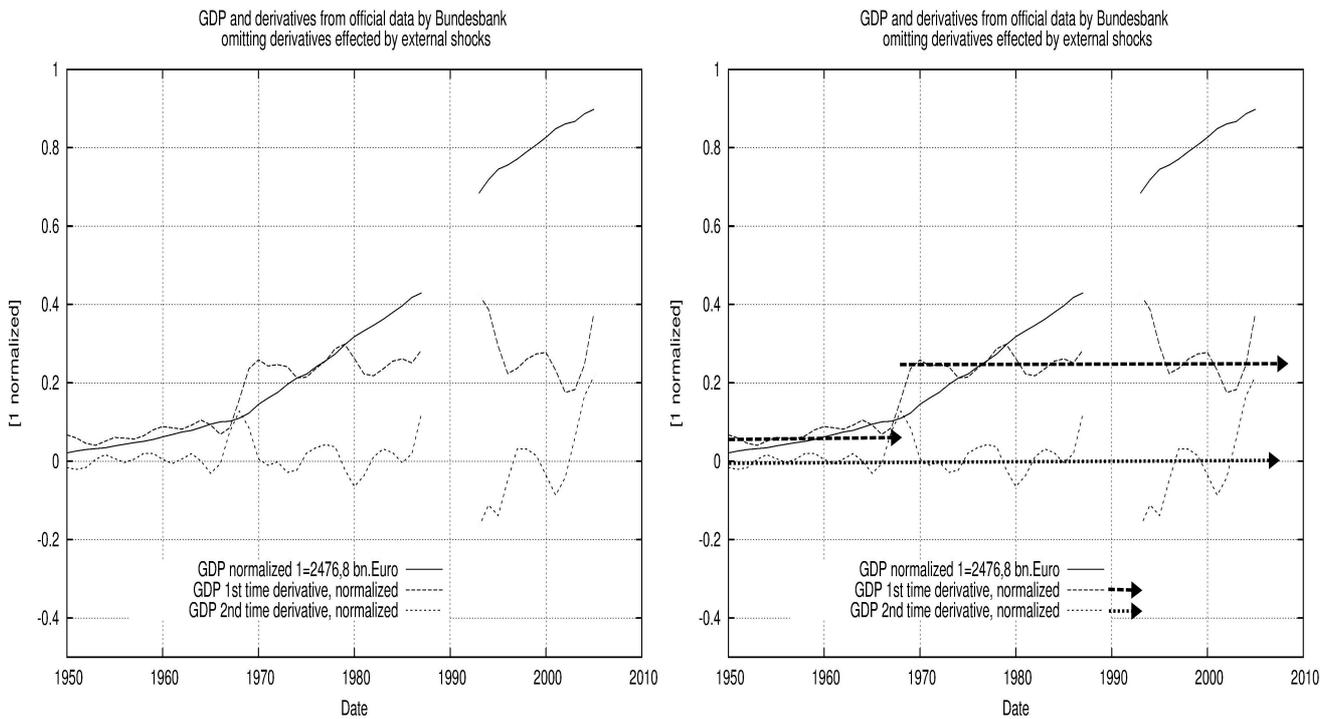

*Fig. 20, left:* Gdp and its 1st and 2nd derivative. The times of externals shocks are omitted. *Fig. 17, right:* Overall the second derivative does not depart very much from zero. There is not any significant deviation which can be related to the main educational or technological revolutions. Only in the 1st derivative one can see the very effect of the phase change in1967, as the economy changed from GDP driven to Capital driven economy: there was clearly an increase of growth at the time of that change.

$$dY = f_1(Y,L,I,K,E,T); \quad dL = f_2(L,Y,I,K,E,T); \quad dI = f_3(I,Y,L,K,E,T);$$
$$dE = f_4(E,Y,L,I,K,T); \quad dT = f_5(T,Y,L,I,K,E); \quad dK = f_6(K,Y,L,I,E,T)$$

which then have to be derived from fundamental consideration (general nonlinear theory) and/or from additional statistical well funded assumptions about micro-parameters (linear theory).

The next step but must and will be introducing at least Labor *L* to the side of real goods giving

$$M_G V_G + M_W V_W + M_I V_I = H_G P_G + H_L P_L + H_I P_I$$

as the quantity equation to be dealt with. Here we have to split the two sectors into their main products $H_R P_R = H_G P_G + H_L P_L$ with Labor index *L* and $M_R V_R = M_G V_G + M_W V_W$ with Wages index *W* and all other products summed under the index *G*.

As the richness and analytical power of the equations of field theory grow exponentially with every additional financial or real product introduced, we can concentrate in this article only on the *minimum non trivial splitting of macro-economy* which is given in principle by

$$M_R V_R + M_I V_I = H_R P_R + H_I P_I$$

here. Otherwise the amount of analytical theory and reasoning would exceed a reasonable page limit. But it must be remarked that the needed steps, in linear and non-linear theory as well, are just straight forward and limited only by the needed amount of practical work to do, but not on theory.

# 17 Rules of Public Debt

The British economist David Ricardo (1772 - 1823) described public debt as *"one of the most terrible scourges that have ever been invented to plague the nation"*. So today the level of government debt is regarded as one of the most serious problems of the economy. This kind of debt is not a problem limited to just a few nations, but appears as a general, perpetual and recurring national dilemma.

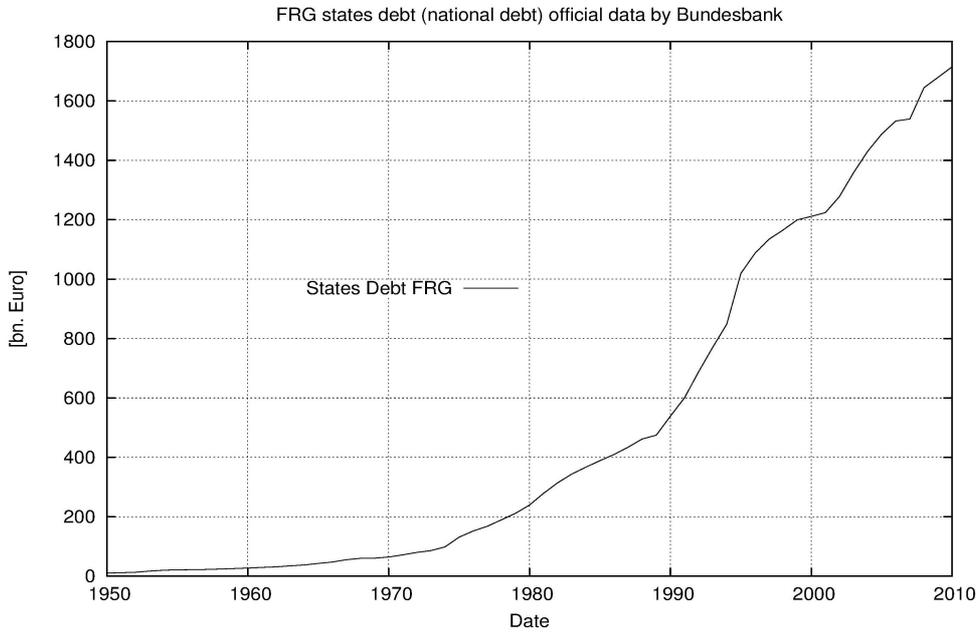

*Fig. 21: Government debt in the FRG 1950 to 2010 according to official data [in bn. €]*

Therefor we consider the special substitution rules for national debt. In our substitution equation with respect to the banks own business, we had already split the financial side off to the two main blocks. The same we can do with the right side $HP$ of the equation

$$\sum_l M_l V_l = \sum_k H_k P_k + \sum_p H_p E_p + \sum_q H_q S_q - \sum_m V_m M_m$$

$$\text{with} \quad i = l + m \quad \text{and} \quad j = k + p + q \tag{17.1},$$

where we have put financial products to the right hand side now including all products. We have to mention here, that also all incomes $E$ and governmental revenues $S$ are broadly speaking commercial products, because they always have to be paid out of the immediate use of capital $\sum_l M_l V_l$ in the final analysis. Government revenues[72] in turn are composed of a variety of taxes $T$, duties $D$ and fees paid $F$, etc.:

$$S = T + D + F + etc. \tag{17.2}$$

With income $E$ we mean all forms of income, including wages W, salaries $S_{alaries}$, pensions $P_{ensions}$, social welfare $S_{welfare}$, and so on. This is also including the income arising in the form of business profits $\Pi_{gross} = \sum_i \alpha_{\Pi i} H_i P_i$, where $\alpha_{\Pi i}$ is the $i$-th profit margin:

$$E = W + S_{alaries} + P_{ensions} + S_{welfare} + \Pi_{gross} \tag{17.3}$$

---

72 It must be said, that also taxes are commercial products, because they make it only possible to provide the necessary state boundary conditions for a prosperous economy and infrastructure ready. Although it is a compulsion of purchase it makes no difference to the balance equation. So it is with pensions, and even with social assistance. In each case, products that must be purchased from various good reasons by the community.

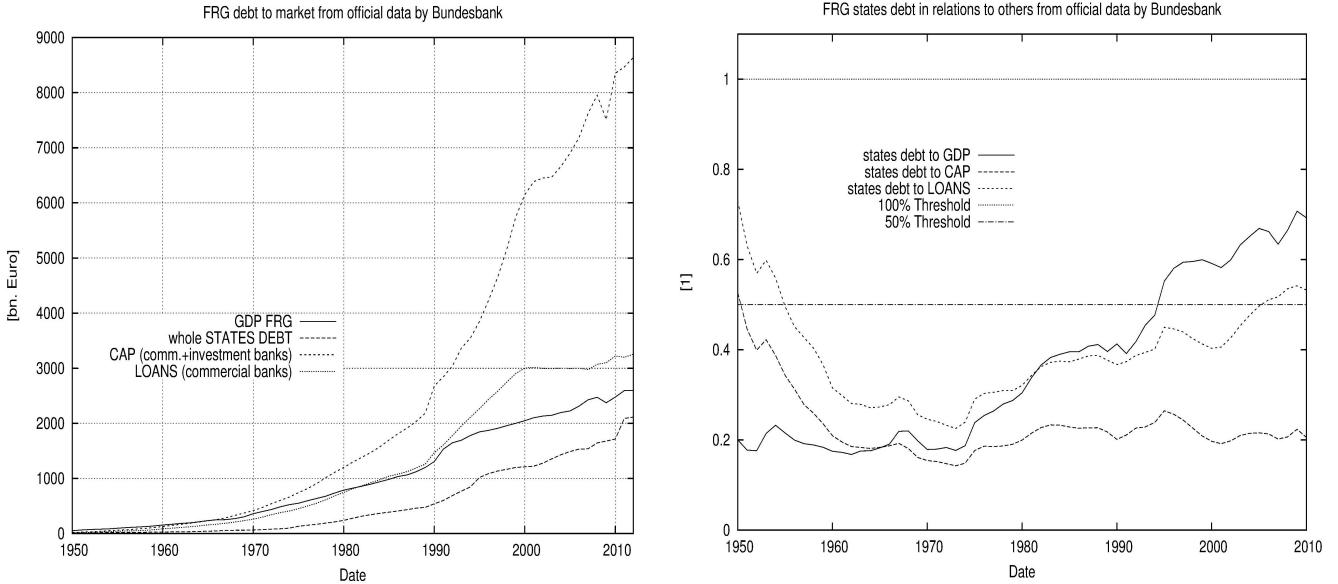

*Fig. 22: Government debt to GDP and Government debt ratio to the total assets. FRG official data from 1950 to 2010. The long term median according to the actual phase of economy was always around 20 %=1/5.*

As between income[73] from labor[74] $E_{p1}^B$ and income without work $E_{p2}^U$ (pensions, social welfare, subsidies, etc.), there is a certain difference, we separate it from even further

$$\sum_l M_l V_l = \sum_k H_k P_k + \sum_{p1} H_{p1} E_{p1}^B + \sum_{p2} H_{p2} E_{p2}^U + \sum_q H_q S_q - \sum_m V_m M_m \qquad (17.4)$$

with $p = p_1 + p_2$. All these products are in competition for the available money $\sum_l M_l V_l$. The most important question is, of course, how the state now uses its revenues

$$\sum_q H_q S_q = \sum_q V_q M_q^S \qquad (17.5)?$$

The particular state of the revenue is now, however, that the state builds up generally no significant reserves, but uses the money raised this year to pay next year's bills. He does it for infrastructure and subsidies of all kinds, including the social obligations, whether in the public service or by awarding contracts to private companies. The total expenditure to be received from taxes, fees and charges from local, state and federal government and spent again, is called the *public expenditure quota*[75]. The macroeconomic balance equation can now be analyzed as follows:

*This year T :*

$$\sum_l^T M_l V_l - \sum_q^T V_q M_q^S = \sum_k H_k P_k + \sum_{p1} H_{p1} E_{p1}^B + \sum_{p2} H_{p2} E_{p2}^U - \sum_m V_m M_m \qquad (17.6)$$

It suggests that government spending initially is negative on the side of the available capital to GDP, because the charges are withdrawn from the total money in circulation and thus the general consumer.

---

73 The individual actions $H_x$ are of course the respective trades $buy/y$. In the case of wages or taxes are this, as usual, the annual number of actions. Actions are purchases, specifically the acts of each payment, which are the representative of a purchase.
74 Instead of the expression "from labor" is more accurate "from general trading activities" should be used.
75 This quota is at the time about 47% of GDP in the FRG. Quotas around 50% are international practice, and the states using quotas above 50%, as Sweden ,are called a "bold" state, and countries such as Switzerland with rates below 40% as a "lean state".

*Next year T+1:*

$$\sum_{l}^{T+1} M_l V_l - \sum_{q}^{T+1} V_q M_q^S + \sum_{q}^{T} V_q M_q^S = \ldots$$
$$\ldots = \sum_{k}^{T+1} H_k P_k + \sum_{p1}^{T+1} H_{p1} E_{p1}^B + \sum_{p2}^{T+1} H_{p2} E_{p2}^U - \sum_{m}^{T+1} V_m M_m \quad (17.8).$$

But the last sum goes back by public expenditure next year practically completely to the volume of trade. And thus almost completely adds to the GDP, because the state uses that money to buy goods and services, public and private. So we may write:

$$\sum_{q}^{T} V_q M_q^S = \underbrace{(\sum_{k}^{T+1} H_k P_k + \sum_{p1}^{T+1} H_{p1} E_{p1}^B + \sum_{q}^{T+1} H_q S_q)}_{\approx 50 \text{ percent}} + \underbrace{\sum_{p2}^{T+1} H_{p2} E_{p2}^U}_{\approx 100 \text{ percent}} \quad (17.9)$$

The difference then is the effect of the perennial nature:

$$\sum_{l}^{T+1} M_l V_l - \sum_{q}^{T+1} V_q M_q^S + \sum_{q}^{T} V_q M_q^S \neq \sum_{l}^{T+1} M_l V_l \quad (17.10)$$

for the multi-year state budget balance is

$$\Delta S_f := \sum_{q}^{T} V_q M_q^S - \sum_{q}^{T+1} V_q M_q^S \neq 0 \quad (17.11).$$

which is in most cases but negative: Since taxation is not popular in any political system, that takes account of public debt rising most frequently with time.

<u>The question now is, what can a state</u> $\Delta S_f < 0$ <u>provide at least in the general good times, anyway?</u>

When he does not want to unduly burden the capacity of economic operators, then he can draw on the natural[76] excesses of the community. This is essentially the savings rate, because this indeed describes the property which is not currently required and is therefore parked on savings accounts. It is therefore that this ratio ultimately can[77] be used as a permanent loan overhang[78].

$$\begin{aligned} \Delta S_f &< 0 \quad \text{deficient state} \\ \Delta S_f &= 0 \quad \text{ideal state} \\ \Delta S_f &> 0 \quad \text{profitable state} \end{aligned} \quad (17.12)$$

However, he does not pay it back completely, but only gradually, with the surplus resulting from each $p_S$

---

76 We call it "natural" as savings are the amount of money which is not spend and thus not needed for consumption and thus also not needed for building up the actual amount of GDP.
77 Can or must? It seems that it must from the internal rules of money creation. It is as banks usually use the opportunity to get cheap money by central banks to create their credits to be sold to the public. There is no need to pay average higher interest rates to the people's savings. Therefore in the end its always the government and its associated central banks which have to create the additional money needed for the technical function of economy. Especially the state has to burden the amount of people's savings with the higher average interest rates for financial products. For example, the savings rate in Germany where in average about 10% a year. In Japan but this rate was up to 25% over decades. Thus it is not so astonishing that Japans national debt of today more then 200% is more than the double of Germany's national debt. Also not surprisingly, the average interest rates on public debts always coincide nearly with the average interests on simple deposit accounts.
78 The need for additional money arises from growth (which approximate equals interest rates and inflation too) and visa versa. Indeed it comes from savings and their demand for interests. From this it is indeed unavoidable and inevitable that national debt increases in time at the same principle rates as the whole of all assets and debts.

remains open. This debt must bear interest at the average rate $p_A$ on government bonds of different maturities:

$$S_{SS}(T) = S_0 + \sum_T [p_S(T) \cdot Y(T)] \cdot (1+p_A)^{T-T_0} \qquad (17.13).$$

The number $S_0$ ist the starting value, which may be zero or the actual states debt at the very beginning of the integration $S_0 = S_{SS}(T_0)$.

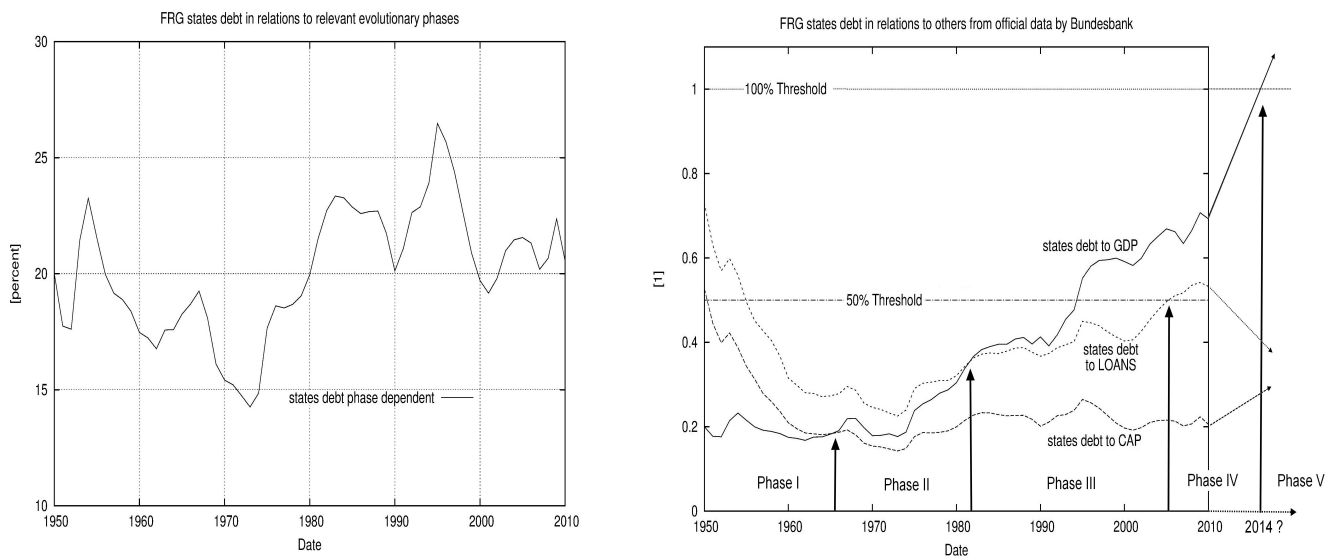

*Fig. 23, <u>Left:</u> Average states debt in relation to the relevant phase of the economy. <u>Right:</u> Debt evolution in relation to GDP and Capital. The evolution of an economy with respect to the national debt (states debt) can be divided into five relevant phases. Phase I: GDP is higher than total capital stock; Phase II: CAP gets greater than GDP; Phase III: States debt to GDP exceeds states debt to Loans; Phase IV: states debt to to Loans exceeds the states quota threshold (here 50%); Phase V: States Debt exceeds 100% of GDP.*

We thus calculate the sum (19.12) from official data and record them against the actual official government debt. The result shows the next figure, where we have set an average approximate interest rate on government bonds of different maturities of $p_A = 0{,}03 = 3$ %. In fact this assumption can explain the fundamentals of the national debt. Deviations[79] are explained by special state spendings, such as the GDR acquisition or stimulus packages or with saving programs.

It is worthwhile to analyze the real public debt of the FRG in their symptomatic course: Surprisingly, we see that the national debt in no way gets out of hand, when considered in relation to the total financial capital stock *K* considered. On the contrary, one can see in Fig. 23, which are the official real numbers, that debt was even very stable at a value of about 20% of the financial capital stock and remains this up to the year[80] 2010.

---

79  In general the development of the interests for governmental bonds do not always coincide with the general interest rate trends in other kinds of assets. This is not least because government bonds are usually seen as relatively safe bonds, while risky assets then make possible higher returns.

80  After then national debt is growing in this relation, since at the high phases of Euro-Crisis the state began to take over very much banks debts, not only from Germany but also from other European banks, in addition to its "natural" public debt rate.

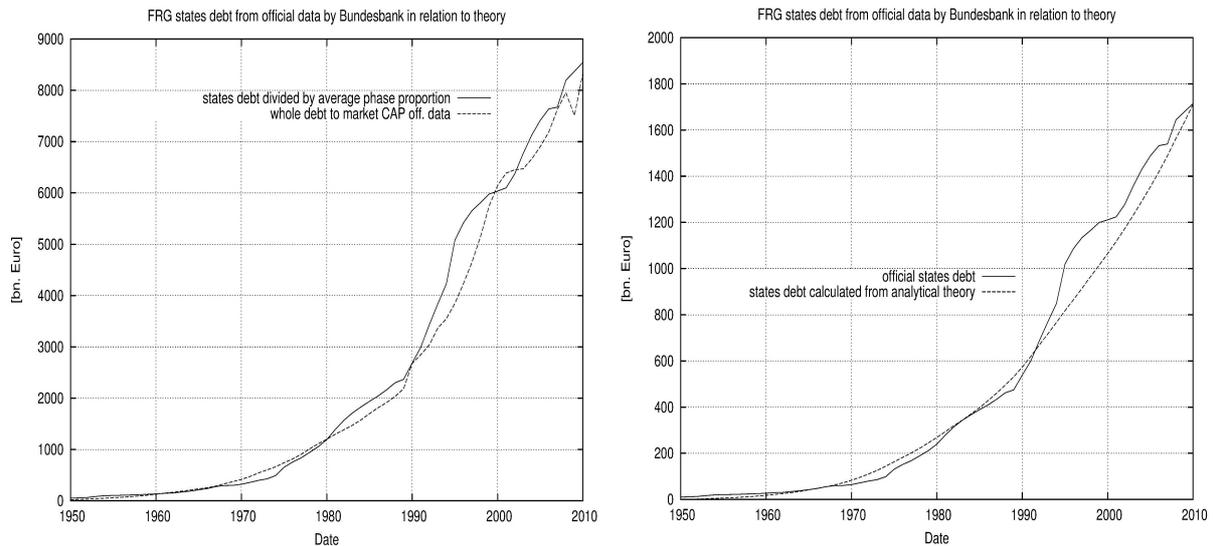

*Fig 24: States debt from official data in comparison with theoretical functions: <u>Left:</u> States debt divided by average proportion (here around 20%=1/5 which means simply multiplying states debt by a constant factor of 5). <u>Right:</u> In comparison with the calculated integrated sum (19.12) of the savings rate (from official data) assuming an average interest rate of 3%.*

Of particular interest is the turning point around the year 1967, when the capital-coefficient for the first time exceeded the value 1 (=100%): Previously, the national debt was about 20% of GDP and also remained constant, while with respect to the total capital stock was considerably higher. In principle, less surprisingly however, that the national debt is based on the general capital development, and less on GDP.

Because the state is an commercial operator like everyone else, and can effectively refinance only with the general capital development. As long as the state's share of the overall debt is kept constant, one can not speak of a high public debt. In fact, the state is based on the resources available to him. These are at the beginning of economic development, as the capital-output ratio $K/Y$ was still less than 1, the driving factor was the real economy. But then increasingly changes to the capital driven economy with increasing $K/Y>1$.

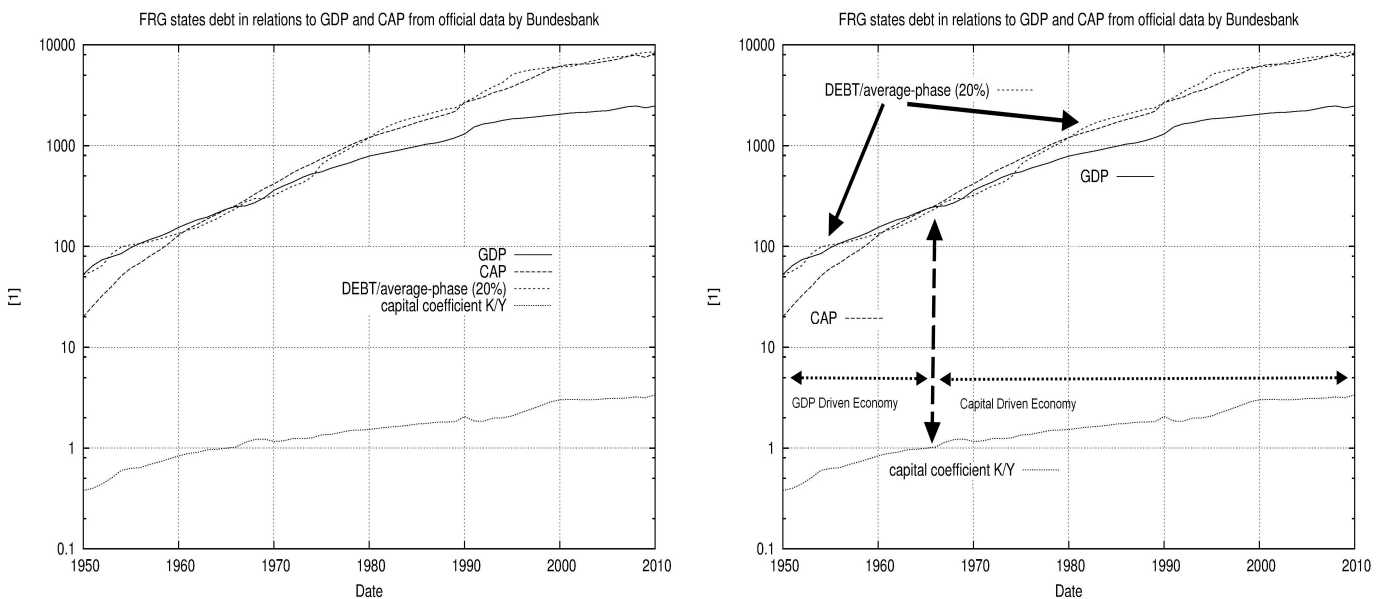

*Fig.25 <u>Left:</u> Official data of GDP and CAP, with states debt times 5 overlaid. Also the capital coefficient K/Y is plotted. <u>Right:</u> The resulting two main phases: Phase I is GDP driven; Phase II is capital driven. The change can be seen at the time when the capital coefficient K/Y exceeds 1=100%. Before this point states debt follows GDP, after this point staes debt follows Capital instead. All figures plotted from official data only.*

The above effect of the phase change is best visible in a logarithmic chart: You can clearly[81] see how the public debt is initially linked to GDP and after exceeding the value of *K/Y = 1* linked to the capital stock. Accordingly, the percentage of public debt $p_{SS}$ then must be correctly measured in each time by the underlying major resources:

$$p_{SS} = \frac{S_S}{Y} \quad \text{if} \quad \frac{K}{Y} < 1 \quad \text{and} \quad p_{SS} = \frac{S_S}{K} \quad \text{if} \quad \frac{K}{Y} > 1 \tag{17.14}.$$

Then there is the following picture: According to (Fig. 23 and 24) the national debt by 2010 in the FRG was always in the average at 20.1% of the economically dominant[82] resource. This value but usually is communicated in relation to GDP. But then such a number essentially just says something to the time it would take the total economic debt to be paid by spending the entire GDP to pay it off. It is also clear that with a total capital stock of around 8,000 billion euros a national debt of 1800 billion (year 2010) is the lower portion of the economic debt. About 80% of the debt[83] is in fact of private nature.

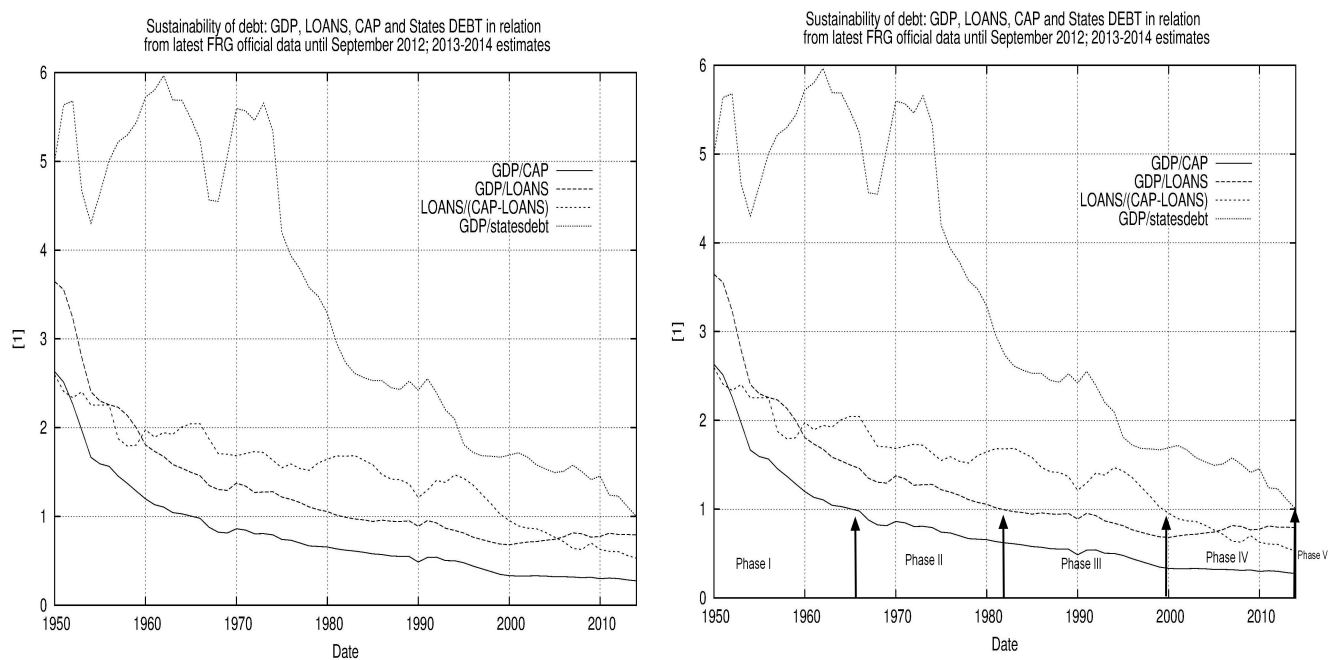

*Fig. 26: Sustainability "Mountain" of the economy. <u>Right:</u> pure Official data; <u>Left:</u> Same plot with explanations. Phase I: GDP greater than total financial Capital-stock; Phase II: Capital-stock is greater than GDP, but Loans to GDP are still smaller than GDP; Phase III: Even Loans to GDP are now greater than GDP; Phase IV: the Commercial banks business (LOANS) gets smaller than Banks-Own-Business; Phase V: States debt gets larger than GDP. The lender of last resort thus gets out of power.*

---

81  To clarify: The above figure is calculated by using real measured values as input to the model equations.
82  As the national debt is measured in euros, they should be in a reasonable manner whatsoever communicated in relation to capital stock.
83  Today money is covered solely to debt. Each property has to face in the same amount of debt. The balance sheets of financial institutions on asset side (assets) and liabilities side (deposits) is always balanced.

# 18 Analysis of Inflation

The official inflation levels are determined by the statistical offices by calculating the yearly percentage change in the value of a special[84] product basket (CPI). To make matters worse, these values are in some countries increasingly effected by the method of hedonic[85] regression. Inflation but is an intrinsic effect of all economies and can be derived from the sources and sinks and thus calculated also theoretically.

The classical quantity equation *MV=HP* is just a singular equation where we but would need at least four independent equations, one for every unknown function. The classical QE but gives at least a tool for some simple structural analysis. We differentiate the quantity equation $P = \frac{KV}{H}$ with respect to the time *t* and then we sort it by price level to get

$$\dot{P} = \frac{V}{H}(\dot{K} + K(\frac{\dot{V}}{V} - \frac{\dot{H}}{H})) \qquad (18.1).$$

Particularly important is the last term in the above equation, because there only the sign is negative and in addition *H* in the inverse square is received. This means that we get a very *critical trading-term* as is

$$i_H := -\frac{KV}{H^2}\frac{dH}{dt} \qquad (18.2).$$

This may have a significant impact on price levels, because under special circumstances $i_H$ may completely reverse the trend, and by reason of its reciprocal value can even well let it "explode". It is further given by a simple transformation using *KV=HP=Y* identity:

$$\dot{P} = \frac{\dot{Y}}{H} - \frac{Y\dot{H}}{H^2} \qquad (18.3)$$

reflecting the fundamental importance of the trading volume for the pricing level. The inflation rate is now a relative value, namely the quotient of temporal price change to the price level. Dividing by P and resorting and by introducing the sign $I := \frac{\dot{P}}{P}$ for the definition of inflation we come to:

$$I = \frac{V}{Y}(\dot{K} - K(\frac{\dot{H}}{H} + \frac{\dot{V}}{V})) \qquad (18.4)$$

---

84 These values are a more or less good approximation to the actual inflation, depending on the chosen basket and the algorithm for calculation. Deflation is in accordance with the case when the basket gets cheaper. However, due to the nature of the investigation depending on the selected basket, that will and must also change over time necessarily, as the consumer desires and habits change, as well as with technological changes and social change does apply, this number is always approximately.

85 Hedonic regression processes lead to reduced rates of inflation and, if applied, to higher official GDP and thus GDP growth. Depending on the scope of hedonic regression, the distorting effects can be enormous. Correctly, governmental institutes should publish nominal raw data in addition to the institute's hedonized figures, as for economic growth modeling just the raw data matter. They do this however rarely, and also the exact method of calculation of the specific shopping baskets and the special hedonizing rates remain mostly in the unknown.
Hedonic regression is used to calculate CPI (inflation) and GDP in the US, GB, AUS and New Zealand since the 1990ies. For example, if the processor speed of computers has increased due to improved techniques from 2,000 MHz to 3,000 MHz, but the prices have remained constant, the increase in frequency now may be taken as a quality improvement of 50 percent. This means that hedonic prices for CPU's fall in the CPI by a third. For calculating hedonized growth or hedonized productivity levels, this process works vice versa. So it is assumed that the real prices in the determination of the value has to be corrected according to the quality improvement of 50 percent upwards too. In addition there is always a subjective decision as to whether there is a qualitative improvement of a product or not. Strictly speaking one would also have to take account of quality losses, but which is almost never done. For economic calculations hedonic regressions can never be of any use, as the demand and supply of goods are just strongly correlated with the availability of free money for the consumer. But this fact does not vary with the quality of products in principle, but does with real prices.

This fundamental equation we may consider briefly. The principle equation $I = f(t) \cdot \dfrac{V}{Y}$ represents the classically known fact that the inflation rate is mainly determined by the ratio of monetary velocity *V* to GDP *Y*. But furthermore the factor $f(t) := \dot{K} - K\left(\dfrac{\dot{H}}{H} + \dfrac{\dot{V}}{V}\right)$ is still effective. This factor modulates the rate of inflation with the development of the financial stock of capital. Thus, the inflation rate gets higher if the capital $\dot{K}$ increases, which reflects the fact that the price of capital is one of the main factors for the prices of goods. It gets however smaller, when a lot of capital *K* is available, multiplied by the relative growth of *H* plus the relative growth of *V*. This is because the last term usually generates a cash inflow to GDP, which reduces the need for capital and thus the price of capital, and thus also for produced and traded goods. From this equation we can now also derive an ODE for what is the fundamentally important trade volume, by solving the inflation differential equation for *H*:

$$\dot{H} = H\left(\frac{\dot{K}}{K} + \frac{\dot{V}}{V}\right) - H^2 \frac{\dot{P}}{KV} \tag{18.5}$$

A closed integration of this ODE is not possible without some additional conditional equations. But for a structural analysis we can do a case distinction. We can now define partial solutions by assuming first of all, that the unstable phase, namely the validity of the quadratic term H has not been reached yet. Disregarding the quadratic term of *H* integrates the expression slightly. The result is analytically the quasi-stable solution:

$$H(t) = c e^{\frac{\dot{K}V + K\dot{V}}{KV}t} = c e^{\frac{(\dot{KV})}{KV}t} = c e^{\frac{\dot{Y}}{Y}t} = H_0 e^{p_W \cdot t} \tag{18.6}$$

The quotient $p_W := \dot{Y}/Y$ is the growth rate of GDP as it is usually communicated, and the constant of integration $c := H_0 = H(T_0)$ is the trading volume at the start of integration. This means at the beginning of an economy that economic growth goes hand in hand with the trading volume actually increasing exponentially at first. The time *t* is counted here from the zero point of the national economy or the first point of calculation. But when after the saturation phase crisis occurs, the quadratic term $H^2$ gets increasingly important. And even may take the lead. By deleting the then much smaller previous terms we can integrate the last term separately, and thus obtain the solution of the quasi-unstable state after the critical time $t \geq T_k$:

$$\Delta H(t' = t - T_k) = \frac{1}{\dot{P}} \frac{KV}{t'} = \frac{Y}{\dot{P}} \frac{1}{t'} = \frac{Y}{I \cdot P} \frac{1}{t'} \tag{18.7}$$

$$\text{with inflation} \quad I = \frac{\dot{P}}{P} \quad \text{and} \quad H = H_k + \Delta H \tag{18.8}.$$

As we may see is that, if deflation *I<0* occurs, then trade *H* will diminish. For the time being greater equal $t \geq T_k$ the critical time we get for the trade volume on the crisis path:

$$\Delta H(t') = \frac{Y}{\dot{P}} \frac{1}{t'} \approx \frac{Y}{\dot{P}} \frac{1}{t' - \ln\left(\dfrac{1}{p_s} - 1\right)/\ln(1 + p_n)} \tag{18.9}$$

In the ultimate crisis, but only there, we get the problem that the price increase which is usually more trade promoting may become but then a devastating effect because of its steep slope: it will lead to a trade decline, which can may be a desastrous effect because of the inverse square term (18.2) of the trade *H*:

$$i_H := -\frac{KV}{H^2} \frac{dH}{dt} = -\frac{\dot{H}}{H^2} Y \tag{18.10}$$

Namely $i_H$ then gets sharply positive, as will be $\dot{H} < 0$ then. The result is a tighter inflation, and thus threatens to refuse additional trade, making the term growing more fast. Also, a simultaneous decline of GDP *Y* cannot stop this, as *Y* enters only linearly. In the very late end of financial crises the term (18.9) thus will lead to the well known hyperinflation catastrophe.

Last we have to introduce an expression for $\dot{P}$ for further analysis. For this we exploit the fact that the price level before the crisis is fully determined by

$$P_R = \frac{Y}{H_0 e^{\frac{\dot{Y}}{Y} \cdot t}} \left( (1-p_s) Y + (1+p_s) \dot{K} \right) \tag{18.11}.$$

It is then given by a longer[86] calculation:

$$\dot{P} = \frac{e^{-\frac{\dot{Y}}{Y} \cdot t}}{H_0 Y} (a \cdot t + b)$$

with $a(t) := Y(\dot{Y}^2 - \ddot{Y}) + \dot{K}(\dot{Y}^2 - Y \ddot{Y}) + p_S \ddot{Y}(Y^2 - \dot{K}) + p_S \dot{Y}^2 (\dot{K} - Y)$

and $b(t) := Y^2 \dot{Y}(1 - p_S) + Y^2 \ddot{K}(1 + p_S) + Y^2 \dot{p}_S (\dot{K} - Y)$ (18.12)

This expression we can now invest in the above formula (18.6) of *H(t)* in order to calculate the volume of trade in normal times. It is therefore given the approximate formula

$$H(t) \approx H_0 \cdot \frac{Y^2}{(a \cdot t + b) \cdot \left( t - \frac{\ln(\frac{1}{p_s} - 1)}{\ln(1 + p_n)} \right) \cdot e^{-\frac{\dot{Y}}{Y} \cdot t}} \tag{18.13}$$

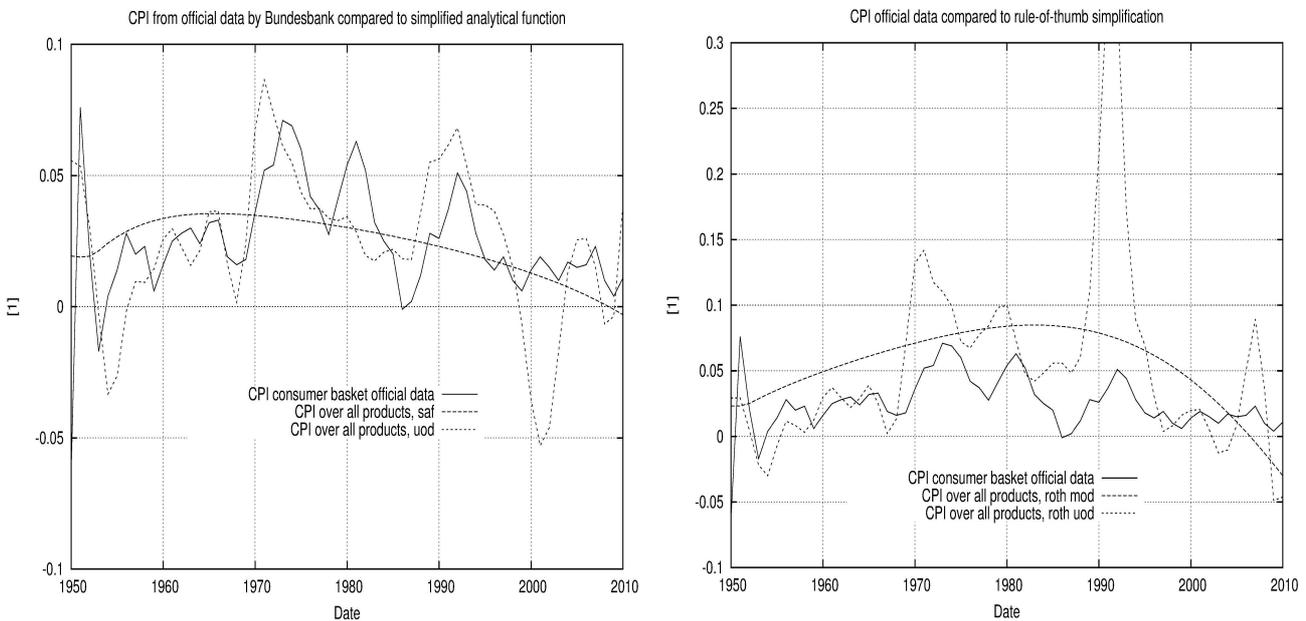

*Fig. 27: Analytical and official data for CPI in comparison.*

The normal rate of inflation in turn is then calculating[87] $I(t) = \dot{P}/P$ analytically, which gives:

---

86 In this case there is some work to do. Therefore one should seek help from analytical Algebra programs like Maple or Maxima.
87 As we consider $P_R$ only, we may set $P_R = P$.

$$I(t) = \frac{1}{Y(1-p_S) + \dot{K}(1+p_S)} \cdot \left( \left( \frac{\dot{Y}^2}{Y} - \ddot{Y} \right) \cdot (1-p_S) + \left( \frac{\dot{K}\dot{Y}^2}{Y^2} - \frac{\dot{K}\ddot{Y}}{Y} \right) \cdot (1+p_S) \right) \cdot t + \ldots$$
$$\ldots + \dot{Y}(1-p_S) + \ddot{K}(1+p_S) + \dot{p}_S(\dot{K}-Y) \right) \tag{18.14}$$

This last expression, although calculable by numerical power of computers easily, is not very manual for analytical examinations. But we may simplify it for further structural analysis. Such as the last term $\dot{p}_S \approx 0$ vanishes for slowly variable savings rates. By regarding that the savings rate is mostly significantly less than 1, it can also be assumed that is $1 \pm p_S \approx 1$ approximately, which results in:

$$I(t) \approx \frac{\left( \frac{\dot{Y}^2}{Y} - \ddot{Y} + \frac{\dot{K}\dot{Y}^2}{Y^2} - \frac{\dot{K}\ddot{Y}}{Y} \right) \cdot t + \dot{Y} + \ddot{K}}{Y + \dot{K}} \tag{18.15}$$

Even further we can assume from experience that the second derivative of growth $\ddot{Y} \approx 0$ is very small, because the growth of the growth is almost very low. But certainly this is not valid for the compound interest $\ddot{K}$, which stems[88] from the ever growing total capital stock. For this, the analytical core inflation $I_c$ may be written as:

$$I_c(t) \simeq \frac{\left( \frac{\dot{Y}^2}{Y} + \frac{\dot{K}\dot{Y}^2}{Y^2} \right) \cdot t + \dot{Y} + \ddot{K}}{Y + \dot{K}} \tag{18.16}$$

We can rearrange the last equation and express it in terms of slip rates of growth as:

$$I_c(t) \simeq p_W^2 \cdot t + \frac{p_W Y + p_v^2 K}{Y + p_v K} \tag{18.17}$$

The second term in the numerator and also in the denominator of the fraction is evidently smaller[89] than the first one. So we may write:

$$I_c(t) \simeq p_w \cdot (1 + p_w) \tag{18.18}$$

The essence is, that in average the core-inflation $I_c \geq p_w$ is in the range of GDP growth but in addition a little bit greater than the growth. A similar outcome we may get from simple assumption. We again look at the principle quantity equation:

*Capital Flow* or **Demand** = $KV = \sum M_i V_i = \sum H_j P_j = HP$ = **Supply** or *Products* or *GDP*

If now one of the both sides of the equation grows, the other has to grow too for consistency, which implies that average interest rates and GDP growth must be in the same order of magnitude $p_v \approx p_w$. Now, as we know GDP is always fast depreciated but assets are mostly not, we have also to demand an inflation, or depreciation of money, $I$ which also must be in the same order of magnitude. Thus we come to the law of approximate similar rates

$$p_v \approx p_w \approx I \tag{18.19}$$

As now investors and thus banks aim for positive net rates, mostly $p_v \geq I$ will apply, and thus we get

$$p_v \geq I \geq p_w \tag{18.20}.$$

---

[88] This is because the increase of the assets or liabilities of the banks grow even more exponentially if the GDP is already stagnating

[89] Due to the fact that mostly GDP and CAP are not so much different. But in the fraction the capital $K$ is always multiplied by an additional rate much smaller than 1 and thus gets in the sum with Y approximately negligible. Note that time here is $t = \Delta t = 1\ year$.

The last effect we may also see when again looking at (17.4). With the identity $H=Y/P$ we can rewrite it as:

$$I = \frac{\dot{K}}{K} + \frac{\dot{V}}{V} - \frac{\dot{H}}{H} \quad (18.21)$$

More simply we may write

$$I = p_v^a + a_V - h_t \quad (18.22).$$

where now is

$p_v^a := \dot{K}/K$ the average interest rate over all assets,
$a_V := \dot{V}/V$ the specific monetary acceleration and
$h_t := \dot{H}/H$ the specific change of the trade volume in time.

For the average asset prices $p_v^a$ we may take as a usable rule the interest rates paid for 1-year governmental bonds. In normal times the monetary acceleration will be small but as we know from (18.2) $h_t$ will be not. Thus follows again that usually rules $I \leq p_v^a$. A good "house number" for inflation in normal times thus will be the mean

$$I \simeq \frac{p_w + p_v^a}{2} \quad (18.23)$$

between GDP growth $p_w$ and average asset interest rate $p_v^a$.

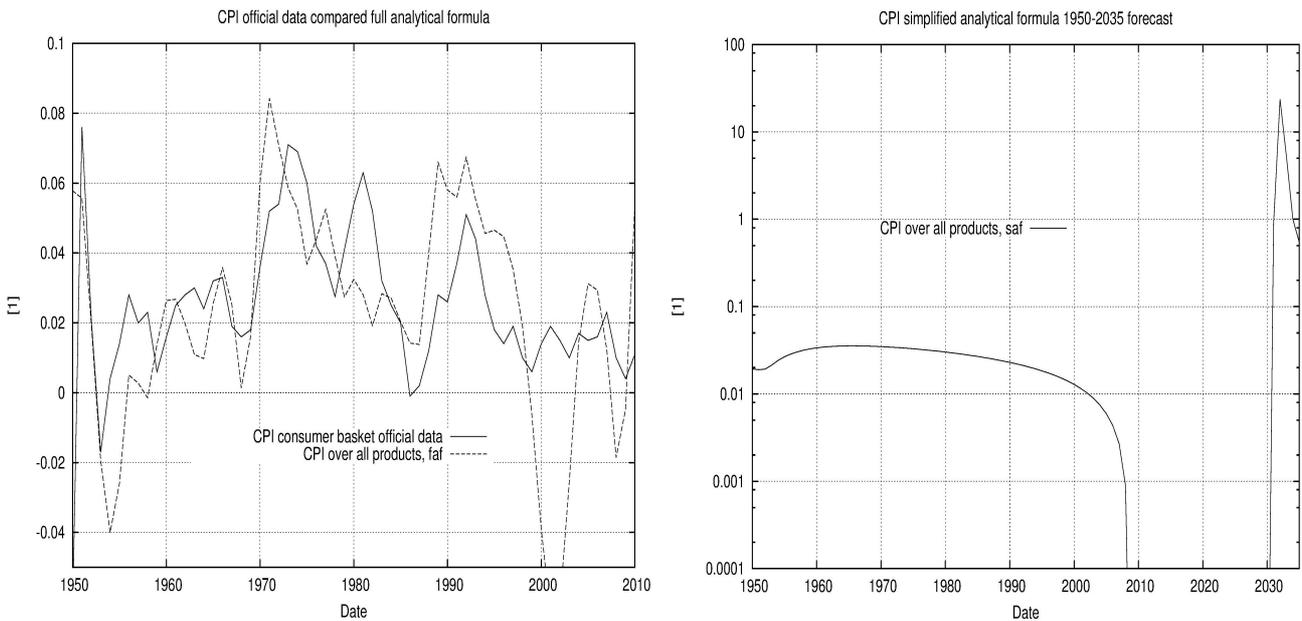

*Fig. 28:. Modeled and "real" statistical inflation. Model data were calculated from the real data of Y and K. The real data are, however, the information provided by the Federal Statistical Office, which were determined from the very special consumer baskets (CPI). The deviations can be related to external shocks e.g. exogenous money, like the Dot-Com-Bubble around year 2000.* **If there is no exogenous money** *(from abroad or by pure money creation by central-banks) the economy will face a long phase of deflation (or very small inflation and small interest rates) which can hold on for about 20 years (right hand plot, 2010-2030) before at the end hyperinflation will occur (around 2030-2032 with inflation rates up to 3000% a year).*

We here make a look ahead to general field theory. Indeed the effect of inflation is the main reason that will make our linear approximated equations nonlinear. As we can see, form $I \simeq \frac{p_w + p_v^a}{2}$ follows also, that $dY/dt \approx I \approx dK/dt =: c_i$ the change rates of Y and K are related to and effected by inflation in a similar way. Thus we have to multiply the basis equation by those derivatives to deal[90] for the effect of inflation. Then we may write in an approximate way:

$$\left(\frac{dY}{dt}\right)^2 \approx c_i^Y \cdot (b_0 + p_B Y - p_n K) \qquad \left(\frac{dK}{dt}\right)^2 \approx c_i^K \cdot (a_0 + p_S Y + p_n K)$$

(18.24),

where $c_i^Y(Y,K,t) \approx c_i^K(K,Y,t)$ is the global correction factor needed for real inflation correction. In general field theory we will see, that indeed the nonlinear equations will have this squared derivatives on the left hand sides making the system of differential equations then strictly nonlinear.

## 19 Effects of Globalization

Globalization is not a fundamentally new invention of the post-Reagan era. However, the increase of the technical possibilities of both, the actual transportation of goods, as well as the practical with the speed of light transportation of intangible benefit and pecuniary papers through the computerized trading is concerned, new possibilities, both quantitatively and qualitatively, have been opened. The main effect of globalization, or more generally internationalization, lies in the fundamental fact to at least temporarily generate an unbalanced foreign trade balance. Such a foreign trade surplus or deficit can be, in principle, both on goods and services in GDP's as related to capital as well. In our system of equations this is considered with the terms $a_0$ and $b_0$ :
$\dot{Y} = b_0 + p_B Y - p_n K$ and $\dot{K} = a_0 + p_S Y + p_n K$. In principle this linear system can be extended to the entire world economy by the following large System of Differential Equations

$$\dot{Y}_i = \sum_j b_{0ij} + p_B Y_i - p_n K_i \quad \text{and} \quad \dot{K}_i = \sum_j a_{0ij} + p_S Y_i + p_n K_i \qquad (19.1)$$

wherein the sizes of $a_{0ij}$ and $b_{0ij}$ considered the mutual interactions of the *i-th* with the *j-th* economy.
$b_{0ij}$ is then the net inflow or outflow of GDP without[91] payment, $a_{0ij}$ then is the net capital inflow or outflow without GDP trade-off. Inflow or outflow can be distinguished by the sign. The world economic system is then solved in principle by the maximal $2 \cdot (i \times i)$ dimensional matrix

$$\dot{\vec{Y}} = \boldsymbol{b_0} \vec{Y} + \boldsymbol{p_B} \vec{Y}_i - \boldsymbol{p_n} \vec{K}_i \quad \text{and} \quad \dot{\vec{K}} = \boldsymbol{a_0} \vec{K} + \boldsymbol{p_S} \vec{Y}_i + \boldsymbol{p_n} \vec{K}_i \qquad (19.2).$$

The occurring determining matrices need not always be diagonal[92]. The practical forecasting abilities of such a model is only limited by the computational effort and the reliable determination of the raw data to fill the coefficient matrices. This considerable hard work we can not do yet, especially because the freely available data bases rarely give reliable input to pay respect to the above defined foreign entanglements.

Instead, we give here an example of two linked *dummy* economies. What we should also be aware of is, that an international exchange of money in trade-off for GDP changes not much in the balance of a national economy ( $a_0$ and $b_0$ are initially defined without direct trade-off). It is given for the usual foreign trade (represented by large identifiers: $B_0$ and $A_0$ ):

$$\dot{Y}_i = B_{0j} + p_B Y_i - p_n K_i \quad \text{and} \quad \dot{K}_i = A_{0j} + p_S Y_i + p_n K_i$$
$$\dot{Y}_j = B_{0i} + p_B Y_j - p_n K_j \quad \text{and} \quad \dot{K}_j = A_{0i} + p_S Y_j + p_n K_j \tag{19.3}$$

---

90 In **linear theory** we deal with inflation by a **chain-correction** of the numerical data. See relating chapter 6.
91 The parameters $b_0$ and $a_0$ reflect on GDP without payments, like foreign donations of crop to a third world country, and on the other side, most recently done in globalized world, to Capital without related GDP Trade, such as foreign investment capital flow into local banks-own-business. But also self-creation of money not related to real goods, like self buying of states debts by central banks, must be considered in such a free parameter. It must be remarked here, that also the depreciation of money (full loss, which means deletion with no replacement, of money) is external or exogenous money just with a negative sign.
92 Thus, for example, in $p_B$ in addition to natural population growth can be incorporated known migration of people. And also other interactions between the economies of the world in the other coefficient matrices, which naturally should also include the impact of the monetary exchange rates around. For all those parameters there must be a continuity equation. So Capital inflow from one to another may be balanced by plus and minus sign here and there. This also is fact for migrations. Just creation out of nothing (like just printing money with no related growth or trade) will have a none zero balance.

If there are only two economies and in the event of a balanced trade is given:

$$\dot{Y}_i = \underbrace{(B_{0j} - B_{0i})}_{=0} + p_B Y_i - p_n K_i \quad \text{and} \quad \dot{K}_i = \underbrace{(A_{0j} - A_{0i})}_{=0} + p_S Y_i + p_n K_i$$
$$\dot{Y}_j = \underbrace{(B_{0i} - B_{0j})}_{=0} + p_B Y_j - p_n K_j \quad \text{and} \quad \dot{K}_j = \underbrace{(A_{0i} - A_{0j})}_{=0} + p_S Y_j + p_n K_j \tag{19.4}$$

where the bracketed values are always zero. And thus results once again in the undisturbed basic equations, as the net effect vanishes. In the case of an unbalanced trade, but this is not zero, and if $j$ sold to $i$ the net quantity $B$, we get:

$$\dot{Y}_i = B + p_B Y_i - p_n K_i \quad \text{and} \quad \dot{K}_i = -A + p_S Y_i + p_n K_i$$
$$\dot{Y}_j = -B + p_B Y_j - p_n K_j \quad \text{and} \quad \dot{K}_j = A + p_S Y_j + p_n K_j \tag{19.5}$$

The resulting imbalance is uncomfortable, because now the capital stock of $i$ decreases with increasing GDP, while the reverse is true at $j$, a rising stock of capital at a reduced offer of goods. In general, one would have to expect a deflation with $i$ and an inflation with $j$. To maintain a general stability that is at least an approximate balance in foreign trade relations is desirable. For the economy $j$ but the problem is surprisingly greater, because it is difficult to replace the missing GDP. On the other hand $i$ has a relatively low problem: it can close the small gap in the balance sheet, which only adds up to her, through additional value-free money creation comparatively easy. This can be money creation through the printing press, or by the formation of new financial derivatives. So we can open the following equation

$$\dot{K}_i = (A + B) - A + p_S Y_i + p_n K_i = B + p_S Y_i + p_n K_i$$

with a GDP of $\dot{Y}_i = B + p_B Y_i - p_n K_i$ \hfill (19.6),

from which the balance of $i$ is compensated without it is to be expected that inflation or deflation occurs over the normal level. It has, however, a different problem: you have to sell financial derivatives. On the one hand, on its own population, but this reduces their purchasing power for its own GDP. Buyers would be better from the outside to absorb these derivatives. Economy $i$ can try to sell these financial products worldwide so that its internal capital increases. In the most unpleasant case now even $j$ buys these derivatives worth $a_0 = (A + B)$, but $j$ has to provide $i$ the corresponding of GDP for this: Due to

$$\dot{Y}_j = -B - (A + B) + p_B Y_j - p_n K_j \quad \text{and} \quad \dot{K}_j = A + (A + B) + p_S Y_j + p_n K_j \tag{19.7}$$

with the amount of currency $|B| = |A|$ equity, then holds:

$$\dot{Y}_j = -3B + p_B Y_j - p_n K_j \quad \text{and} \quad \dot{K}_j = +3A + p_S Y_j + p_n K_j \tag{19.8}$$

Because most of the money creation by $i$ is next-year, the effect is not quite so strong, so that instead of a factor 3 one should estimate[93] about a factor of about 2. The situation but worsens considerably for $j$. The bottom line is: If two economies in mutually strongly unbalanced exchange with each other, then the negative effects are multiplied for the one economy, which exports its GDP to the same economy it buys financial bonds from. Globalization of trade and financial flows can thus quickly have significant consequences for unbalanced economies. The net differences

---

93 Currency effects are also not considered here. These are but of course also financial derivatives, and so can be accounted for as a financial product too.

$$B=\pm(B_{0j}-B_{0i}) \quad \text{and} \quad A=\mp(A_{0j}-A_{0i})=:a_0 \tag{19.9}$$

we can again express as our freely definable $b_0$ and $a_0$ coefficients. The following model calculation, we are now taking the less serious case that only a third country takes such financial securities. This happens for example when a national finance incorporates extra national funds into domestic financial system. This is done within the framework of international stock markets in the modern globalization very easily. It applies in this case, without regard to population growth:

$$\dot{Y}_i=-p_n K_i \quad \text{and} \quad \dot{K}_i=-a_0+p_S Y_i+p_n K_i$$

$$\dot{Y}_j=-p_n K_j \quad \text{and} \quad \dot{K}_j=a_0+p_S Y_j+p_n K_j \tag{19.10}$$

In the graph below we first see, just for comparison with the second graph, the case $a_0=0$ without such an interaction: Both economies take the usual course of their development of the GDP (the capital stock is in the graph given only for $Y_1$, here we are essentially interested in both GDP's). We assume in the model that the second economy $Y_2$ starts 25 years later, and is only half as strong as $Y_1$. The maximum of the GDP's of $Y_2$ remains well below the maximum of $Y_1$, as was to be expected.

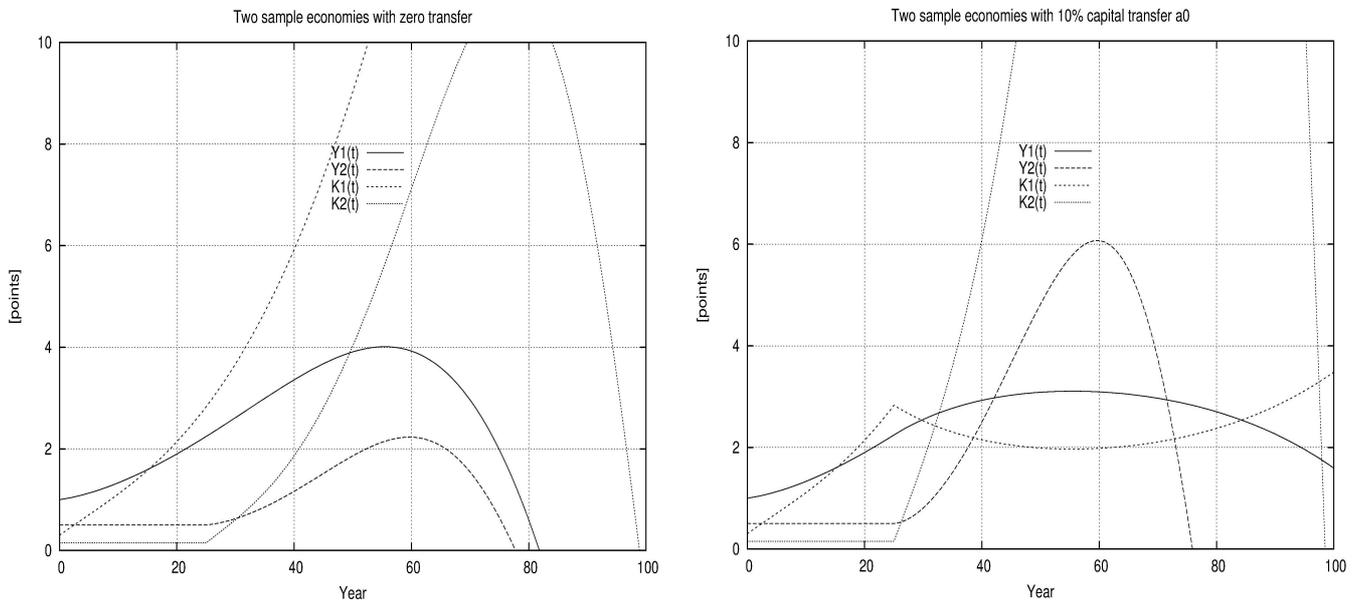

*Fig. 29: Two sample economies with and without capital transfer in comparison. Due to export of 10% capital per year economy $Y_1$ can be maintained very much longer than usual possible. Sample economy $Y_2, K_2$ starts 25 years later and has half the power of $Y_1, K_1$. Due to capital transfer it gets more powerful as $Y_1$ at its maximum, but collapses earlier.*

In the second graph we see the significant effect of such a capital transfer in an amount equal to one tenth of the GDP of $Y_1$: Now, while $Y_2$ dramatically increases and $Y_1$ even gets overhauled, the GDP $Y_1$ stagnates at a pretty good level. But also evident is, that now $Y_2$ crashes earlier and far more violent, because of the disproportionately high stock accumulation. At the same time can make it $Y_1$ over much longer periods than usual to keep "afloat". And this will happen as long as $Y_1, K_1$ can internationalize its surplus funds by selling them abroad.

# 20 Remark on Money Theory and Quantity equation

Financial products may play a double role: some times they can be thought to be a product of trade as also as a complement of money. Indeed the notions about *"money"* and its effects on the economy differs very much between different economists and economic schools. The monetary base is *"highly liquid money"*, which means *cash or short term assets* which can be very fast converted to *cash*. Cash to buy something, and thus is short term relevant to inflation or deflation. *Long term assets*, or even derivatives, but can also be converted to cash, which but will need some more or less time consuming intermediate steps to be transferred by selling them to someone else. Classical macroeconomics thus believes that only *short term money* is responsible for effects on the real economy while long term assets (and thus Banks own Business) does not play any role (*"classical dichotomy"*) for the real economy.

Also at least a part of real goods are taken as a kind of money, which can be seen looking at the notion of *"capital stock"* which includes as capital also capital bounded to goods, e.g. machinery, a firm bought and which could be, at least in principle, be re-transfered into money by selling them. But this classical notion is, in terms of theoretical handling of macro-economy, not well[94] founded: As buying and selling *goods* is already accounted in the measure of GDP, and thus <u>accounting it as Capital is doubling it and thus getting sources and sinks unbalanced</u>.

So the classical theory about different "kinds" of money sometimes confuses more than it explains, one thus we again should go *"back to the roots"* to clear up some very fundamentals about money. So we should do some general redefinition of the notion of money:

<u>Definition (20.1):</u>

**Definition of a <u>General Good</u>**   $H_G$  (commodity, article, product, wares): A Good is anything that can be an object of a trade.

**Definition of <u>General Money</u>**   $M_G$  (Cash, Assets, Capital): It's every gadget which enables someone to make a trade.

We do this General redefinition as it is easy to see, that both gadgets, goods or money, play a double role. A good can also be taken as money, e.g. *"I give you five cows for your one car"*, which means here cows and cars are both goods and money and vice versa. The same of course applies for money-gadgets, *"I give you 5000 Dollars cash for your rice future derivative"*. Every cross trade between the four samples gadgets, cash, cows, cars, derivatives, are also possible.

Indeed we have to consider the difference between **virtual and physical goods**, and **virtual and physical money**: *Real money* is an exemption, e.g. gold coins are such a thing. Mostly but there is *virtual money* like cash (FIAT-money) and practically every long or short term assets are as well just papers and guarantees and thus virtual money. Vice versa there are *real goods* like cars, but also Labor and services are real goods. But indeed today *virtual goods* are also not an exemption: e.g. derivatives like rice futures, stock papers and so on, which are sold for cash exactly like cars or houses too.

Virtual money or goods are *worth for itself* at the best some useless pennies for the paper they are

---

94  The classical notion of *"capital-stock"*, including machinery, real estate etc., but may be well founded for **micro**-economics. As such assets owned by a firm may be transferred to cash (coming from outside) which then can be used for investments by the firm itself. But macro- and micro-economic reasoning doesn't fit here, <u>as micro-economics handles open systems while macro-economics has to deal with (quasi-)closed system</u>, which makes a large difference in balancing rules.

written on. But also *worth* even billions of dollars as it is a claim on goods, which are assured by the state and its financial, judical and law enforcement agencies. Regardless how, where, when and why some virtual money (or assets convertible to money) is created, it is always a by law enforcement assured claim on actual[95] real goods available for trade . So we should do another general redefinition of the notion of money:

Definition (20.2):

**Definition of a Virtual Good** $H_V$ : A Good that has no material worth in comparison to its claimed worth (e.g. a derivative, assets)

**Definition of Virtual Money** $M_V$ : It's every gadget which enables someone to make a trade, but which has no material worth in comparison to its claimed worth (e.g. Cash, Assets)

**Definition of a Physical Good** $H_P$ : A Good that has immediate material worth (e.g. a house)

**Definition of Physical Money** $M_P$ : Every gadget having immediate material worth which enables someone to make a trade (e.g. physical gold, but also Real Goods of any kind).

Corollary (20.3):

*Every **Virtual Good** is also a Virtual Money, and every **Real Good** is also a Real Money. The wording just depends on the special main usage of those gadgets.*

Corollary (20.4):

***Money is neither worth nor wealth. Money is just a claim on worth and thus wealth.***

We will put now the principle meaning of the above definitions into some maths:

Our usual notation is $M_R V_R + M_I V_I = H_R P_R + H_I P_I$ , which means on the left hand side we have money $M_R$ which is available for trade in the real economy (loans) plus money $M_I$ which is available for the trade of any further financial product, e.g. derivatives. On the right hand side this is balanced by real products $H_R$ and the financial products $H_I$ .

Now we will rearrange this dependencies with respect to the double role those four gadgets play in economy:

$$M_R V_R + M_I V_I = H_R P_R + H_I P_I \to ...$$
$$... \to F(M_V) = M_R V_R + M_I V_I - H_I P_I = H_R P_R = f(H_R = M_R) \quad (20.5)$$

---

95 Indeed money claims on actual GDP, which means those in present or near future produced goods and services. Virtual goods may be saved and even increase due to compound interest, but real goods cannot be saved. They depreciate very fast in the macro-economic average: Services and labor depreciate to zero at once(!) after being served, food in days, cars in years, houses in decades. But the three first samples make the bulk indeed.

This now says, that we have put all virtual money $F(M_V)$ on one side and the balance of real goods $f(H_R)$, which may be seen also as the function of real money $f(M_R)$, on the right hand side. Also virtual products $H_I$ are virtual money or products $H_V$ or $M_V$.

So we come to the notion in general as $M_R V_R + M_I V_I - H_V P_V = H_R P_R$, but as from definition, *virtual money and virtual goods are the same balanced by sign*, so we have to write:

$$money\ available\ for\ real\ economy\ =\ M_R V_R - 2 \cdot H_V P_V = H_R P_R = Y \qquad (20.6)$$

The effect is simply that to much investment products will depress GDP at some time. Not at every time, as a first view onto the equation may suggest. *This is not the case* as the classical economics often used form $MV = HP$ as a rule of thumb is a *basically flawed notion*. It has to be mentioned of course as a time dependent equation of change

$$\frac{d}{dt}(M_R V_R + M_I V_I) = \frac{d}{dt}(MV) = \frac{d}{dt}(HP) = \frac{d}{dt}(H_R P_R + H_I P_I) \qquad (20.7)$$

which is *fundamentally another statement*. The quantity equation doesn't mean that both sides are equal, it means that the *growth of both sides* over time is equal. Using the notion $MV = HP$ is just a special time snap $(MV)_{T_0} = (HP)_{T_0}$ at a given present time $t := T_0$ and may be used with good reasons only as a rule of thumb. So we may now have a look at the derivatives of the quantity equations:

$$M_R V_R + M_I V_I = H_R P_R + H_I P_I \rightarrow \frac{d}{dt}(M_R V_R + M_I V_I) = \frac{d}{dt}(H_R P_R + H_I P_I)\ !$$

Because of the double role of $H_I P_I = H_V P_V = M_V P_V$ we can rearrange and write

$$\underbrace{\frac{d}{dt}(H_R P_R)}_{GDP\ Growth} = \underbrace{\underbrace{\frac{d}{dt}(M_R V_R)}_{loans\ interest} + \underbrace{\frac{d}{dt}(M_I V_I - H_I P_I)}_{investment\ banks\ interest}}_{net\ business\ rate - p_n K\ of\ financial\ business} \qquad (20.8)$$

which then resembles the first main differential equation for *Y* found earlier in this book from reasoning just *coupled sinks and sources*. The main obstacle now is the term $\frac{d}{dt}(M_I V_I - H_I P_I)$ which *in classical macroeconomics is usually misleadingly omitted* and set to zero. Which stems from the flawed notion of $M_I V_I - H_I P_I \approx 0$ as one thinks of Banks Own Business just always canceling out by balance. But this is entirely wrong, as not the balance in the bulk (at a arbitrary chosen time slice, which indeed cancels out) but only the changing rates (interest rates) are important to the evolution of an economy over time.

And indeed $\frac{d}{dt}(M_I V_I - H_I P_I)$ is <u>usually not zero</u> and may be mostly and in the bulk average negative[96]. A fact which one will not see at a first view. It stems from the time delay in the "worth" of the financial product $H_I P_I$ and the money paid for it $M_I V_I$ which results in

$$\frac{d}{dt}(M_I V_I - H_I P_I) \neq 0 \tag{20.9}$$

in general. Crisis will occur if the GDP growth at critical time $t = T_C$ tends to zero. Which then says

$$\frac{d}{dt}(M_R V_R) + \frac{d}{dt}(M_I V_I - H_I P_I) = 0_{t=T_C} \Rightarrow \underbrace{\frac{d}{dt}(M_R V_R)}_{interest\,CBB} = \underbrace{\frac{d}{dt}(H_I P_I - M_I V_I)}_{interest\,BoB} \tag{20.10}$$

that crisis occurs when the interest amount of Commercial Banks Business (CBB) gets equal the interests form Banks own Business (BoB). Assuming not very different interests in the bulk average between the both types of business, this means CBB=BoB in the bulk. Thus we come to the fact found out also by the principle system of differential equations that crisis occurs when the relation of CBB (loans) to the bulk of financial assets

$$p_{rel} = \frac{Loans}{whole\ of\ all\ assets} \leq \frac{1}{2} \quad \text{gets less equal to 50\%.} \tag{20.11}$$

On the other side we have the effect that the double accountance for financial products in BoB leads then to

$$Whole\ of\ all\ assets_{t=T_C} = |M_R V_R| + |M_I V_I| + |H_I B_I| = 3 \cdot |M_R V_R| \approx 3\,Y(T_C) \tag{20.11}$$

which again resembles what we could see from analytical and statistical reasoning as well. If there were created to much claims on real goods[97] then there are only few possibilities left to solve the resulting allocation[98] crisis: Politics in anyway will have to reduce[99] those claims. Otherwise workers and entrepreneurs will lose more and more of their claim on the products they created to the financial industry. In the very last end, when everyone who owns claims actual wants to realize them, then hyperinflation will occur as prices have to rise exponentially to comply with much to much claims.

---

96 It is the simple fact that an investor always buys a <u>financial</u> product if he assumes that it is more worth (HP) than the price (MV) to pay for it. At least in the near future, as otherwise nobody would make such a deal. There are more comfortable ways to loose money.
97 Which are indeed not much more than the actual trade able GDP.
98 Allocation Crisis but depends also on the fact, that money isn't distributed (nearly) equal on the people, but very much concentrated in few hands. The question how this situation arises can also be asked analytically when Labor *L* is introduced into the field equations. This will be on display in further work.
99 Quantitative Easing by Central banks is no way out. Indeed QE even keeps and further growths those claims. Politics mostly tries to finance QE by stagnating or cutting incomes, more taxes and less public expenditures. Which then but undermines consuming power and thus GDP. When in the very end the common believe in the effectiveness of such policies gets widely lost, the currency, but also often society, will vanish in hyperinflation, revolutions and wars.

# 21    General Field Theory of Macroeconomics

The first part dealt with the Special Macroeconomic Fieldtheory (SMF) which is a practically usable, and thus in the broadest sense, ready-to-use-theory[100]. For the use of SMF, we need only the values $p_B, p_S, p_v, p_{rel}$ for empirically. These can be determined both from statistical data, as well as from suitable adapted analytical functions to generate them. From which then the behavior of World macro-economy can be computed in real Dollar and Cents consistently.

But it would be nice, if we could also derive such parameters self-consistently out of theory. The purpose of the thus ensuing General Macroeconomic Field-Theory (GMF) is to

- proof generality and validity of the special field theory on fundamental grounds

- describing non-linear behavior which always exists between dependent concurrent forces

- further investigation of the fundamental structures and invariants of the economy

- calculating and predicting self-consistently the main important micro-factors of macro-economy.

It must be remarked here, that general nonlinear field theories are always a long-term subject for analytical research. This stems from the fact, that systems of non-linear differential equations are always an analytical hard task and often cannot been treated with simple mathematical standard tools.

***In this sense the following chapters may be understood and used just as an <u>incomplete mathematical toolbox</u> for ideas and further research.***

For generality we will show, that the theory can be described within the framework of the "*mother-of-all-theories*", the *Noether-Theorems*. Despite the fact that the thus nonlinear differential equations are solvable at least numerically with today's computing power, the much more valuable analytical solutions are almost very difficult, if only possible, to integrate. To solve the main macroeconomic parameters, we have to find in this fundamental framework at least as much invariants (symmetries) as unknown functions are to be derived. In principle this should be widely accessible.

It's obvious that a size, such as population growth $p_B$ and also the savings quota $p_S$, are not easy to predict analytically. Those parameter functions are very much influenced by cultural and psychological factors hardly to predict. The other variables however should be deductible as part of economic modeling with good accuracy from first principles. Thus for example, as for it is clear that not only the price level for goods, but also for nominal interest rates on assets and liabilities are functions of supply and demand for such financial products. So me may easily suggest

$$p_v \approx c(t) \cdot \frac{Y}{K} \qquad (21.1),$$

for the average interest rate. Because the demand for loans comes from the current GDP, and the more capital is available, the sooner this will reduce the average interest rate. In a complete self-consistent theory, as should also enter into the difference between the interest rate for real loans (consumptive and investment purposes; commercial bank model) and bank proprietary trading and derivatives (banks own business, investment banking), because they are not in average necessarily identical and are in accordance with their specific laws to model. Despite hardly to find out cultural and psychological factors, the savings rate and also the population growth may be modeled approximately by analytical functions, mainly as a function of the GDP's and its growth and less important from K and its derivatives:

$$p_S \approx f(Y, \dot{Y}, K, \dot{K}, ...) \quad \text{and/or} \quad p_B \approx g(Y, \dot{Y}, K, \dot{K}, ...) \qquad (21.2).$$

---

[100] Because SMF is already sufficient to describe the essential problems of macro-economies. It is able to make predictions for the development of GDP and its substitutes as a function of financial capital stocks worldwide. This can be achieved but only using statistically significant numbers which must be accrued in the international statistical offices on a matchable basis. So the numbers for the total capital stock and GDP and few other required ratios, and then one needs just a powerful database server and of course some software development. As an initial value problem we need no very long time-lines, but just reliable data of the actual last year(s). The better the information about the national parameters, the more precisely one can predict the future. A high accuracy is achievable in principle, but a lot work is to be done. This should not be underestimated, because many statistical agencies determine their figures incomplete or qualitatively deficient for our purpose. Moreover, these figures are mostly not published as nominal numbers, but presented already in statistical processing. In the worst case we have to deal with hedonic processed data. So there will still be much to do in the econometric science.

This is, because people need not only saving surpluses (from $Y$ ), but also confidence in the future potential of the system (from $\dot{Y}$ ). Not the same but similar dependencies are to be assumed for the population growth. Of course in most practical cases we will just use official statistical data for this functions, except of the cases where such cultural dependencies are of greater interest. In parallel an analytical solution for the relative share $p_{rel}$ or $p_r$ of the commercial bank model and the investment banking model should essentially depend on

$$p_r \approx f(\frac{Y}{K}, ...) \qquad (21.3)$$

the reciprocal value of the capital-coefficient Y/K, as this is a measure of demand in the banking business. As a first guess of such non-linear differential equations, one may suitable take $p_v \approx c(t) \cdot \frac{Y}{K}$ and $p_r \approx d(t) \cdot \frac{Y}{K}$ and then fit it using $c(t)$ and $d(t)$ to the measured data. Then finally invest these functions into the basic equations, thus giving

$$\dot{Y} = p_B(t)Y - (c(t)\frac{K}{Y} - d(t)\frac{Y}{K})K \qquad ??$$
$$\dot{K} = a_0(t) + f(\dot{Y}, Y, t)Y + (c(t)\frac{K}{Y} - d(t)\frac{Y}{K})K \qquad ?? \qquad (21.4).$$

Then one could check such derived equations for practical suitability. But this procedure is not advisable generally as it will not guarantee for letting invariants indeed invariant and thus possibly breaking prominent balancing rules. Much safer and analytically more useful however is the derivation of such relationships from very elementary principles.

## 22 Energy and Euler-Lagrange-Equations

The calculus of variations[101] implies in principle that real physical systems move on stable orbits:

$$\delta \int_{q(a)}^{q(b)} L(t, q, \dot{q}) \delta t = 0 \qquad (22.1)$$

The variation $\delta$ means a variation of the orbital parameters on the track $q(a) \rightarrow q(b)$ . The execution of the variation leads to the Euler-Lagrange equations

$$\frac{d}{dt}\frac{\partial L}{\partial \dot{q}} - \frac{\partial L}{\partial q} = 0 \qquad (22.2)$$

which are a condition for the solution. The positions $q$ and velocities $\dot{q}$ are the so-called generalized coordinates, which need not necessarily be the usual location $x$ and speed $v = \dot{x}(t)$ like often used in mechanical physics. In practical cases, one needs to apply the algorithm to a global conserved quantity of the system under consideration in order to derive the equations determining the relevant variables. Each invariant corresponds to a symmetry of the system. The Euler-Lagrange-Equations then allow to calculate from these symmetries the equations of motion of the underlying system. Our most fundamental conserved quantity now is given by the quantity equation $KV = HP$ and thus

$$L = KV - HP = const. \qquad (22.3).$$

---

[101] First I would like to put a quote here: "By generalizing Euler's method of variational calculus, discovered by Lagrange [1736-1813], he showed how to write in a single line the basic equation for all the problems of analytical mechanics." (Quote by Carl Gustav Jacob Jacobi, 1804-1851).

From Noether-theorems[102] it is known that invariance under time translations always has the importance of an energy. This is now trivial by using the most simplest form of the quantity equation and thus we get

$$\frac{d}{dt}(KV - HP) = 0 \quad \text{with}$$

$$L = KV - HP =: T - U$$

$$\text{with} \quad T := KV \quad \text{and} \quad U := HP \qquad (22.4)$$

That is, the *kinetic energy* of our system is given by $T = KV$ and through $U = HP = Y$ the *potential energy* is given. The four Euler-Lagrange equations are in the most simplest case providing only the mathematically correct though, but useless, trivial solution identically zero:

$$\frac{d}{dt}\frac{\partial L}{\partial \dot{q}_i} - \frac{\partial L}{\partial q_i} = 0 \quad \text{and} \quad K \equiv 0 \ \wedge \ V \equiv 0 \ \wedge \ H \equiv 0 \ \wedge \ P \equiv 0 \qquad (22.5)$$

The reason is the theorem that follows from $L \equiv 0$ that the total energy

$$E_G = q' L_{q'} - L = const.$$

remains constant over time. In our system, but $L$ is not identically zero, and our total energy $E_G = T + U$ increases with time.

For all nontrivial solutions one has to divide the quantity equation into suitable sub-variables which accounts for their mutual dependencies differentially. Also it has to be considered, that e.g. GDP value-free money creation (or money free GDP-creations, e.g. donations) also lead to more so-called anholonom Lagrangians, which are characterized by additional external sources[103] with $Q_e \neq 0$ :

$$\frac{d}{dt}\frac{\partial L}{\partial \dot{q}_i} - \frac{\partial L}{\partial q_i} = Q_{ie} \qquad (22.6)$$

In the general case it is to sum over all *j* products of the real economy and *k* products in the financial industry (with *i=j+k*).

---

[102] Emmy Noether (1882-1935) was an influential mathematician known for her groundbreaking contributions to abstract algebra and theoretical physics. She is described as the most important woman in the history of mathematics. In physics, Noether's theorem explains the fundamental connection between symmetry and conservation laws. Noether's (first) theorem states that any differentiable symmetry of the action of a physical system has a corresponding conservation law (proved 1915, published 1918). The action of a physical system is the integral over time of a Lagrangian function. From this the system's behavior can be determined by the principle of least action and this principle has become a fundamental tool of modern theoretical physics.

[103] One can identify this external sources with the so-called "external shocks".

# 23 The General Quantity Equation

Next we have to define the full and thus General Quantity Equation (GQE). This can be achieved by mentioning that we have in general *m* monetary flows to pay for *k* trades of any products.

$$\begin{pmatrix} V_{11} & V_{12} & ... & V_{1m} \\ . & . & ... & . \\ . & . & ... & . \\ . & . & ... & . \\ V_{n1} & V_{n2} & ... & V_{nm} \end{pmatrix} \cdot \begin{pmatrix} M_1 \\ . \\ . \\ . \\ M_m \end{pmatrix} = \begin{pmatrix} P_{11} & P_{12} & ... & P_{1k} \\ . & . & ... & . \\ . & . & ... & . \\ . & . & ... & . \\ P_{n1} & P_{n2} & ... & P_{nk} \end{pmatrix} \cdot \begin{pmatrix} H_1 \\ . \\ . \\ . \\ H_k \end{pmatrix} \qquad (24.1)$$

The norm of this matrices and vectors are to be defined for consistency by the following summation-norm:

$$[\![X]\!]_s = \sum_n X_{n1} + ... + \sum_n X_{ns} \qquad (24.2)$$

The GQE includes all money and all trades of an economy and also there complicated interdependencies. For example buying a car means also buying a lot of screws, cables, paint and so on, but also means buying labor and financial products like credits the car builder uses. Of course the full GQE is just of theoretical use. In practice all the billions of time-dependent elements of the matrices will never be known[104] exactly. In most cases we will have to find and to separate *"clean"* products and flows. This means to deal with $X_{ij}=0$ for $i \neq j$ and eventually $m=k$ which then will make the main matrices quadratic and also diagonal for much easier analytical derivations:

$$\begin{pmatrix} V_{11} & 0 & ... & 0 \\ 0 & V_{22} & ... & 0 \\ 0 & 0 & ... & 0 \\ 0 & 0 & ... & 0 \\ 0 & 0 & ... & V_{nn} \end{pmatrix} \cdot \begin{pmatrix} M_1 \\ . \\ . \\ . \\ M_n \end{pmatrix} = \begin{pmatrix} P_{11} & 0 & ... & 0 \\ 0 & P_{22} & ... & 0 \\ 0 & 0 & ... & 0 \\ 0 & 0 & ... & 0 \\ 0 & 0 & ... & P_{nn} \end{pmatrix} \cdot \begin{pmatrix} H_1 \\ . \\ . \\ . \\ H_n \end{pmatrix} \Leftrightarrow \sum_{i=1}^n (M_i V_i - H_i P_i) = 0 \qquad (24.3).$$

# 24 Deriving the basic linear ODE's from fundamentals

First we will investigate the simplest non-trivial splitting of the economy, which means to sum it up into the two main parts of a capital driven economy, the so-called "real economy" *R* and the financial or "investment economy" *I*:

$$\begin{pmatrix} V_R & V_{RI} \\ V_{IR} & V_I \end{pmatrix} \cdot \begin{pmatrix} M_R \\ M_I \end{pmatrix} = \begin{pmatrix} P_R & P_{RI} \\ P_{IR} & P_I \end{pmatrix} \cdot \begin{pmatrix} H_R \\ H_I \end{pmatrix} \qquad (24.1).$$

This gives by virtue of the sum-norm $[\![.]\!]_s$ :

$$M_R V_R + M_I V_{RI} + M_R V_{IR} + M_I V_I = H_R P_R + H_I P_{RI} + H_R P_{IR} + H_I P_I$$

which can be rearranged to:

$$M_R(V_R + V_{IR}) + M_I(V_I + V_{RI}) = H_R(P_R + P_{IR}) + H_I(P_I + P_{RI}) \qquad (24.2)$$

---

[104] We but should mention theoretically, that if known, indeed every evolution of the economy could be calculated and predicted exactly. So it should be mentioned, that the function $V_{ij}(t)$ would include not only the monetary flows between contractors, but even the written or unwritten contractual money flows in time for the future (e.g. redemption plans etc.).

Now we may assume "cleaner" products by symmetry and/or small valued mixed coefficients

$$(V_{RI} \approx -V_{IR} \ \wedge \ P_{RI} \approx -P_{IR}) \ \vee \ (V_{RI}, V_{IR} \approx 0 \ \wedge \ P_{RI}, P_{IR} \approx 0)$$

from which follows:

$$M_R V_R + M_I V_I + V_{IR}(M_R - M_I) = H_R P_R + H_I P_I + P_{IR}(H_R - H_I) \tag{24.3}$$

Now we define the differences as $M_{RI} := M_R - M_I \ \wedge \ H_{RI} := H_R - H_I$ which leads to

$$M_R V_R + M_I V_I + M_{RI} V_{IR} = H_R P_R + H_I P_I + H_{RI} P_{IR} \tag{24.4}.$$

This equation can be interpreted as:

*loans to real economy + banks own business + interests flow from R to I =
production by loans + financial products + real econ. interest drain from R to I*

Now we invest Euler-Lagrange relations and define in a first step

$$\dot{q} = V \ \wedge \ q = P \tag{24.5},$$

Although this is not entirely correct, as the units $1/y$ and $cur./buy$ in standard quantity equation disagree, which should not happen for the variational calculus as units [] should be $[dt \cdot dP/dt] = [V]$ equal. Thus we denote this a little badly defined quantities with a $b$ index.. Now we do

$$\frac{d}{dt} \frac{\partial L}{\partial \dot{q}_b} - \frac{\partial L}{\partial q_b} = 0$$

with $\quad L := M_R V_R + M_I V_I - H_R P_R - H_I P_I + M_{RI} V_{IR} - H_{RI} P_{IR} \tag{24.6}$

which results in

$$\frac{d}{dt} \frac{\partial L}{\partial V_R} - \frac{\partial L}{\partial P_R} = 0 \quad \text{which gives} \quad \dot{M}_R = H_R$$
$$\frac{d}{dt} \frac{\partial L}{\partial V_I} - \frac{\partial L}{\partial P_I} = 0 \quad \text{which gives} \quad \dot{M}_I = H_I$$
$$\frac{d}{dt} \frac{\partial L}{\partial V_{IR}} - \frac{\partial L}{\partial P_{IR}} = 0 \quad \text{which gives} \quad \dot{M}_{RI} = \dot{M}_R - \dot{M}_I = H_R - H_I \tag{24.7}.$$

The last equation also means $\dot{M}_I = H_I - H_R + \dot{M}_R$ or $\dot{M}_R = \dot{M}_I - H_R + H_I$. Now we get by just adding this three equations

$$\dot{K} = \frac{d}{dt}(M_R + M_I + M_{RI}) = H_R + H_I + H_{RI} = 2 H_R$$

and by subtracting them we get also

$$-\dot{K} = -\frac{d}{dt}(M_R + M_I + M_{RI}) = 2 H_I$$

Subtracting and adding this two equations trivially results in
$$\dot{K} = H_I + H_R \tag{24.8}$$
This now means by some handsome redefinitions

$$\dot{K} = d \cdot Y + [e \cdot K(interest(Y)) - c \cdot K(loans(Y))] \tag{24.9},$$

as $H_I$ can be interpreted as the savings of the economy to banks, and $H_R$ as the net result of the loans and interests from the GDP. The constants $d,e,c$ are just factors due to our badly defined units of the standard Quantity Equation, by which we get the principle form of the basic model derived in the first chapters of this article. The second equation is just trivial, as it means

$$\dot{K} = \frac{d}{dt} MV = \frac{d}{dt} HP = \dot{Y} \quad \text{which can be written as} \quad \dot{Y} = \dot{M} V + M \dot{V}$$

which now can be interpreted as
$$\dot{Y} = f \cdot K (loans(Y)) - g \cdot K (interests(Y)) \qquad (24.10)$$

with $f,g$ again some constant factors resulting from the bad definition of the standard. As we see is seemingly $e = -f$ and $c = -g$ and also we may set $d := p_S$ by definition.

The term $e \cdot K(interest(Y)) - c \cdot K(loans(Y))$ is the sum-effect of giving loans to the real economy on one hand, and to get interests in return on the other hand. The net effect now can be summarized in our net-business-rate by some sorting and redefinition:

$$e \cdot K(interest(Y)) - c \cdot K(loans(Y)) = ...$$
$$... = e \cdot \alpha(t) K - c \cdot \beta(t) K = (e \cdot \alpha - c \cdot \beta) K =: P_n(t) K \qquad (24.11)$$

This brings then back our basic system of differential equations from the first part of the article

$$\dot{Y} = -p_n K \quad \wedge \quad \dot{K} = p_S Y + p_n K \quad ,$$

which just had to be shown.

## 25 Field Equations of Economics

Next we will derive the main field equations of economics. For generality we first will see how we have to deal with the physical units in our main equation. The units $1/y$ and $cur./buy$ in standard quantity equation disagree, which should not happen for the variational calculus as units [.] have to be $[dP/dt] = [V]$. Thus try to define them in some better useable[105] quantities to fit smoothly into variational calculus. Instead of the usual dimension of money [currency] we use the more meaningful money dimension of „purchasing power" in dimension of *buy/cur* . The inverse dimension then is the "value of goods" in dimension of *cur/buy*. Thus we define *VoG:="Value of Goods"* and *VoM:="Value of Money"* with VoM = 1/ VoG which leads to the helping functions of units:

$$[V] = 1/y \cdot cur/buy = cur/(buy \cdot y) = [dP/dt]$$
$$M \cdot (V \cdot VoG) = (H \cdot VoG) \cdot P \quad \text{or} \quad M \cdot (V/VoM) = (H/VoM) \cdot P$$
$$[H \cdot VoG] = buy/y \cdot cur/buy = cur/y = [dM/dt] \quad .$$

By our definition rules:
$$v_{pp} := VoG = \frac{1}{VoM} \qquad (25.1)$$

Thus we get:
$$M \cdot (V \cdot v_{pp}) = (H \cdot v_{pp}) \cdot P \quad \text{or} \quad M \cdot V_{eff} = H_{eff} \cdot P$$

with $V_{eff} := v_{pp}(t) V$ and $H_{eff} := v_{pp}(t) H$ from which we again get the matrices:

---

105 A sample from physics is location and velocity, and resulting acceleration from forces:
$$[x] = m \quad [v] = [dx/dt] = m/sec \quad [a] = [d^2 x/dt^2] = m/sec^2$$

$$\begin{pmatrix} v_{pp}V_R & v_{pp}V_{RI} \\ v_{pp}V_{IR} & v_{pp}V_I \end{pmatrix} \cdot \begin{pmatrix} M_R \\ M_I \end{pmatrix} = \begin{pmatrix} P_R & P_{RI} \\ P_{IR} & P_I \end{pmatrix} \cdot \begin{pmatrix} v_{pp}H_R \\ v_{pp}H_I \end{pmatrix}$$

or

$$\begin{pmatrix} V_{Reff} & V_{RIeff} \\ V_{IReff} & V_{Ieff} \end{pmatrix} \cdot \begin{pmatrix} M_R \\ M_I \end{pmatrix} = \begin{pmatrix} P_R & P_{RI} \\ P_{IR} & P_I \end{pmatrix} \cdot \begin{pmatrix} H_{Reff} \\ H_{Ieff} \end{pmatrix} \quad (25.2).$$

This leads like before to:
$$M_R V_{Reff} + M_I V_{Ieff} + M_{RI} V_{IReff} = H_{Reff} P_R + H_{Ieff} P_I + H_{RIeff} P_{IR} \quad (25.3).$$

This equation seemingly is just:

$$\textit{effective}\,[\,\textit{loans real econ.} + \textit{banks own busn.} + \textit{interests}\,] = ...$$
$$... = \textit{effective}\,[\,\textit{loans} + \textit{financial products} + \textit{interest drain}\,]$$

Now we again invest Euler-Lagrange using the now corrected units:

$$\dot{q}_1 := V_{\textit{eff}} \quad \wedge \quad q_1 := P \quad \text{and} \quad \dot{q}_2 := H_{\textit{eff}} \quad \wedge \quad q_2 := M \quad (25.4).$$

Applying[106] the Lagrangian as before finally gives the nonlinear differential equation

$$\left(\left(\frac{dM_R}{dt}\right)^2 - \left(\frac{dM_I}{dt}\right)^2\right) \cdot (V_I^2 - V_R^2) = (H_R^2 - H_I^2) \cdot \left(\left(\frac{dP_I}{dt}\right)^2 - \left(\frac{dP_R}{dt}\right)^2\right) \quad (25.5)$$

which is now the *Basic Nonlinear Quantity Equation of Economics with Lagrangian applied.* We call it basic, as it splits up in just the two main components *R* and *I*. Of course it can be split up to any number of components as one needs, but splitting up *at least into the two main parts* is essential for deeper analysis. Even without explicit solving this equation can be investigated some further. As now we have to show, that this nonlinear differential equation is identical to the common known classical quantity equation *MV=HP* in the linear quasi-constant case. For this we have to assume, that the banks own business *I* does play no big role of influence to the GDP, and thus all *I*-values can be set to be zero. Thus we get just one component left: $\dot{M}_R^2 \cdot V_R^2 = H_R^2 \cdot \dot{P}_R^2$ ,

---

106 The same derivation as before applies, fortunately the helping functions on units drop out: Now we have

$$\frac{d}{dt}\frac{\partial L}{\partial \dot{q}_i} - \frac{\partial L}{\partial q_i} = 0 \quad \text{with} \quad L := M_R V_{Reff} + M_I V_{Ieff} - H_{Reff} P_R - H_{Ieff} P_I + M_{RI} V_{IReff} - H_{RIeff} P_{IR} \quad ,$$

which results in: $\frac{d}{dt}\frac{\partial L}{\partial V_{Reff}} - \frac{\partial L}{\partial P_R} = 0$ which gives $\dot{M}_R = H_{Reff} = v_{pp} H_R$ ; $\frac{d}{dt}\frac{\partial L}{\partial V_{Ieff}} - \frac{\partial L}{\partial P_I} = 0$ giving

$\dot{M}_I = H_{Ieff} = v_{pp} H_I$ and $\frac{d}{dt}\frac{\partial L}{\partial V_{IReff}} - \frac{\partial L}{\partial P_{IR}} = 0$ giving $\dot{M}_{RI} = H_{Reff} - H_{Ieff} = v_{pp} H_{RI}$ . The last

equation is equal to $\dot{M}_I = \dot{M}_R - v_{pp} H_{RI}$ or $\dot{M}_R = \dot{M}_I - v_{pp} H_{RI}$ and so we get $\frac{d}{dt}\frac{\partial L}{\partial H_{Reff}} - \frac{\partial L}{\partial M_R} = 0$

which gives $-\dot{P}_R = V_{Reff}$ ; $\frac{d}{dt}\frac{\partial L}{\partial H_{Ieff}} - \frac{\partial L}{\partial M_I} = 0$ giving $-\dot{P}_I = -V_{Ieff}$ ; $\frac{d}{dt}\frac{\partial L}{\partial H_{RIeff}} - \frac{\partial L}{\partial M_{RI}} = 0$ giving $-\dot{P}_{IR} = -V_{IReff}$ . Addition /subtractions/ multiplication like before of the main equations results in

$\dot{M}_R = H_{Reff} = v_{pp} H_R$ and $\dot{M}_I = H_{Ieff} = v_{pp} H_I$ . By addition/ substraction we get

$\frac{d}{dt}(M_R + M_I) = v_{pp}(H_R + H_I)$ and $\frac{d}{dt}(M_R - M_I) = v_{pp}(H_R - H_I)$ ; $-\dot{P}_R = V_{Reff}$ plus and minus

$-\dot{P}_I = -V_{Ieff}$ results in $\frac{d}{dt}(P_R + P_I) = v_{pp}(V_I - V_R)$ and $\frac{d}{dt}(P_I - P_R) = v_{pp}(V_R + V_I)$ . The

products of those give $\dot{M}_R^2 - \dot{M}_I^2 = v_{pp}^2(H_R^2 - H_I^2)$ and $\dot{P}_I^2 - \dot{P}_R^2 = v_{pp}^2(V_I^2 - V_R^2)$ from which we finally get

$(\dot{M}_R^2 - \dot{M}_I^2)(V_I^2 - V_R^2) = (H_R^2 - H_I^2)(\dot{P}_I^2 - \dot{P}_R^2)$ .

from which we now can simply take the root which gives in the linear approximation

$$\dot{M}_R \cdot V_R = \pm H_R \cdot \dot{P}_R \quad .$$

The last equation now is a more realistic *differential form of the classical Quantity Equation* of Economics

$$V \cdot \frac{d}{dt} M \simeq \pm H \cdot \frac{d}{dt} P \tag{25.6}.$$

As, if further more the values $M$ and $P$ are assumed to be quasi-constants in time as can be done for small differences $dt = t_1 - t_0$ approximately, one can multiply this equation by the then needless constant $dt$. Also one can choose the more meaningful positive solution[107] by which we get the *Classical QE* $MV \simeq HP$ which is known since centuries but will apply only approximate in all quasi-constant situations.

## 26 Complex General Field Equations

As systems of linearized differential equations are much easier to solve, the question thus is, when may I linearise the strictly nonlinear differential equation and how far can I go with it? As we could show in the first part of the article, the linear case is given for all times until deep crisis occurs, but after this it has to be indeed questioned. For this reason, we start further investigations with the nonlinear quantity equation, from which we will get some more insights into the main economic behavior.

$$(\dot{M}_R^2 - \dot{M}_I^2) \cdot (V_I^2 - V_R^2) = (H_R^2 - H_I^2) \cdot (\dot{P}_I^2 - \dot{P}_R^2) \tag{26.1}$$

may be written more simpler with the use of complex vectors $Z = A + iB$. Such complex numbers have the usual euclidic metric defined by using definition of the conjugated complex

$$\check{Z} := A - iB \quad \text{with} \quad |Z| = \sqrt{Z \cdot \check{Z}} = \sqrt{(A + iB)(A - iB)} = \sqrt{A^2 + B^2} \tag{26.2}$$

But instead of the usual euclidic norm we define now a special *p*-norm for the *k*-dimensional pseudo-complex vector $X_C$ :

$$[X_C]_p := \sqrt{X_1^2 - \sum_{k=2}^{n} X_k^2} \tag{26.3}.$$

This pseudo-complex number is best represented by using a so-called pseudo-riemanian metric represented here by the special Minkowsy-tensor

$$\eta_{ij} = \begin{pmatrix} 1 & 0 & \ldots & 0 & 0 \\ 0 & -1 & \ldots & 0 & 0 \\ . & . & . & . & . \\ . & . & . & . & . \\ . & . & . & . & . \\ 0 & 0 & \ldots & -1 & 0 \\ 0 & 0 & \ldots & 0 & -1 \end{pmatrix} \tag{26.4}$$

which thus gives the rule for the length of our pseudo-complex vector $X_C = (X_1, X_2, \ldots, X_n)$ of any dimension $n$ by the norm

---

[107] The classical QE gives just approximate locally valid results. For time dependencies we have to use the differential form and one may define an *altered classical quantity equation* as

$V \cdot \frac{d}{dt}(M_R + M_I) \simeq H \cdot \frac{d}{dt}(P_R + P_I)$ . We have used here the *simeq*-sign which is just to demonstrate the approximate state of any of such linear equations.

$$[\boldsymbol{X_C}]_p = \sqrt{\boldsymbol{X_C^T} \cdot \eta_{ij} \cdot \boldsymbol{X_C}} = \sqrt{X_1^2 - X_2^2 - ... - X_n^2} \qquad (26.5)$$

So we can redefine our two-dimensional basic vectors to

$$\begin{array}{cc} \boldsymbol{M_C} = (M_R, M_I) & \wedge \quad \boldsymbol{V_C} = (V_R, V_I) \\ \wedge \quad \boldsymbol{H_C} = (H_R, H_I) & \wedge \quad \boldsymbol{P_C} = (P_R, P_I) \end{array}$$

$$\text{with} \quad [\boldsymbol{X_C}]_p = \sqrt{X_R^2 - X_I^2} \qquad (26.6).$$

By applying the square-root we finally get

$$[\dot{\boldsymbol{M}}_C]_p \cdot [\boldsymbol{V}_C]_p = \pm [\boldsymbol{H}_C]_p \cdot [\dot{\boldsymbol{P}}_C]_p \qquad (26.7)$$

the much simpler looking *Pseudo-Complex Differential Quantity Equation*. From this arrangement we can see the fundamentals of economics now more clearly. As we have four vectors of the same principal riemanian-complex form $\boldsymbol{X_C} = X_R + (i) \cdot X_I$, we see that each side of the equation builds up its own footage area. This can be seen by the fact, that the scalar product of two vector is given by

$$\bar{c}_1 \cdot \bar{c}_2 = |\bar{c}_1| \cdot |\bar{c}_2| \cdot \cos(\alpha) \quad = \text{spanned footage}.$$

Thus the effect of GQE is the fact, that the area spanned by the money supply $\dot{\boldsymbol{M}}_C \cdot \boldsymbol{V}_C$ and the area spanned by merchandise $\boldsymbol{H}_C \cdot \dot{\boldsymbol{P}}_C$ has always to be equal over time:

$$\left[\frac{d}{dt} \boldsymbol{M}_C\right]_p \cdot [\boldsymbol{V}_C]_p = \pm [\boldsymbol{H}_C]_p \cdot \left[\frac{d}{dt} \boldsymbol{P}_C\right]_p \qquad (26.8)$$

Using elementary vector calculus we can rewrite this equation as

$$\frac{\dot{\boldsymbol{M}}_C \cdot \boldsymbol{V}_C}{\boldsymbol{H}_C \cdot \dot{\boldsymbol{P}}_C} = \frac{\cos\alpha(t)}{\cos\beta(t)} \qquad (26.9)$$

where $\alpha$ is the angle between the *R*- and *I*-components on the money side, and $\beta$ the angle between the *R*- and *I*-components on the merchandise side. As the vectors $\boldsymbol{X}_C$ may have generally any dimension *n*, we can say that (27.9) is the *Fundamental Geometric Law of Standard Substitutional Economics*.

By some new greek-symbols for our definition of the *p*-norm pseudo-complex vectors of any dimension the **nonlinear GQE** is written as:

$$\left[\frac{d}{dt}\mathbf{M}\right]_p^2 \cdot [\|\boldsymbol{\Lambda}\|_s]_p^2 = [\mathbf{H}]_p^2 \cdot \left[\frac{d}{dt}\|\boldsymbol{\Pi}\|_s\right]_p^2 \qquad (26.10)$$

Here $\mathbf{M}, \mathbf{H}, \Lambda$ and $\prod$ are our pseudo-complex matrices with matrix-summation norm $s$, vector norm $p$. The sub-index $t$ denotes the time-derivative *d/dt* respectively. Looking ad those subscripts we see, that the perfect symmetry is broken now. Although on both sides the summation-index-sum *spp,t* agrees, the implications[108] for *M* and *H* are but indeed different. For further tensorial algebra, a convenient way of writing the field Equations of Economy is using the geometric tensor $g_{ij}$ which writes the *relinearized*[109] *field equations* in the so-called *"Einstein-notation"* as:

---

[108] This symmetry-breaking between *H* and *M* is indeed the most important fact for economic growth on one hand, and the later in time coming intrinsic crisis of capital driven substitutional economies on the other hand.

[109] For most calculation outside crisis times the linearized field equations are convenient and adequate. Thus we investigate here just the usual positive solution of our general non-linear equation.

$$g_{ij}\,\boldsymbol{M}^{C}\cdot\boldsymbol{\eta}_{ij}\,\boldsymbol{\Lambda}_{s}^{C}-\boldsymbol{\eta}_{ij}\,\boldsymbol{H}^{C}\cdot g_{ij}\,\boldsymbol{\Pi}_{s}^{C}=Q_{ij}^{e} \quad \text{with} \quad g_{ij}=\frac{d}{dt}\cdot\boldsymbol{\eta}_{ij} \tag{26.11}.$$

The subscipt *s* denotes the *summation-norm* and the superscript *C* means that the vectors are split up into there *i* different *complex* contributions. External contributions $Q_{ij}^{e}\equiv 0$ are given in the case of closed, or $Q_{ij}^{e}\simeq 0$ for quasi-closed national economies. The resulting *j* coupled differential equations have to be solved for adequate constraints at last.

## 27   Introducing Labor and Wages

Labor and wages, which are responsible for the main consumption power of an economy, are an important sample step. To incorporate wages (*W*) accompanied labor (*L*) into the equations, this can be easily done by just splitting up e.g. the money supply to the real economy $M_R = M_G + M_W$ into the supply for buying goods (*G*) and buying labor (wages: *W*). Systems of splittings can be done to any desired dimension and thus econometric precision, at the price of much more complexity of course. Introducing labor we now have to define vectors of dimension *n=3* :

$$\boldsymbol{M}^{C}=\begin{pmatrix}M_G\\M_W\\M_I\end{pmatrix} \quad \text{and} \quad \boldsymbol{H}^{C}=\begin{pmatrix}H_G\\H_L\\H_I\end{pmatrix} \tag{27.1}.$$

This results in

$$\begin{vmatrix}\frac{d}{dt}&0&0\\0&\frac{-d}{dt}&0\\0&0&\frac{-d}{dt}\end{vmatrix}\cdot\begin{pmatrix}M_G\\M_W\\M_I\end{pmatrix}\begin{pmatrix}1&0&0\\0&-1&0\\0&0&-1\end{pmatrix}\cdot\begin{pmatrix}V_G&V_{GW}&V_{GI}\\V_{WG}&V_W&V_{WI}\\V_{IG}&V_{IW}&V_I\end{pmatrix}=\ldots$$

$$\ldots=\begin{pmatrix}1&0&0\\0&-1&0\\0&0&-1\end{pmatrix}\cdot\begin{pmatrix}H_G\\H_L\\H_I\end{pmatrix}\begin{vmatrix}\frac{d}{dt}&0&0\\0&\frac{-d}{dt}&0\\0&0&\frac{-d}{dt}\end{vmatrix}\cdot\begin{pmatrix}P_G&P_{GL}&P_{GI}\\P_{LG}&P_L&P_{LI}\\P_{IG}&P_{IL}&P_I\end{pmatrix}$$

which results finally in

$$\begin{pmatrix}\dot{M}_G V_G+\dot{M}_W V_{WG}+\dot{M}_I V_{IG}\\ \dot{M}_G V_{GW}+\dot{M}_W V_W+\dot{M}_I V_{IW}\\ \dot{M}_G V_{GI}+\dot{M}_W V_{WI}+\dot{M}_I V_I\end{pmatrix}-\begin{pmatrix}H_G \dot{P}_G+H_L \dot{P}_{LG}+H_I \dot{P}_{IG}\\ H_G \dot{P}_{GL}+H_L \dot{P}_L+H_I \dot{P}_{IL}\\ H_G \dot{P}_{GI}+H_L \dot{P}_{LI}+H_I \dot{P}_I\end{pmatrix}=0 \tag{27.2}.$$

So we get for our sample here the *linear special field equations* as

$$\begin{aligned}F_1&=\dot{M}_G V_G+\dot{M}_W V_{WG}+\dot{M}_I V_{IG}-H_G \dot{P}_G-H_L \dot{P}_{LG}-H_I \dot{P}_{IG}\\ F_2&=\dot{M}_G V_{GW}+\dot{M}_W V_W+\dot{M}_I V_{IW}-H_G \dot{P}_{GL}-H_L \dot{P}_L-H_I \dot{P}_{IL}\\ F_3&=\dot{M}_G V_{GI}+\dot{M}_W V_{WI}+\dot{M}_I V_I-H_G \dot{P}_{GI}-H_L \dot{P}_{LI}-H_I \dot{P}_I\end{aligned} \tag{27.3}$$

which are to be solved simultaneous and consistently. To solve the system, one has to find some constraints on the different parameters $V_{ij}$ and $P_{ij}$ resulting from further symmetries or other meaningful or manifest constraints. One important constraint here will be the fundamental geometric law of macroeconomics, which means the fundamental manifestation of balancing rules resulting in equal footprints on both sides over time.

The full solution of the resulting system of this three coupled differential equations and deriving the constraints is not in the scope here, as it takes lengthy discussions. Meanwhile in the following chapters we will find some additional symmetries in addition to the main symmetry of quantity equation. As for full self-consistent solutions, we will have to find at least as much invariants as unknown functions there are to be solved.

## 28   The Predator-Prey Symmetry

Real solution will break the latest when GDP has fallen to zero. But to examine analytically the formal structure of such system of equations, we may simply continue to run the equations beyond their natural endpoint at $Y(t_{End})=0$. In the following figure we can see the resulting effect: The solutions for *K* and *Y* will oscillate somehow out of phase around the zero line. It does however increase the maximum amplitude of both functions in time exponentially.

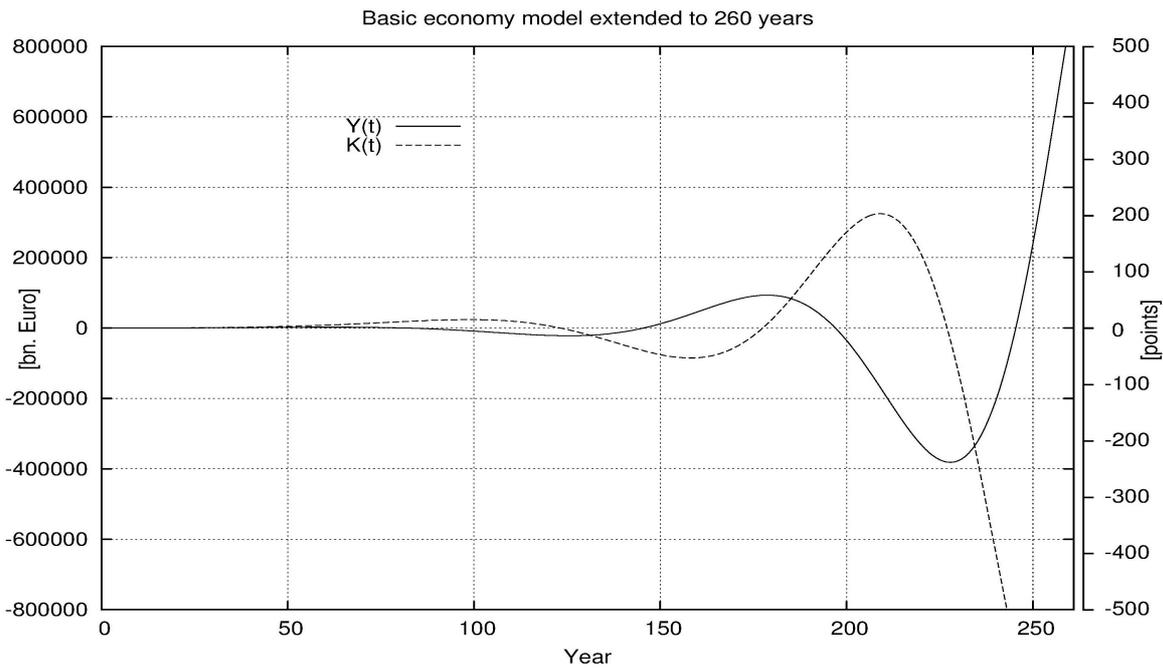

*Fig. 30: The solution of the basic system for 260 years in time ignoring the natural stop at Y=0.*

Such phase-shifted solutions are typically found in so-called "predator-prey" models e.g. in theoretical biology. The most prominent one is the Lotka-Volterra model which was already established in the 19th-century to explain the dynamics of populations of the animal kingdom.

The Lotka-Volterra equations are:
$$\dot{N}_1 = \epsilon_1 N_1 - \gamma_1 N_1 N_2$$
$$\dot{N}_1 = -\epsilon_2 N_2 + \gamma_2 N_1 N_2 \qquad (28.1)$$

where two populations, the prey-population $N_1$ and its predators $N_2$ are in competition and mutual dependency.

On the one hand, the predators eat and decimate their prey at the rate $\epsilon_1$, on the other hand, the spoils should go never out to survive. The decline of prey now results in a phase-shifted decrease in predators $\epsilon_2$, the prey population which in turn gives time to recover. The killing of prey is simulated by $\pm\gamma_i N_1 N_2$, i.e. the respective direct encounter between the two populations which at a certain rate for the predator is positive (*eat*) and negative for the prey (*die*).

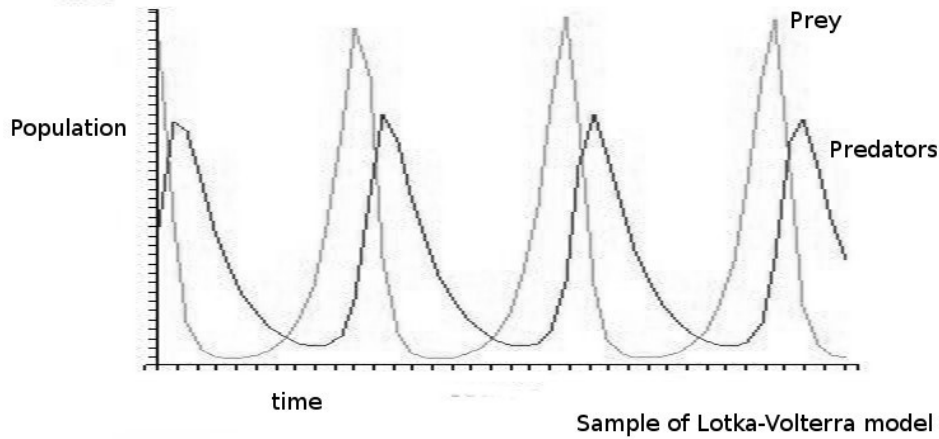

Fig. 31: Trajectory of a Lotka-Voltera system (Image Source: Wikipedia.it).

Now the basic system of a national economy is of a quite similar structure:

$$\dot{Y} = p_B Y - p_n K$$
$$\dot{K} = p_n K + p_S Y \qquad (28.2)$$

for the "meeting" of capital and GDP has different effects on GDP and financial capital stock. In particular $p_n$ is the result of the encounter $p_n K \approx (\alpha Y) \cdot K$. Similarly, however, are savings $p_S$ because it depends on an appropriate encounter $p_S Y \approx (\beta Y) \cdot K$ of the two. Thus in principle we can write:

$$\dot{Y} = p_B Y - \alpha Y K$$
$$\dot{K} = -0 \cdot K + (\alpha + \beta) Y K \qquad (28.3)$$

The term $-0 \cdot K \equiv 0$ we have just inserted in order to demonstrate the same type of structure here. The difference lies in the very different structure of the reproduction rates of the two populations. The first term of the second equation, we can also replace by net exports, which would ideally be zero, but often is not:

$$\dot{Y} = p_B Y - \alpha Y K$$
$$\dot{K} = a_0 (K_A) + (\alpha + \beta) Y K \qquad (28.4)$$

The different problem is ergo, that the growth rates $p_B$ and $a_0$ are not fed back[110] to each other. This does not result in a converging, but a as we will see, in divergent behavior.

We thus want to investigate this further. Now you can also simply indicate that GDP is a function of capital. This means instead of the two functions *Y(t)* and *K(t)* in time *t* space, we can now draw the derived one function *Y* in *K*-space *Y(K)*. So we change our reference room, to come up with a simpler but more meaningful representation. For this, we just have to plot the real data of *Y* and *K* and the basis model data as well to show for comparison. At first glance, this appears to be a parable function in *K*. In fact it can be approximated with a high confidence level for the FRG with the quadratic regression

$$Y(K[bn.€]) = -2.86 \cdot 10^{-5} K^2 + 0.519 K + 196.18 \quad [bn.€/y] \qquad (28.5).$$

This regression is a practical rule-of-thumb to us, from which we get the GDP *Y* to be expected at an existing financial capital stock of *K(t),* here for example in the FRG.

---

110 This evident structural difference means, at first view funny sounding, that a foreign investor who delivers some capital (predator) to an economy would have to deliver also enough prey (workers and consumer families) as "food" to make the investment sustainable in the long term.

The rule of thumb

$$Y(K) \approx -a_K^N K^2 + b_K^N K + c_K^N \qquad (28.6)$$

can be determined for every country $N$ and can be used for every country without large external net premiums slightly. The coefficients are sorted according to their meaning, here for the FRG:

$$c_K^{BRD} = 197{,}9 \, [bn. \, €/y] \qquad \text{GDP-Offset (} Y_0(K_0) \text{)}$$
$$b_K^{BRD} = 0{,}5174 \, [1/y] \qquad \text{Average Capital Efficiency}$$
$$a_K^{BRD} = 2{,}852 \cdot 10^{-5} \, [1/(bn. \, € \cdot y)] \qquad \text{Repressional Capital Coefficient} \qquad (28.7).$$

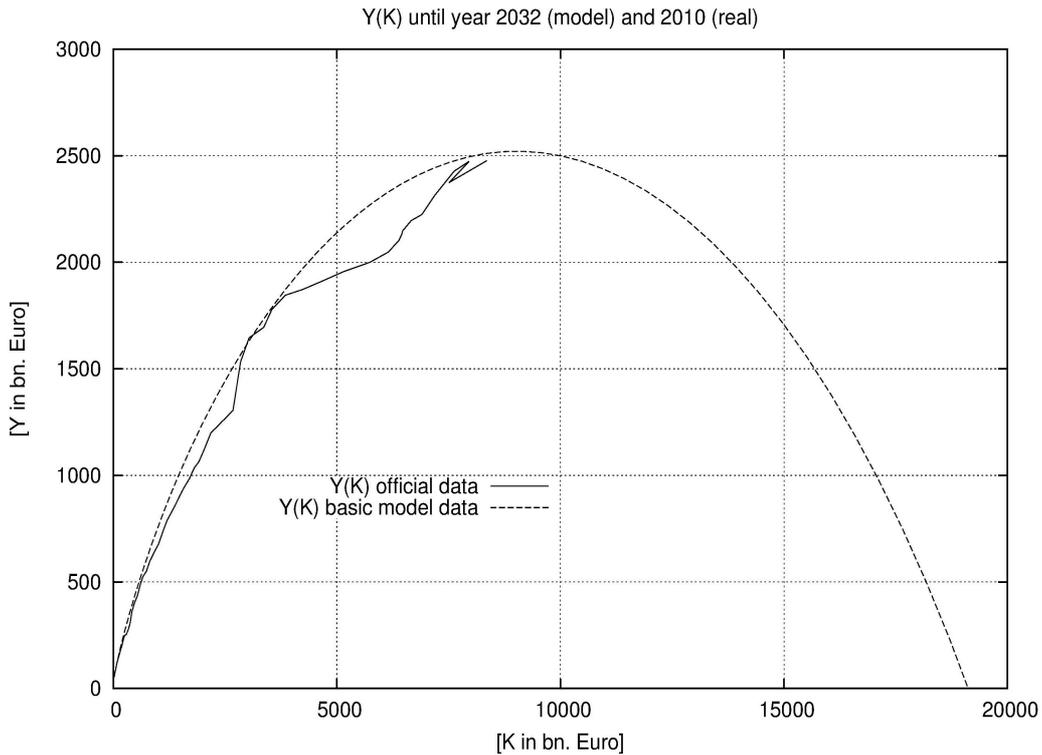

*Fig. 32: GDP Y as a function of the capital stock K. That is, the function Y(K) instead of Y(t),K(t). Both axes in units of currency in billions of € from 1950 to 2010. Model data for the FRG are from 1950 to about 2032 in the future.*

The constant $c_K^{BRD}$ describes in principle the GDP at the beginning of the census, compounded from $T_0$ with inflation. We already know the constant $b_K^{BRD}$ which is nothing more than the average marginal capital (debt) efficiency, which was for the FRG about 51.7 %. So in the long term about one half of capital goes into the real economy, the other half into the building up of the capital. The constant $a_K^{BRD}$ now describes the negative feedback by the with time increasing amount of return on capital, as the pressure on the GDP ultimately resulting from the compound interest will grow in time. We call it, because of its low absolute but due to the entrance to the square of the capital increasing weight, the *Capital Repression Coefficients* $a_K^N$ with its unit $\frac{1}{cur \cdot y}$. From the regression equation $Y(K) = -a_K^N K^2 + b_K^N K + c_K^N$ follows that the minimums of the solution are given for $Y(K_E) = 0$ which results in

$$K_E = \frac{b_K^N}{2 a_K^N} \pm \frac{1}{2 a_K^N} \sqrt{(b_K^N)^2 + 4 a_K^N c_K^N} \qquad (28.8).$$

With the values for the FRG we get $K_E = -375 \quad \vee \quad K_E = 18516$ bn. €/year, the latter of which is the maximum positive value of financial capital stock which could be reached. But only by the GDP depressed to zero. The maximum of the development of the GDP is given by the derivative equaling zero, which gives

$$\frac{dY}{dK}(K_{max})=0 \quad \text{at} \quad K_{max}=\frac{b_K^N}{2a_K^N} \tag{28.9}$$

For the FRG this will be at $K_{max} \simeq 9071$ bn. €. This is the maximum[111] value of capital raising ability of the FRG until the turn down of GDP-development arises. Especially the final value $K_E$ is in the extent of theoretical nature, as we would assume that until finite decline all market participants would behave unaffected by the development.

But at the latest after exceeding the maximum $Y(K_{max})$ this will be increasingly unlikely as then the artificial interventions into the monetary and economic system will be increasing. This leads to the relevant variables $a_0$ and $b_0$ getting more and more important in the basic equations and in reality as well.

## 29 Spiral Symmetry of Economy

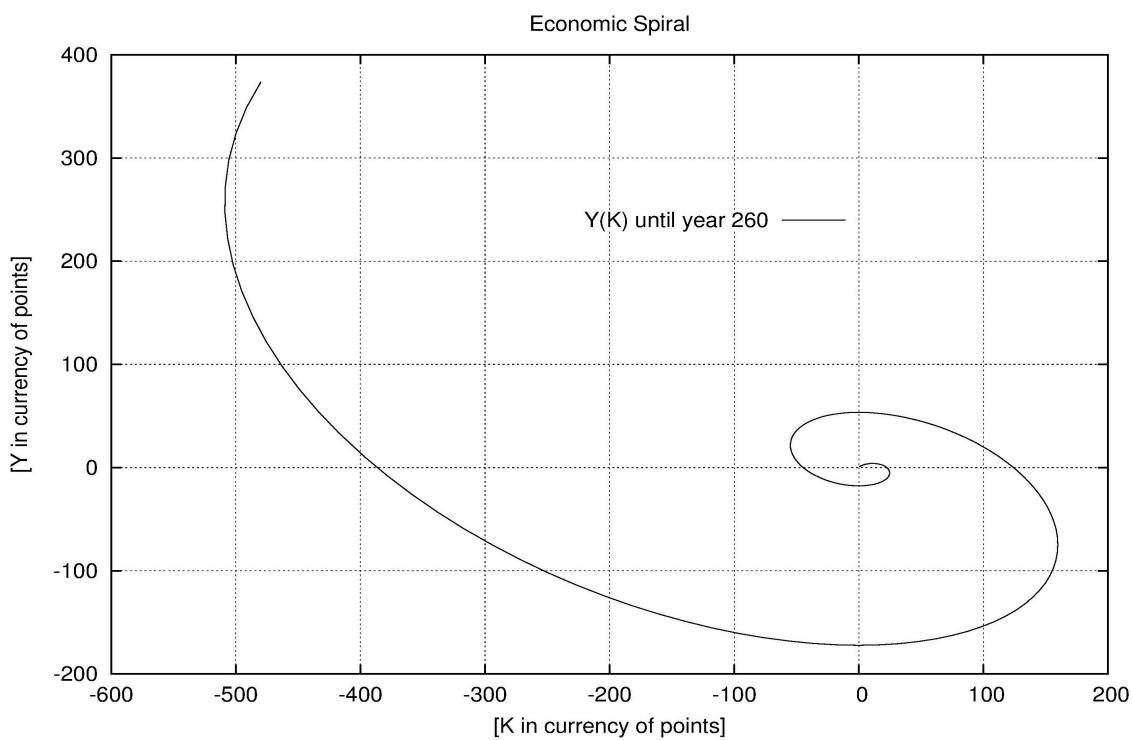

*Fig.: 33: Plot of Y(K) for 260 years, ignoring the realistic stop at Y=0. The real economy (as can be seen in the preceeding graph) is the very small first positive bow in the first quadrant from the start at center point 0.*

In fact, *Y(K)* is no parable but a <u>logarithmic spiral</u>, as we can see in a slightly to 260 years[112] lengthened integration. The real part is at the beginning of the spiral in the first quadrant. Such logarithmic spirals[113] are subject to certain symmetries.

---

111 In current prices and assuming a closed nationaleconomy.
112 Of course one may integrate over any period of time, then the spiral is only growing wider.
113 The logarithmic spiral has several unique characteristics, so it was one of her biggest fans, Jakob Bernoulli, who referred to it as *spiral mirabilis* ("miracle spiral") . The logarithmic spiral is invariant with respect to a central dilation by the factor exp(kγ) with simultaneous rotation of angle γ.

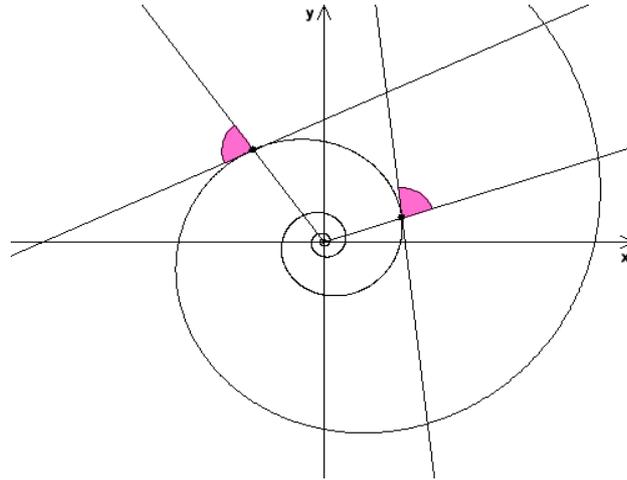

*Fig. 34: General features of a Logarithmic spiral (Image source: Wikipedia)*

With each rotation it increases the distance from its center (pole) by the same factor. Also each straight line through the pole intersects the spiral at the same angle, for which reason it is also known as equiangular spiral. It can be written in polar coordinates as

$$r(\phi) = a \exp(k\phi)$$

or in Cartesian coordinates as

$$x(\phi) = r(\phi) \cos(\phi)$$

$$y(\phi) = r(\phi) \sin(\phi) \tag{29.1}$$

Here $k = const.$ is the constant pitch of the spiral with pitch angle $\tan(\alpha) = k$. The name stems from the implicit representation

$$\phi = \frac{1}{k} \ln\left(\frac{r}{a}\right) \tag{29.2}$$

where $r/a$ is the so-called normalized radius. Elegantly the logarithmic spiral can be represented in the complex plane:

$$\omega(t) = z^t \quad \text{with} \quad z^t = \exp(t \ln z)$$

which will prove

$$z^t = \exp(t(\ln|z| + i \arg z)) = |z|^t \exp(it \arg z)$$

and can finally be written as:

$$z^t = |z|^t (\cos(t \arg z) + i \cdot \sin(t \arg z)) \tag{29.3}$$

With each turn then the radius increases by the constant factor

$$r(\phi + 2\pi) = a \exp(k(\phi + 2\pi)) = \exp(2\pi)^k \cdot r(\phi) \tag{29.4}$$

The logarithmic spiral is also invariant under a rotation and expansion by the constants $\exp(2\pi)^k$. The inversion $r \to 1/r$ leads to a reversal of the direction of rotation and mirroring the curve of the *y*-axis. In the case *a = 1*, it is only the reflection. For *k=0*, we obtain the circle as a limiting case of a logarithmic spiral with the equiangular intersection angle of $\pi/2$ = 90 degrees.

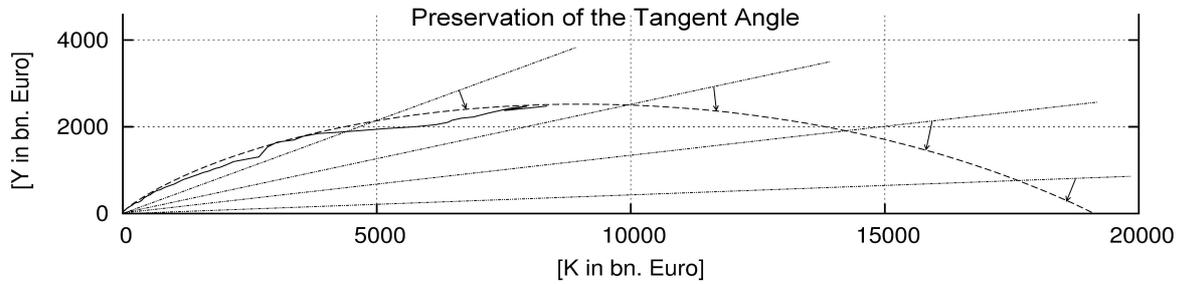

*Fig. 35: Plot of Y(K) first quadrant (realistic part). The invariance of the tangent angle can be seen here in a roughly approximate to scale plot.*

First, we exploit the symmetry of the constant tangent cutting angle $\alpha = const.$ . Our pole vector is $r = (K, Y)$ and thus the tangent vector is $r_T = (1, \partial_K Y)$ . Thus follows by rule of the scalar-product

$$r \cdot r_T = |r| \cdot |r_T| \cdot \cos \alpha \tag{29.5}$$

from which follows now the *'Economic Pythagoras'* :

$$\cos \alpha =: C = \frac{K + Y Y_K}{\sqrt{(K^2 + Y^2)(1 + Y_K^2)}} =: L(K, Y, Y_K) \tag{29.6}$$

From this invariant we can now solve the Euler-Lagrange equation:

$$\frac{d}{dK} \frac{\partial L}{\partial Y_K} - \frac{\partial L}{\partial Y} = 0 \tag{29.7}$$

The results are the three solutions

$$L_1 : Y_K = 0 \wedge Y = Y, K = K$$
$$\text{and} \quad L_{23} : Y = (Y_K \pm \sqrt{(-3Y_K^2 - 4)}) \frac{K}{2} \wedge Y_K = Y_K, K = K \tag{29.8}.$$

The importance of the first $L_1$ is trivial, it simply states that if no growth takes place through capital investment, then neither capital nor economy[114] grows. The other solutions can be resolved meaningfully related to the growth rate resulting from capitalization:

$$Y_K = \frac{Y \pm i \cdot \sqrt{(3Y^2 + 4K^2)}}{2K} = \frac{Y}{2K} \pm \frac{i}{2K} \cdot \sqrt{(3Y^2 + 4K^2)} \tag{29.9}$$

The importance of the real part is that the growth rate with respect to $K$ decreases as $dY/dK = Y/2K$ does. This means that the marginal benefit of capital decreases as $\frac{dY/dt}{dK/dt}$ with time, a fact which is confirmed by the experience and was treated as part of the special theory already. The absolute amount of the complex vector

$$|Y_K| = \frac{1}{2K} \cdot \sqrt{(\pm Y^2 + 3Y^2 + 4K^2)} \tag{29.10}$$

results in two possible amounts:

$$|Y_K| = \sqrt{(1 + \frac{Y^2}{2K^2})} \quad \text{or} \quad |Y_K| = \sqrt{(1 + \frac{Y^2}{K^2})} \tag{29.11}$$

---

114 Economic stagnation in the absence of credit trade.

The value $Y_K$ is now linked by the quantity equation with the average monetary velocity:

$$Y = \hat{H}\hat{P} = K\hat{V} \quad , \text{ and thus applies } \quad \partial_K Y = \hat{V} \tag{29.12}.$$

In the General Theory it proves useful to investigate the correlation of *Y* with *K* in the complex plane. Thus it is the complex vector

$$Z := K + iY = (\Re(Z), \Im(Y)) = (K, Y) \tag{29.13}.$$

It now runs the imaginary part in the *Y(K)*-diagram along the vertical *y*-axis and the real part along the horizontal *x*-axis. On the course of the real and imaginary part of the solution $\hat{V}$, we can already derive some typical properties for the average monetary velocity: So we can see the typical U-shaped curve of the average monetary velocity by the imaginary part corresponding to the real economy component. Unlike the real part that corresponds to the proportion of the financial capital stock, where the velocity decreases and eventually becomes even negative.

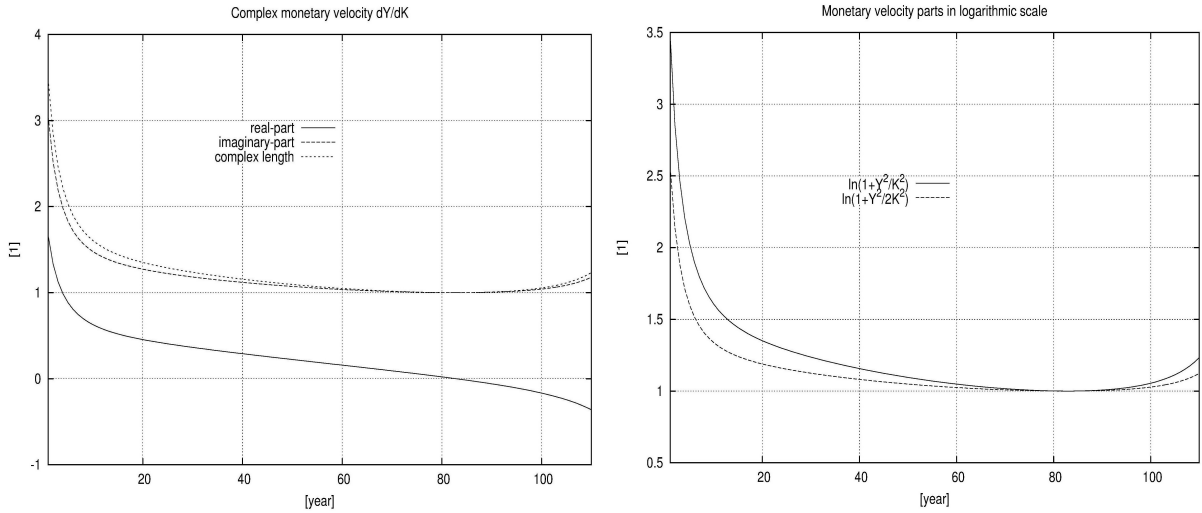

*Fig. 36: (left): Real part, imaginary part and magnitude of the complex monetary velocity. Fig. 42 (right): The amounts of the two average values of the complex monetary velocities in logarithmic scale. The two solutions differ just in little.*

It correspond to the empirical facts that could be observed in recent crisis that the banks due to lack of returns no longer trusted each other. And therefore, funds withheld, instead of selling loans. The real economy on the other hand is stagnating first, but only then to re-tighten. Which can be explained by the fact that first the general economic situation warns the money to be restraint. But at the end of the growing crisis money will be exchanged increasingly into real values. While the different development of the two contributions $\hat{V}(MV)$ and $\hat{V}(HP)$ is clear, but in the real economy, however, the sum of share is the load-bearing component.

With *L=0* all powers will result in zero too, so that the higher-orders $L^n = 0$ have their effects with

$$L^n = \left(\frac{K + YY_K}{\sqrt{(K^2+Y^2)(1+Y_K^2)}}\right)^n = \frac{n(K+YY_K)^{(n-2)}((K^2+Y^2)(1+Y_K^2))^{(-n/2)}}{((K^2+Y^2)(1+Y_K^2))} \cdots$$

$$\cdots \cdot [Y^3(n-1-Y_K^2) - K^3 Y_K(1-Y_K^2) + Y^2 K Y_K(1-2n+Y_K^2) + Y K^2 Y_K(n-1-Y_K^3)]$$

$$\cdots = n \frac{(K+YY_K)^{(n-2)}}{((K^2+Y^2)(1+Y_K^2))^{(1+\frac{n}{2})}} \cdots$$

$$\cdots \cdot [Y^3(n-1-Y_K^2) - K^3 Y_K(1-Y_K^2) + Y^2 K Y_K(1-2n+Y_K^2) + Y K^2 Y_K(n-1-Y_K^3)] \tag{29.14}$$

and thus giving

$$L^n = n \cdot L \cdot [(K+YY_K)^{(n-3)}((K^2+Y^2)(1+Y_K^2))^{(\frac{1-n}{2})}] \ldots$$
$$\ldots \cdot [Y^3(n-1-Y_K^2) - K^3 Y_K(1-Y_K^2) + Y^2 K Y_K(1-2n+Y_K^2) + Y K^2 Y_K(n-1-Y_K^3)] \quad (29.15).$$

One can rewrite it for further investigation. At least one of the factors of the last equation must be zero, and then the ancillary claims

$$(K+YY_K)((K^2+Y^2)(1+Y_K^2))^{(\frac{1-n}{2n-6})} = 0 \quad (29.16)$$

or

$$\frac{Y^2}{K}(n-1-Y_K^2) - K^2 Y_K(1-Y_K^2) + Y Y_K(1-2n+Y_K^2) + K Y_K(n-1-Y_K^3) = 0 \quad (29.17)$$

are justifiable. The first equation is nothing but our original equation, the second however, can be exploited further. The latter but can not be resolved for $Y_K$ elementary. For *n=1* there is a fairly complicated basic solution

$$Y_K^{(1)} = \frac{1}{6\gamma K} \cdot (4K^2 + K^2(8Y+2\gamma) - 8Y^2 + 2\gamma Y + \gamma^2) \quad (29.18)$$

with the abbreviations

$$\alpha = 8K^6 + K^4(24Y-108) - K^2(12Y^2+108Y) - 28Y^3$$
$$\beta = -12K^{10} + K^8(81-48Y) + K^6(162Y-18Y^2) + \ldots$$
$$\ldots + K^4(60Y^3 + 81Y^2 - 3Y^4) + K^2(42Y^4 - 6Y^5) + 9Y^6$$
$$\gamma = 6K(\alpha + 12\sqrt{(\beta)})^{1/3} \quad (29.19)$$

which is, however, of little benefit. The more interesting solutions up from *n=2* can not be solved elemental. But one can make an approximation for the inner region of the solution, as we know that there is $Y_K^2 \approx 1$ in normal times. Thus one can justify for $n \geq 2$ the need of the following approximation:

$$\frac{Y^2}{K}(n-2) + 2Y Y_K(1-n) + K Y_K(n-1-Y_K) \approx 0 \quad (29.20)$$

which results in the two approximate solutions:

$$Y_{K_{1,2}}^{(n)} \approx \frac{Y}{K}(1-n) \pm \sqrt{Y^2(n^2-n-1) - \frac{Y}{K}(n^2-2n+1) + \frac{n^2-2n+1}{4}} + n - \frac{1}{2} \quad (29.21)$$

Using the abbreviation of the capital coefficient as $K/Y := \delta K$ which results in

$$Y_{K_{1,2}}^{(n)} \approx \frac{1}{\delta K}(1-n) \pm \sqrt{\frac{K^2}{\delta K}(n^2-n-1) - \frac{1}{\delta K}(n^2-2n+1) + \frac{n^2-2n+1}{4}} + n - \frac{1}{2} \quad (29.22)$$

wherein the first of the higher orders is given by *n=2*:

$$Y_{K_{1,2}}^{(n)} \approx \frac{-1}{\delta K} \pm \sqrt{\frac{2K^2}{\delta K} - \frac{1}{\delta K} + \frac{1}{4}} + \frac{3}{2} \quad (29.23)$$

Although this approximation is not a very good one, it clearly shows however a similar trend as in the previously known solution of the special theory. The difference lies in the last section, where the crisis is already in the lower phase: The mean velocity increases significantly at the end, instead of decreasing. The reason for this is that in the last phase, as already discussed, for the preservation of quantity equation a high $\hat{V}$ is necessary. Whether it's the fact that one is pumping fresh money into the system, or that investors will convert more unprofitable money into real property values. To include the higher order terms, we can summarize them in a linear combination of solutions, in rough analogy to the Taylor series

$$\psi = \sum_n \frac{1}{n!} \cdot \psi(L^n) \tag{29.24}$$

and because of

$$\psi(Y_{K_{1,2}}^{(n)}) \approx \sum_n \frac{1}{n!}[\frac{1}{\delta K}(1-n) \pm \sqrt{\frac{K^2}{\delta K}(n^2-n-1) - \frac{1}{\delta K}(n^2-2n+1) + \frac{n^2-2n+1}{4}} + n - \frac{1}{2}] \tag{29.25}$$

we get the results for larger $n \gg 1$ in a further approximation:

$$\psi(Y_{K_{1,2}}^{(n)}) \approx \sum_n \frac{1-n}{n! \delta K} \pm \frac{n}{n!}\sqrt{\frac{K^2}{\delta K} - \frac{1}{\delta K} + \frac{1}{4}} + \frac{n-\frac{1}{2}}{n!}$$
$$\ldots \approx \sum_n \frac{1}{(n-1)!}(\frac{-1}{\delta K} \pm \sqrt{\frac{K^2}{\delta K} - \frac{1}{\delta K} + \frac{1}{4}} + 1) \tag{29.26}$$

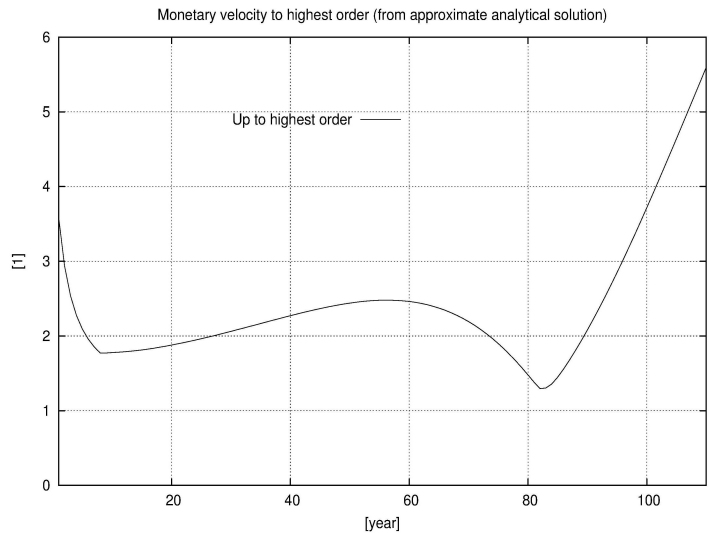

*Fig. 37: The sum of the contributions up to the highest order to the above-discussed approximations. (The nine undefined values at the beginning of the approximate solution (years 1 to 9), due to negative root argument have been interpolated in this sample.) The now non-linear (approximative) calculation shows, in contrast to the simpler linear calculation, the self-consistent occurrence of hyperinflation as a consequence of increasing monetary velocity in the very end after some years of deflation (which resulted from decreasing monetary velocity).*

The second factor is now adjusted for $n \to \infty$, and therefore we can isolate the total faculty. Because of the known exponential series representation $\exp(x) = \sum_i \frac{x^i}{n!}$ the sum total of the rest can be approximated[115] to

$$\psi(Y_{K_{1,2}}^{(n)}) \approx (e-5/2)(\frac{-1}{\delta K} \pm \sqrt{\frac{K^2}{\delta K} - \frac{1}{\delta K} + \frac{1}{4}} + 1) \tag{29.27}$$

The four approximations (+, +), (-, -), (+, -), (-, +) can be used to estimate higher order effects on the average value of $Y_K = \hat{V}$, wherein the (+, +) solutions results are shown in the curve above.

---

[115] Since $\exp(1) = \sum 1/n!$ the terms for n=0 to 2 must be subtracted. This results in $e - 5/2 \approx 0{,}218$.

# 30 Rotational-Stretching-Symmetry

If we want to transfer the position vector from the *Y-K*-space currently used into the usual (*Y,K*)-*t*-space at the time $t_0$ to the time $t_1$, then we can achieve this with the help of a rotational expansion. The rotational extension symmetry is the same symmetry, as occurs in the multiplication of imaginary numbers:

$$z_1 \cdot z_2 = r\, e^{i\phi} \cdot \rho e^{i\psi} = r\rho\, e^{i(\phi+\psi)} \tag{30.1}$$

Our position vectors are as shown above, just

$$r_0 = (K_0, Y_0) \quad \text{and} \quad r_1 = (K_1, Y_1) \tag{30.2},$$

where we have set *Y* in the imaginary direction, i.e. in general

$$r = \rho e^{i\phi} = (K^2 + Y^2) \cdot (\cos\phi + i\sin\phi) \tag{30.3}$$

for the position vector. The expression $\cos\phi + i\sin\phi$ is often associated with the *cis*-function $\operatorname{cis}\phi := \cos\phi + i\sin\phi = \exp(i\phi)$.

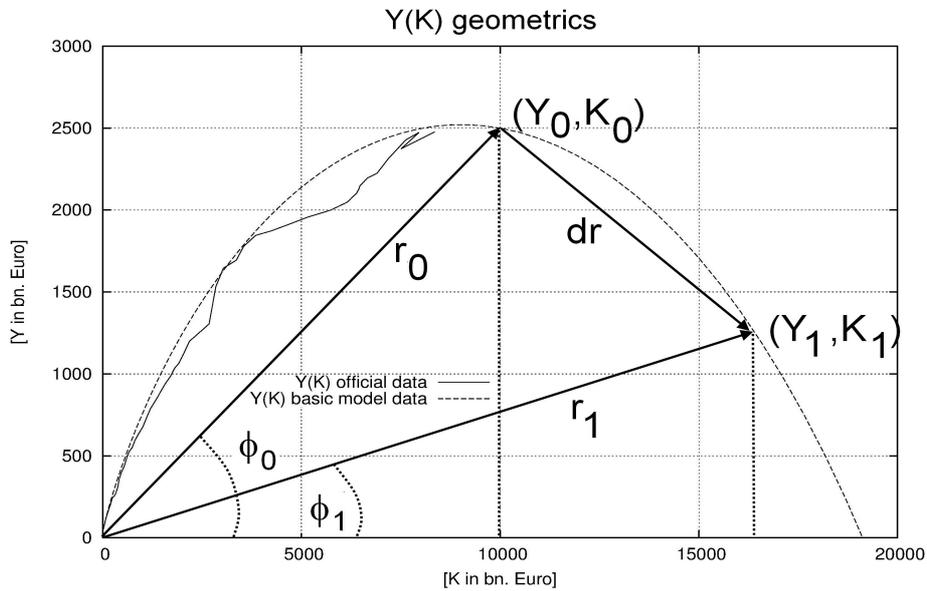

*Fig.38: The geometric situation in the Y-K-Space used for further analytical calculations.*

It is now apparent manner:

$$r_1 = r_0 + dr = r_0 \cdot a \exp(-ib\phi) \tag{30.4}$$

or

$$r_1 = (K_1^2 + Y_1^2)\operatorname{cis}\phi_1 = ((K_0 + dK_0)^2 + (Y_0 + dY_0)^2)\operatorname{cis}\phi_1 = ...$$
$$... = ((K_0 + 1)^2 + (Y_0 + Y_K)^2)\operatorname{cis}\phi_1 = ...$$
$$... = (K_0^2 + Y_0^2)\, a \exp(-ib(\phi_0 - \phi_1)) = (K_0^2 + Y_0^2)\, a \operatorname{cis}(b(\phi_0 - \phi_1)) \tag{30.5}$$

with the implicit Logspiral-representation $\phi_i = \dfrac{1}{b}\ln\left(\dfrac{r_i}{a}\right)$. The effect of rotation and expansion is evidently composed both of inflation or pricing $a_I := a$ and trading gain $b_H := b$. So we get

$$a_I = \frac{(K_0+1)^2+(Y_0+Y_K)^2}{K_0^2+Y_0^2} \operatorname{cis}(\phi_1+b(\phi_0-\phi_1)) \tag{30.6}$$

and

$$b_H = \frac{i}{\phi_0-\phi_1}\left(\phi_1-\ln\left(a\frac{K_0^2+Y_0^2}{(K_0+1)^2+(Y_0+Y_K)^2}\right)\right) \tag{30.7}.$$

We see here already that $b_H$ moves on its own axis of GDP, while $a_I$ affects the GDP and capital axis equally. Because inflation affects both the capital and the value of GDP both measured in money. The implicit representation

$$\phi_i = \frac{1}{b_H}\ln\left(\frac{r_i}{a_I}\right) \quad \text{results in} \quad b_H = \frac{1}{\phi_i}\ln\left(\frac{r_i}{a_I}\right) \quad \text{and} \quad a_I = r_i\exp(-b_H\phi_i) \;.$$

Inserting brings:

$$b_H\left(1-i\frac{\phi_0}{\phi_0-\phi_1}\right) = \frac{i}{\phi_0-\phi_1}\left(\phi_1-\ln(dr)+\ln\left(\frac{K_0^2+Y_0^2}{(K_0+1)^2+(Y_0+Y_K)^2}\right)\right) \tag{30.8}$$

The effect of the helical symmetry is that now enters the commercial growth in the real *and* imaginary parts as well.

***The dynamics of economic development thus corresponds always to a rotation and expansion, so that a change in trade is always accompanied by a price change due.***

The division of two complex numbers is given by $re^{i\phi}/\rho e^{i\psi} = r/\rho \cdot e^{i(\phi-\psi)}$, and thus can[116] be written:

$$\begin{aligned}
b_H &= (1/2-1/2\,sign(dr))\pi/((\phi_0-\phi_1)(1+\phi_0^2/(\phi_0-\phi_1)^2))-(\phi_1-\ln(|(dr)|)+...) \\
&(...+\ln((K^2+Y^2)/((K+1)^2+(Y+Y_K)^2)))\phi_0/((\phi_0-\phi_1)^2*(1+\phi_0^2/(\phi_0-\phi_1)^2))+... \\
&...+i\cdot((\phi_1-\ln(|(dr)|)+\ln((K^2+Y^2)/((K+1)^2+(Y+Y_K)^2)))/...) \\
&.../((\phi_0-\phi_1)(1+\phi_0^2/(\phi_0-\phi_1)^2))+... \\
&...+(1/2-1/2\,sign(dr))\pi\phi_0/((\phi_0-\phi_1)^2(1+\phi_0^2/(\phi_0-\phi_1)^2)))
\end{aligned} \tag{30.9}$$

Because of $sign(dr)=1$ and $|dr|=dr$ at $dr>0$ it follows:

$$b_H = (\phi_1-\ln(dr)+\ln\left(\frac{(K^2+Y^2)}{(K^2+2K+1+Y^2+2YY_K+Y_K^2)}\right))\frac{(-\phi_0+i\cdot(\phi_0-\phi_1))}{(2\phi_0^2-2\phi_0\phi_1+\phi_1^2)} \tag{30.10}$$

and

$$H(r) = H(r_0)\cdot b_H \tag{30.11}.$$

This implicit representation we have to convert into a suitable explicit time dependence. For $a_I$ results in the meantime:

$$a_I = \frac{(K_0+1)^2+(Y_0+Y_K)^2}{K_0^2+Y_0^2}\operatorname{cis}\left(\phi_1+\frac{\phi_0-\phi_1}{\phi_0}\ln\left(\frac{dr}{a_I}\right)\right) \tag{30.12}$$

and further to resolve complex for $a_I$ takes after some merging:

---

[116] Of course today one uses the help of analytical software like Maple and others.

$$a_I = \rho^2 \exp\left(d\phi \frac{\psi_2}{\psi_1}\right) \cos\left(\phi_0 \frac{\psi_2}{\psi_1}\right) + \frac{i}{r_0^2} \cdot \sin\left(\phi_0 \frac{\psi_2}{\psi_1}\right) \quad (30.13)$$

with the abbreviations:

$$\psi_0 := \phi_0(\phi_1 + \ln(dr)) \quad , \quad \psi_1 := 2\left(\phi_0^2 - \phi_0\phi_1 + \frac{\phi_1^2}{2}\right) \quad ,$$

$$d\phi := \phi_0 - \phi_1 \quad , \quad r_0^2 := K_0^2 + Y_0^2 \quad ,$$
$$r_1^2 := (K_0 + 1)^2 + (Y_0 + Y_K)^2 \quad , \quad \rho^2 := r_1^2 / r_0^2$$

$$\text{and} \quad \psi_2 := \psi_0 - d\phi \ln(\rho^2) - \phi_1 \ln(dr) \quad (30.14).$$

In accordance with this the inflation or price-formation is

$$I(r) = I_0(r_0) \cdot a_I \quad (30.15).$$

Then there is the transformation back to the usual time dimension. For this we assume the mathematical relationship

$$\phi_i(t) = \omega_i t + \omega_0 = \frac{\pi/2}{T_{END} - T_0} \cdot (T_0 - t) + \pi/2 \quad (30.16).$$

The angular velocity $\omega_i$ we can determine by use of the start and end time of the model economics. Because the angle was 90 degrees = $\pi/2$ at the beginning, and 0 at the end in our vector space of *Y(K)*-representation. Therefore, one may write as we start counting at $T_0 = 0$ :

$$\phi_i(t) = \omega_i t + \omega_0 = \frac{-\pi/2}{T_{END}} \cdot t + \pi/2 \quad \text{and} \quad \omega_i = \frac{-\pi/2}{T_{END}} \quad (30.17).$$

This is $T_{END}$ the time when the GDP dropped to zero in basis theory. With

$$H(r) = H(r_0) \cdot b_H(\phi_i(t)) \quad \text{and} \quad P(r) = P_0(r_0) \cdot a_I(\phi_i(t)) \quad (30.18)$$

one can now determine in principle the behavior of price formation and trading volume in the usual time dimension.

## 31 Commutator Symmetry

For some calculations about the evolution of average interest rates we can make use of commutator symmetry. We make here a mathematical bond in quantum mechanics. Numbers are subject to some simple group symmetries, especially the commutativity property $a \cdot b - b \cdot a = 0$ or abbreviated as $[a, b] = R$ . The brackets [] are called the *"commutator"*, and the right side of the equation *"commutator-residue"*. This residue *R* must be but not zero in the general[117] case. Therefore let us look again on the derivation of the previous one to get an impression on how to procede in principle on this issue. When considering the quantity equation in the linear SMF, we have the relationship

$$V(t) = \left((1 - p_s) Y(t) + (1 + p_s) \frac{dK(t)}{dt}\right) \frac{Y(t)}{K(t)}$$

to derive the average monetary velocity. This structure enables us to resort something to

$$KV = ((1-p_s)Y + (1+p_s)d_t K)Y$$

or

$$(\mathring{KV}) \cdot Y := ((1-p_s)Y + (1+p_s)d_t K) \cdot Y = KV$$

or

$$(V - (1+p_S)d_t) \cdot K = \frac{(1-p_S)Y}{P} \cdot P \tag{31.1}$$

---

[117] If the commutator-residue is zero, it is called Abel's numbers, if it is not equal to zero, whereas called non-Abelian numbers. Elementary particles are represented in particular by such non-Abelian numbers. Operators are, as the name suggests, some mathematical gadgets that do anything when applied to other numbers, say, in the case of differential operators that make a differentiation.

We can now write in *operator* notation:

$$\mathring{V} \cdot K = \mathring{H} \cdot P \tag{31.2}$$

with the velocity and trade operators defined by

$$\mathring{V} := V - (1+p_s)d_t \quad \text{and} \quad \mathring{H} := \frac{(1-p_s)Y}{P} \tag{31.3}$$

From e.g. quantum mechanics we know the simple coherences of such elementary symmetries as $[\mathring{x}_i, \mathring{p}_j] = i\hbar\delta_{ij} \wedge [\mathring{x}_i, \mathring{x}_j] = 0 = [\mathring{p}_i, \mathring{p}_j]$ as known about the properties of atoms and their constituents. This means that the *KV*-operator defined here is applied to *Y* so that the *KV*-product is produced.
Thus we get:

$$[K\mathring{V} - \mathring{V}K]Y = (1+ps)(KY_t - K_t Y) \tag{31.4}$$

The commutator-residue should we expect equal zero as in the economy we use always finite abelian numbers. So either $p_S = -1$, which almost never happens, or specify:

$$\frac{K}{Y} = \frac{\dot{Y}}{\dot{K}} \quad \text{or} \quad \dot{K} = \dot{Y} \cdot \frac{Y}{K} \tag{31.5}$$

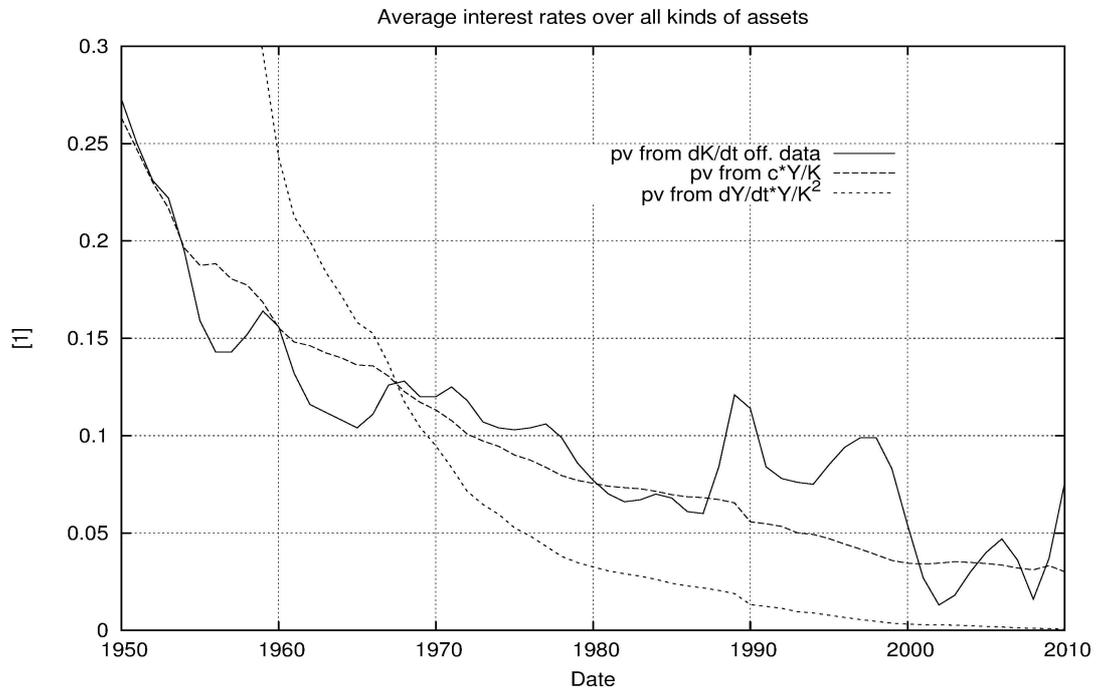

*Fig. 39: Approximate average interest rate of an economy by real data and by two theoretical arguments as well. It shows the fact, that without the bank's own business one can not achieve the actually higher interest rates after phase change (1967). On the other side, before phase change one could have had achieved much higher returns on investments due to investments solely in the real economy.*

Now $\dot{K}/K$ is precisely the nominal return on capital (and derived from that also namely $p_v$ or $p_n$ ):

$$p_{nominal} = \dot{Y} \cdot \frac{Y}{K^2} \qquad (31.6)$$

which gives the desired results in relation to the _average_ return on all assets of an economy in the linear case.

This interest rate we can now compare with the supply-demand estimate at the beginning of the second part of this article. In the last graph we see the result. The real change in the total nominal capital stock $dK/dt$, is roughly equivalent to the nominal interest rate $p_v$ over all kind of assets. The curve (21.1) with the choice $c = 1/8 = const.$ comes very close to it. This corresponds to the fact that in the underlying function, also for financial products, there is an effective supply-demand ratio. The curve (31.6) on the other hand, goes faster to zero than the real numbers do. The reason for this lies in our last very simplified derivation in which only $p_S$ occurs but not $p_n$. The investment business is therefore _not included_. Therefore (31.6) intersects the curve of the real numbers at the phase change in 1967, and then falls off more quickly.

This is, since after phase change one can not achieve then actually higher interest rates without the bank's own business. Before phase change but the interest rates would have been even higher, due to the same effect in the other direction, as then loans to the real economy would have been the very better choice.

This first degree analysis was made for the bulk of all real and investment assets. Further analysis will have to split up for at least both principle kind of assets which should lead for different interest rates prediction and thus also to the different distribution of money into the two main parts of financial business. Which then should give an analytical sustained equation for the relation (4.5) $p_{rel}(t)$ .

# 32  Conclusions and Outlook

Large economies such as Germany, are an extremely complex entity in which a wide array of important subjects are hardly predictable in detail. Nevertheless, but the two key variables of an economy behave surprisingly unperturbed by most of them, which is the gross domestic product (GDP) $Y$ and the total financial capital stock (assets) $K$ which tend to grow in a manifest coupled manner over time. But as experience and pure theory shows as well, growth will come to stagnation and serious crisis at least once per century as a seemingly unavoidable intrinsic consequence of any economy driven by credit and interests coupling its GDP growth to the growth of Capital.

Traditional growth models showed to be unsatisfactory, especially regarding the fact to predict and explain the (historically regular) arise of financial crisis. The reasons why stems from economic reasoning and mathematical-physical reasoning as well:

Economic Reasoning:

*Classical Dichotomy:* It must be clear that there is *no ad hoc reason* to exclude the so called Banks Own Business or Investment-banking from the overall balance of an economy. As every product also financial products have their influence on the motion of money, savings and substitution with other possible consumptions or investments. As the balance sheet of the financial business e.g. in Germany was in 1950 about a third of GDP, in 2010 but it was more than three times the GDP which is a relative increase by a factor of approximate 10, this cannot ad hoc be regarded as being of no questionable influence to the GDP.

*Micro-economy:* A lot of confusion in modelling growth in macro-economy stems from the fact, that micro- and macro-economic concepts are interwoven mostly unquestioned. But those concepts differ much, as micro-economy handles open systems and macro-economy handles (quasi-) closed systems. So for a firm, which is very small in comparison to the national economy, the surroundings of the firm may be regarded to be an infinite pool and source of money, goods, consumers and traders. But in a closed system like a national economy (quasi-closed) or the worlds economy (closed), we have to deal with the principle effects of a closed system. Which means that every current must be balanced by sinks and sources. Giving at one point always means taking away at another point in the system due to substitutional effects.

Mathematical Reasoning:

Although it is seemingly a simple guess that GDP $Y$ gives reason for the rise of Capital (Assets) $K$, the other way around, that capital <u>always</u> needs interests stemming in last consequence from the GDP, is widely denied. Mostly accepted is just the fact, that (only) loans to the GDP give rise to GDP growth, and thus also to Capital growth through repaying plus interests as well. But whatever model construction one prefers, this ultimately means that we have to deal mathematically with <u>at least</u> *a system of two linear independent but <u>coupled</u> differential equations* for $Y$ and $K$. This is just a minimum requirement. Every construction with less than this cannot give reliable results in the long term. E.g. the textbook Solow–Swan model, but is set such that it must be an asymptotically Cobb-Douglas-production function of the kind $Y(t) = K^\alpha(t)[A(t)L(t)]^{1-\alpha}$ with $0 < \alpha < 1$ called the elasticity of output in respect to capital and $A \cdot L$ the effective Labour part of the economy. There exist a lot of similar models which are <u>but all of the same principle kind</u>, which is a single equation for $Y$ of the form $Y(t) := f(K(t), ...)$. Such singular models may make sense only for open systems like in micro-economics or at the best as an statistical approach for a short term extrapolation in macro-economics.

We have showed that avoiding this two principle flaws in classical growth theory of macroeconomics leads, by concepts of field theory, directly to a consistent analytical model of macro-economy. For practically exact calculations and predictions in "Dollars and Cents" of the worlds economies then just is needed the availability of some reliable econometric main numbers, which must be provided from the statistical institution of the nations on a common basis.

Essential for the macro-economy will be the consideration of the *total capital coefficient K/Y* . The effect of too much capital has been discussed by a lot of authors, and even without mathematics the logic lies close as the absorbing ability of the GDP for loans is not infinite. Using all existing assets in the form of direct loans to GDP at a today total capital ratio of more than 3 would mean that the GDP would have to be implemented at least every four months by crediting fully again. At present Capital therefore preferably will be used as an investment instrument traded in the so-called *"banks own business"* or *"investment banking"*. For the needs of the GDP are already more than covered, and the lack of demand for loans will result in low capital prices (interests). This means that the attainable yields in the commercial bank model are rather lower than they are remaining in the commercial banking sector. As we could show, financial crisis of an economy will occur when (a) the ratio of the whole financial stock to the GDP exceeds 3 and (b) the share of Commercial Banks Business on the whole banks business gets lower van 0,5=50%. From that one can follow that it is also a sign for a coming crisis, when (c) the amount of loans into the economy firstly rises up to about 150%.

We could show that the typical time for a (quasi-) <u>closed</u> economy from the start at point zero (currency and debt depreciation/reform, mostly done after large wars) to the first signs of crisis will be around 50 to 60 years. The main obstacle of a capital driven economy is that only as long as the savings quota of the economy can pay for the interests of the whole of all assets of an economy crisis can be avoided. If the growth of capital (which by banks balance sheets equals to the economies over all debt) exceeds the growth of paying power through the growth of real economy but crisis will occur. It will not end until the balance between capital and real goods is established again.

Although classical growth theory is fundamentally flawed, we could show that macroeconomics at the whole is not. So many of the classical knowledge could be fundamentally substantiated using field theoretic tools or can be fixed at least with some subtle corrections. Today the nations statistical institutes are due to classical growth theory very much focused on GDP data and much less on assets data. For future exact calculation but there is the need for much more and precise data especially on assets. So today in a lot of nations asset data are only available in parts. And even in Germany, were data availability and quality is good, there is some lack on one question regarding assets data: Who pays the bill (interests)? So we have to know not only the bulk amount of assets but we also must address the question which assets in banks balance sheets are buyed and sold abroad, which means which national GDP has to pay the effective interest rates. This question will be prominent for countries like the US or Japan, which have a large part of financial business abroad. Regarding all data, especially today's GDP data is that we need the raw data for exact calculations too, as already statistically altered data, especially if hedonized, can get misleading or in the worst case being even useless for reliable calculations.

Next Developments and Outlook:

It must be remarked that field theory of macro-economics has no principle limits for further developments, but it has still a lot of work to do. Hopefully this article will encourage some more scientists to adapt field theory and its tools in larger extend to economic analysis and research. Some example issues to be worked on in the near future are:

<u>Introducing Wages and Labour to the equations:</u> This will be the next step and is but just straight forward working with the mathematical tools showed herein. Besides the fact that the product Labour is the most important part of GDP which must be analysed in depth, especially for the large social and political aspect behind it and the need to understand its long-term evolutions and rules.

<u>Inequality:</u> Inequality is also one of the main economic obstacles in modern times. Although inequality is needed as a driving force for economic benefit and thus for growth, to much inequality regularly leads to economic and thus social destructions which in the worst case will lead to revolutions and civil wars. It is crucial to understand how inequality arises and to address questions like: How much inequality is needed? What is, and how is to avoid, to much counter productive inequality? Indeed those question relate to Wages and Labour and the distribution of productivity growth onto the main parts of the real and the financial economy, and on their real and investment parts and their interdependent substitutions.

<u>Investment Banking:</u> also called Banks-Own-Business is a financial technical tool which growths with time and with growing *K/Y*-ratio. Although the theoretical handling of it in the average bulk gives already good results for more exact calculations, and also for better understanding of its positive and negative aspects as well, one has to analyze the different kinds of assets and their special relations to GDP in more detail. For example the effect of stock market products or derivatives like CDO's are presumably much different.

<u>Interest Rates:</u> Interest rates may be analyzed, understood and derived self-consistently from non-linear theory.

<u>Arbitrage:</u> For understanding international relations the effects of arbitrage in currency business must be analyzed theoretically in detail.

<u>Econometry:</u> This means deriving the needed international economy data from existing databases as much and with as good quality as possible. Hopefully sensitizing economists at econometric institutions to establish a common sense on needed data for growth prediction in the future, regarding quality and special kind of data as well.

<u>Practical Guide:</u> This article focusses on the analytical theory and tools mainly. For the use in day-to-day routine a guide with practical instructions and examples will be issued.

For any question, hint, consideration or support regarding this article don't hesitate to contact me.

# Used Data and Tables

The original tables usually have monthly or annual values until 1998 in Deutsche Mark (DM) and from 1999 in EURO. For the numerical values of the DM availability were uniformly converted into EURO. All EURO amounts are expressed in units of billions of euros. The following table shows the values on an annual basis. In the first column of the tables calendar years are indicated.

**Time Series OU0308 (FRG):** Banks balance sheet total (up to December 1998: Volume of business) / All bank categories [*Bilanzumme (bis Dez. 1998 Geschäftsvolumen) / Alle Bankengruppen; Allgemein: Geschäftsvolumen = Bilanzumme zuzüglich Indossamentsverbindlichkeiten aus rediskontierten Wechseln, den Kreditnehmern abgerechneten eigenen Ziehungen im Umlauf sowie aus dem Wechselbestand vor Verfall zum Einzug versandte Wechsel.*]
General: billed business volume = total assets plus endorsement liabilities arising from rediscounted bills of exchange in circulation drawn by the borrowers as well as from the stock exchange before maturity for the collection of bills sent. Source:

**Time Series OU0115 (FRG):** Lending to domestic non-banks / Total / MFIs / All categories [*Kredite an inländische Nichtbanken/ insgesamt / mit Wertpapieren und Ausgleichsforderungen / Alle Bankengruppen*]. Source:

**Time Series: GDP of the FRG**
Source: Destatis, in units of billion Euros.

**Time Series: Savings rate FRG**
Source: in units of natural percent, for example *0,042=4,2 %.*

**Time Series: Population FRG**
Source: http://www.statistikportal.de/Statistik-Portal/de_zs01_bund.asp in units of thousands. Beginning with 1991 including the population of the GDR.

**Time Series: National States Debt FRG**
Source: http://www.miprox.de/Wirtschaft_allgemein/BRD-Oeffentliche-Verschuldung2003.htm
in units of billion of Euros.

**Time Series: Inflation rate (CPI) for the FRG**
Source: http://vbf-online.de/files/Entwicklung%20Teuerung%20(Inflation)%20und%20Verschuldung.pdf
in units of percent.

**Time Series: Target 2**
http://www.eurocrisismonitor.com with data at
http://www.eurocrisismonitor.com/Intra_Eurosystem_balances.xlsx

| YEAR | ASSETS (bn.€) | LOANS (bn.€) | GDP (bn.€) | States Debt (bn.€) | Savings natural [1] | Population in 1000 | CPI in % consumer |
|---|---|---|---|---|---|---|---|
| 1950 | 19,966 | 14,418 | 52,582 | 10,530 | 0,042 | 50958 | -6,4 |
| 1951 | 25,688 | 18,151 | 64,528 | 11,450 | 0,040 | 51435 | 7,6 |
| 1952 | 32,585 | 22,820 | 73,820 | 13,000 | 0,066 | 51864 | 2,1 |
| 1953 | 40,235 | 28,412 | 79,282 | 17,000 | 0,076 | 52454 | -1,7 |
| 1954 | 51,098 | 35,400 | 85,102 | 19,790 | 0,081 | 52943 | 0,4 |
| 1955 | 61,050 | 42,284 | 97,252 | 20,960 | 0,074 | 53518 | 1,4 |
| 1956 | 68,721 | 47,620 | 107,513 | 21,470 | 0,066 | 53340 | 2,8 |
| 1957 | 80,588 | 52,562 | 117,162 | 22,450 | 0,088 | 54064 | 2 |
| 1958 | 91,713 | 58,880 | 125,891 | 23,780 | 0,091 | 54719 | 2,3 |
| 1959 | 107,401 | 69,091 | 137,888 | 25,360 | 0,094 | 55257 | 0,6 |
| 1960 | 128,913 | 85,594 | 154,531 | 27,000 | 0,092 | 55958 | 1,6 |
| 1961 | 149,404 | 97,774 | 169,336 | 29,190 | 0,099 | 56589 | 2,5 |
| 1962 | 166,453 | 109,923 | 184,192 | 30,880 | 0,095 | 57247 | 2,8 |
| 1963 | 187,110 | 123,128 | 195,219 | 34,310 | 0,107 | 57865 | 3 |
| 1964 | 208,307 | 139,006 | 214,516 | 37,730 | 0,118 | 58587 | 2,4 |
| 1965 | 233,208 | 156,623 | 234,426 | 42,800 | 0,129 | 59297 | 3,2 |
| 1966 | 254,427 | 170,871 | 249,230 | 47,550 | 0,123 | 59793 | 3,3 |
| 1967 | 287,338 | 187,240 | 252,396 | 55,320 | 0,118 | 59948 | 1,9 |
| 1968 | 330,941 | 208,755 | 272,254 | 59,870 | 0,131 | 60463 | 1,6 |
| 1969 | 374,232 | 235,683 | 304,774 | 60,280 | 0,138 | 61195 | 1,8 |
| 1970 | 417,525 | 261,932 | 360,062 | 64,370 | 0,147 | 61001 | 3,6 |
| 1971 | 471,972 | 297,728 | 399,626 | 71,790 | 0,144 | 61503 | 5,2 |
| 1972 | 541,310 | 343,159 | 435,719 | 79,810 | 0,153 | 61809 | 5,4 |
| 1973 | 601,733 | 380,392 | 485,290 | 85,800 | 0,147 | 62101 | 7,1 |
| 1974 | 662,341 | 410,709 | 525,211 | 98,370 | 0,156 | 61991 | 6,9 |
| 1975 | 742,411 | 451,085 | 550,175 | 131,100 | 0,162 | 61645 | 6 |
| 1976 | 814,827 | 500,337 | 596,479 | 151,700 | 0,145 | 61442 | 4,2 |
| 1977 | 906,577 | 549,459 | 635,583 | 167,960 | 0,132 | 61353 | 3,7 |
| 1978 | 1014,980 | 612,087 | 677,905 | 189,590 | 0,133 | 61322 | 2,75 |
| 1979 | 1110,888 | 684,107 | 736,256 | 211,620 | 0,139 | 61439 | 4,1 |
| 1980 | 1200,339 | 746,360 | 787,307 | 239,590 | 0,131 | 61658 | 5,4 |
| 1981 | 1295,882 | 812,753 | 824,523 | 278,960 | 0,135 | 61713 | 6,3 |

| Year | | | | | | | |
|------|---------|---------|---------|---------|-------|-------|------|
| 1982 | 1383,304 | 866,870 | 858,880 | 314,340 | 0,129 | 61546 | 5,2 |
| 1983 | 1470,925 | 923,241 | 896,913 | 343,440 | 0,113 | 61307 | 3,2 |
| 1984 | 1576,195 | 977,931 | 940,561 | 366,850 | 0,118 | 61049 | 2,5 |
| 1985 | 1699,219 | 1041,329 | 982,883 | 388,690 | 0,117 | 61020 | 2 |
| 1986 | 1812,879 | 1079,506 | 1035,567 | 409,550 | 0,125 | 61140 | -0,1 |
| 1987 | 1913,793 | 1123,253 | 1063,492 | 433,990 | 0,127 | 61238 | 0,2 |
| 1988 | 2033,947 | 1190,694 | 1121,588 | 461,700 | 0,131 | 61715 | 1,2 |
| 1989 | 2183,621 | 1260,982 | 1198,675 | 474,890 | 0,125 | 62679 | 2,8 |
| 1990 | 2677,021 | 1467,730 | 1304,656 | 538,650 | 0,137 | 63726 | 2,6 |
| 1991 | 2845,313 | 1606,547 | 1532,251 | 600,210 | 0,129 | 80275 | 3,7 |
| 1992 | 3037,938 | 1775,660 | 1645,875 | 687,790 | 0,127 | 80975 | 5,1 |
| 1993 | 3365,366 | 1953,407 | 1694,302 | 770,210 | 0,121 | 81338 | 4,4 |
| 1994 | 3549,477 | 2112,059 | 1779,470 | 848,540 | 0,114 | 81539 | 2,8 |
| 1995 | 3848,664 | 2265,083 | 1845,668 | 1019,210 | 0,110 | 81818 | 1,8 |
| 1996 | 4233,336 | 2436,688 | 1872,128 | 1087,170 | 0,105 | 82012 | 1,4 |
| 1997 | 4650,663 | 2582,357 | 1909,671 | 1134,660 | 0,101 | 82057 | 1,9 |
| 1998 | 5149,833 | 2746,418 | 1956,699 | 1165,800 | 0,101 | 82037 | 1 |
| 1999 | 5740,741 | 2904,491 | 2000,200 | 1199,800 | 0,095 | 82164 | 0,6 |
| 2000 | 6148,318 | 3003,705 | 2047,500 | 1211,400 | 0,092 | 82260 | 1,4 |
| 2001 | 6386,110 | 3014,114 | 2101,900 | 1223,930 | 0,094 | 82440 | 1,9 |
| 2002 | 6452,299 | 2997,225 | 2132,200 | 1277,630 | 0,099 | 82537 | 1,5 |
| 2003 | 6470,882 | 2995,622 | 2147,500 | 1358,120 | 0,103 | 82532 | 1 |
| 2004 | 6663,797 | 3001,257 | 2195,700 | 1430,100 | 0,104 | 82501 | 1,7 |
| 2005 | 6903,169 | 2995,100 | 2224,400 | 1488,300 | 0,105 | 82438 | 1,5 |
| 2006 | 7187,714 | 3000,706 | 2313,900 | 1532,600 | 0,106 | 82315 | 1,6 |
| 2007 | 7625,737 | 2975,731 | 2428,500 | 1539,500 | 0,108 | 82218 | 2,3 |
| 2008 | 7956,390 | 3071,088 | 2473,800 | 1644,500 | 0,117 | 82002 | 1 |
| 2009 | 7509,829 | 3100,099 | 2374,500 | 1679,750 | 0,111 | 81802 | 0,4 |
| 2010 | 8352,276 | 3220,947 | 2496,200 | 1715,000 | 0,114 | 81752 | 1,1 |
| 2011 | 8466,671 | 3197,757 | 2592,600 | 1870,000 | 0,104 | 81844 | 2,3 |
| 2012 | 8314,596 | 3220,356 | 2645,000 | 2067,000 | 0,103 | 82000 | 2 |

Note: german data notation is comma (,) instead of point(.) and vica versa to US. Please change if needed. E.g. the first number 19,996 thus means in US notation 19.966 ergo 19 bn. 966 mio. Euro.

Note: savings quota in natural notion here e.g. 0,099 = 9,9% ; inflation row I is in %, use 2,3%=0,023 in calculations,.

Note: the data row B to row I are the usefully needed data; row from K on are derived data.

Note: For basis model one needs row C,D,E,G,H

Note: No data about "foreign paid assets" which could be subtracted if available for more exact calculations

Sources:

B:http://www.bundesbank.de/Navigation/DE/Statistiken/Zeitreihen_Datenbanken/Makrooekonomische_Zeitreihen/its_details_value_node.html?tsId=BBK01.OU0308

C:http://www.bundesbank.de/Navigation/DE/Statistiken/Zeitreihen_Datenbanken/Makrooekonomische_Zeitreihen/its_details_value_node.html?tsId=BBK01.OU0115

D: http://www.destatis.de/jetspeed/portal/cms/Sites/destatis/Internet/DE/Content/Publikationen/Fachveroeffentlichungen/VolkswirtschaftlicheGesamtrechnungen/Inlandsprodukt/InlandsproduktsberechnungVorlaeufig2180140108004,property=file.pdf

E: http://www.miprox.de/Wirtschaft_allgemein/BRD-Oeffentliche-Verschuldung2003.htm und http://www.steuerzahler.de/Verschuldung/1233b477/index.html

F: http://www.bundesbank.de/Navigation/DE/Statistiken/Zeitreihen_Datenbanken/Makrooekonomische_Zeitreihen/its_details_value_node.html?tsId=BBK01.JJA327

G: https://www.destatis.de/DE/ZahlenFakten/Indikatoren/LangeReihen/Bevoelkerung/lrbev03.html

H: http://www.inflationsrate.com and www.destatis.de